\documentclass[a4paper,fleqn,usenatbib]{mnras}

\usepackage[english]{babel}
\usepackage{graphicx}
\usepackage{fleqn}
\usepackage{amssymb}
\usepackage{amsmath}
\usepackage{booktabs}
\usepackage{threeparttable}
\usepackage{enumerate}
\usepackage{enumitem}
\usepackage{multirow}
\usepackage{morefloats}
\usepackage{comment}
\usepackage{bm}
\usepackage{txfonts} 

\usepackage{pifont}

\renewcommand{\S}{Section}

\newcommand{\F}{Fig.}
\newcommand{\Fs}{Figs}
\newcommand{\Eq}{Equation}
\newcommand{\eq}{Equation}

\newcommand{\ve}[1]{\mathbf{#1}}
\newcommand{\unit}[1]{\hat{\mathbf{#1}}}

\newcommand{\msun}{\mathrm{M}_\odot}

\newcommand{\au}{\textsc{au}}

\newcommand{\yr}{\mathrm{yr}}
\newcommand{\gyr}{\mathrm{Gyr}}
\newcommand{\hz}{\mathrm{Hz}}

\newcommand{\init}{\mathrm{init}}

\newcommand{\pd}{\mathrm{p}}
\newcommand{\sd}{\mathrm{s}}

\newcommand{\pgpc}{\mathrm{Gpc^{-3}}}
\newcommand{\pyr}{\mathrm{yr^{-1}}}

\newcommand{\AppICs}{Supplementary Appendix A}
\newcommand{\AppGenOne}{Supplementary Appendix B}
\newcommand{\AppGenTwo}{Supplementary Appendix C}

\newcommand{\eff}{\mathrm{eff}}
\newcommand{\ligo}{\mathrm{LIGO}}
\newcommand{\lisa}{\mathrm{LISA}}

\newcommand{\bse}{\textsc{BSE}}

\begin{document}

\title[Repeated BH and NS mergers in quadruples]{First and second-generation black hole and neutron star mergers in 2+2 quadruples: population statistics}
\author[Hamers et al.]{Adrian S. Hamers$^{1}$\thanks{E-mail: hamers@mpa-garching.mpg.de}, Giacomo Fragione$^{2,3}$, Patrick Neunteufel$^{1}$, and Bence Kocsis$^{4}$\\
$^{1}$Max-Planck-Institut f\"{u}r Astrophysik, Karl-Schwarzschild-Str. 1, 85741 Garching, Germany \\
$^2$Center for Interdisciplinary Exploration \& Research in Astrophysics (CIERA), Evanston, IL 60202, USA\\
$^3$Department of Physics \& Astronomy, Northwestern University, Evanston, IL 60202, USA\\
$^4$Rudolf Peierls Centre for Theoretical Physics, Clarendon Laboratory, Parks Road, Oxford OX1 3PU, UK}

\date{Accepted 2021 July 20. Received 2021 July 03; in original form 2021 March 05}

\label{firstpage}
\pagerange{\pageref{firstpage}--\pageref{lastpage}}
\maketitle

\begin{abstract} 
Recent detections of gravitational waves from mergers of neutron stars (NSs) and black holes (BHs) in the low and high-end mass gap regimes pose a puzzle to standard stellar and binary evolution theory. Mass-gap mergers may originate from successive mergers in hierarchical systems such as quadruples. Here, we consider repeated mergers of NSs and BHs in stellar 2+2 quadruple systems, in which secular evolution can accelerate the merger of one of the inner binaries. Subsequently, the merger remnant may interact with the companion binary, yielding a second-generation merger. We model the initial stellar and binary evolution of the inner binaries as isolated systems. In the case of successful compact object formation, we subsequently follow the secular dynamical evolution of the quadruple system. When a merger occurs, we take into account merger recoil, and model subsequent evolution using direct $N$-body integration. With different assumptions on the initial properties, we find that the majority of first-generation mergers are not much affected by secular evolution, with their observational properties mostly consistent with isolated binaries. A small subset shows imprints of secular evolution through residual eccentricity in the LIGO band, and retrograde spin-orbit orientations. Second-generation mergers are $\sim 10^7$ times less common than first-generation mergers, and can be strongly affected by scattering (i.e., three-body interactions) induced by the first-generation merger. In particular, scattering can account for mergers within the low-end mass gap, although not the high-end mass gap. Also, in a few cases, scattering could explain highly eccentric LIGO sources and negative effective spin parameters.
\end{abstract}


\section{Introduction}
\label{sect:introduction}
Massive stars (birth masses $\gtrsim 8\,\msun$) are the progenitors of neutron stars (NSs) and black holes (BHs). Recent detections of gravitational waves (GWs) from merging BHs and NSs by LIGO and Virgo (e.g., \citealt{2016PhRvL.116x1103A,2016PhRvL.116f1102A,2017PhRvL.118v1101A,2017ApJ...851L..35A,2017PhRvL.119n1101A,2017ApJ...848L..12A,LIGO_O3_Catalog,LIGO_O3_rates}) have spiked interest in the evolution of these events' progenitor systems. As the individual progenitor stars of these compact objects reach physically large radii, the inefficiency of angular momentum loss through GW radiation implies a merger time exceeding the Hubble time in many cases, or a pre-compact object merger in others. Any proposed progenitor scenario must therefore provide a mechanism to place the merging binary in a closer configuration than undisturbed, isolated binary evolution predicts.

A number of channels have been proposed, which can be divided into three main channels, each of which have subchannels. One main channel involves interactions such as mass transfer and common-envelope (CE) evolution in isolated field binaries \citep[e.g.,][]{1973NInfo..27...70T,1992ApJ...386..197T,1993MNRAS.260..675T,1997AstL...23..492L,2002ApJ...572..407B,2003MNRAS.342.1169V,2007PhR...442...75K,2012ApJ...759...52D,2013ApJ...779...72D,2014ApJ...789..120B,2016Natur.534..512B,2017arXiv170607053B,2017NatCo...814906S,2018MNRAS.474.2937C,2019MNRAS.482.2234G,2019MNRAS.490.3740N,2020arXiv201016333B}. A second channel involves dynamical interactions; a subset of which relies on secular interactions in triple-star systems that excite orbital eccentricities \citep[e.g.,][]{2011ApJ...741...82T,2013MNRAS.430.2262H,2017ApJ...841...77A,2017ApJ...836...39S,2017ApJ...846L..11L,2018ApJ...863...68L,2018arXiv180506458A,2018A&A...610A..22T,2019ApJ...881...41L,FragioneLoeb2019,FragioneKocsis2020}. Another subset of the dynamical channel involves strong gravitational interactions in dense stellar systems such as globular clusters \citep[e.g.,][]{1993Natur.364..423S,2000ApJ...528L..17P,2006ApJ...637..937O,2014MNRAS.441.3703Z,2015PhRvL.115e1101R,2016PhRvD..93h4029R,2016MNRAS.463.2443K,2016MNRAS.459.3432M,2017ApJ...840L..14S,2018ApJ...855..124S,2018ApJ...853..140S,FragioneKocsis2018,2018PhRvD..97j3014S,2018PhRvL.120o1101R,FragioneLoeb2020,MartinezFragione2020,2021MNRAS.502.3879F}, and galactic nuclei which could contain supermassive BHs \citep[e.g.,][]{2012ApJ...757...27A,2014ApJ...781...45A,2015ApJ...799..118P,2016MNRAS.460.3494S,2016ApJ...828...77V,2016ApJ...831..187A,2017ApJ...846..146P,2018MNRAS.477.4423A,2018ApJ...864..134R,2018ApJ...853...93R,2018ApJ...865....2H,2018ApJ...860....5G,2018ApJ...856..140H,2019MNRAS.488...47F,2019MNRAS.483..152A,2020ApJ...897...46F}. Lastly, BH and NS mergers could be gas assisted in active galactic nuclei which have large-scale accretion disks around a central supermassive BH \citep[e.g.,][]{2017ApJ...835..165B,2020MNRAS.494.1203M,2020ApJ...898...25T,2020arXiv201009765S,2021ApJ...907L..20T}.

A variant of the dynamical channel which has received recent attention involves hierarchical quadruple systems. Triple systems are known for the possibility of the inner orbit to become highly excited in eccentricity due to a gravitational dynamical phenomenon known as von Zeipel-Lidov-Kozai (ZLK) oscillations (\citealt{1910AN....183..345V,1962P&SS....9..719L,1962AJ.....67..591K}; see \citealt{2016ARA&A..54..441N,2017ASSL..441.....S,2019MEEP....7....1I} for reviews). However, it has been shown that hierarchical quadruple systems (either in the `2+2' configuration -- two binaries orbiting each other -- or `3+1' configuration -- a triple orbited by a fourth body) can give rise to similar but also more complicated behaviour, and that eccentricity excitation in such systems can be more efficient compared to triples \citep{2013MNRAS.435..943P,2015MNRAS.449.4221H,2016MNRAS.461.3964V,2017MNRAS.470.1657H,2018MNRAS.476.4234F,2018MNRAS.474.3547G,2019MNRAS.483.4060L,2019MNRAS.486.4781F}. This is highly important for BH and NS mergers, not only because more efficient eccentricity excitation (leading to compact object mergers) occurs for a larger parameter space in quadruples, but also since massive stars (in the field, and with primary star masses $\gtrsim 20 \, \msun$) tend to occur in high-multiplicity systems such as triples and quadruples. For example, \citet{2017ApJS..230...15M} find that, among O-type stars, the multiplicity fraction is $(35 \pm 3)\%$ for triple stars, and $(38\pm11)\%$ for quadruple stars, showing that triples and quadruples significantly outnumber both single and binary stars among massive stellar systems. Quadruple-star systems can also be of relevance for other astrophysical phenomena, such as producing short-period binaries \citep{2019MNRAS.482.2262H} and Type Ia Supernovae as a result of mergers of white dwarfs \citep{2018MNRAS.476.4234F,2018MNRAS.478..620H}. 

In the context of NS and BH mergers, quadruple systems are not only interesting in light of their ubiquity among massive stellar systems and because of their propensity to induce coalescence of compact objects through secular evolution, but also in light of recent LIGO/Virgo detections of compact objects in the `lower mass gap' range between $\sim 2$-$5\,\msun$ (e.g., \citealt{2011ApJ...741..103F}) such as GW190814 \citep{2020ApJ...896L..44A}, mergers with a high mass ratio such as GW190412 \citep{2020arXiv200408342T}, and mergers involving unusually massive BHs in the electron-positron pair-instability mass gap (GW190521; \citealt{2020PhRvL.125j1102A}). Mass gaps in the BH mass function are predicted by current stellar evolution models. The high-mass gap results from pulsational pair-instabilities affecting the massive progenitors, which leads to large amounts of mass being ejected whenever the pre-explosion stellar core is approximately in the range 45-65 $\msun$, leaving a BH remnant with a maximum mass around 40-50 $\msun$ \citep{heger2003,woosley2017}. The low-mass gap is related to the explosion mechanism in a core-collapse supernova \citep[SN; see][]{belc2012,2012ApJ...749...91F}. 

In 3+1 quadruple systems, mass-gap mergers could arise if two NSs in the innermost orbit merge due to secular evolution forming a low-mass-gap object, and the resulting object (now in a triple system) merges due to continued secular evolution with another BH in the system \citep{2020ApJ...888L...3S}. The mass ratio and spin properties of GW190412 might be explained by a similar scenario in 3+1 systems \citep{2020ApJ...898...99H}. Other ways to produce mergers in the mass gaps is through repeated mergers in star clusters or AGN disks \citep[e.g.,][]{AntoniniGieles2019,BaibhavGerosa2020,2020ApJ...895L..15F,FragioneSilk2020,MapelliSantoliquido2020,YangGayathri2020,TagawaKocsis2021}.

\begin{figure*}
\center
\includegraphics[scale = 0.7, trim = 0mm 0mm 0mm 0mm]{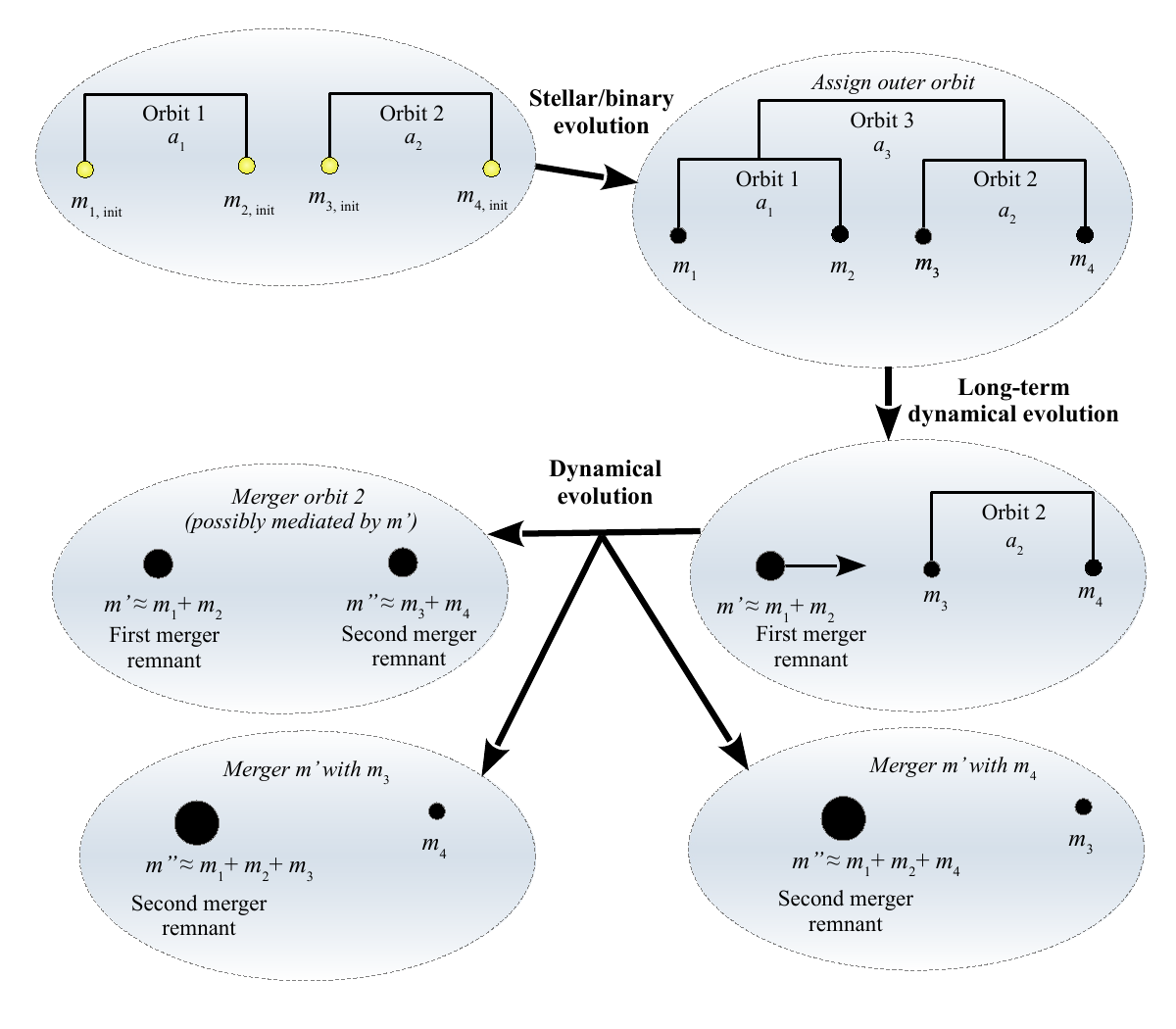}
\caption{A sketch of the scenario considered in this paper. A 2+2 stellar quadruple is assumed to evolve successfully to a 2+2 compact object quadruple (a strong assumption). After a first merger induced by secular evolution, the remnant object receives a recoil velocity into the direction of the companion binary. Subsequently, the first merger remnant could perturb the companion binary and accelerate its coalescence. Alternatively, the remnant object could merge directly with one of the objects in the companion binary. }
\label{fig:cartoon}
\end{figure*}

Quadruple systems of the other type (2+2) may also play a role in explaining mass-gap mergers or the spin properties of merging compact objects. Recently, \citet{2020ApJ...895L..15F} explored the possibility that the merger remnant of an inner binary in a 2+2 quadruple receives a recoil kick from the anisotropic emission of GWs and, coincidentally, in the direction of its companion binary, triggering a dynamical interaction in which another (second-generation) merger could occur. Several possibilities for mergers exist in this scenario, sketched in \F~\ref{fig:cartoon}. After a first merger induced by secular evolution in the 2+2 quadruple system, the remnant object could perturb the companion binary and accelerate its coalescence. Alternatively, the remnant object may merge directly with one of the objects in the companion binary. 

In \citet{2020ApJ...895L..15F}, a number of numerical examples were considered of the remnant compact object being propelled into the general direction of the companion binary (as a consequence of GW merger recoil). Here, we provide a more systematic study in which we include stellar and binary evolution prior to compact object formation, long-term gravitational dynamical evolution after compact object formation, and a shorter phase of gravitational interaction after a first merger occurred in one of the inner binaries. Also, we include in our modelling the spin evolution of the compact objects, which allows us to make predictions for spin-related properties in this channel. Although this work is more extensive compared to \citet{2020ApJ...895L..15F} in the sense that we include the evolution of the system before the first merger, we still make approximations by splitting the evolution in distinct phases in which stellar/binary evolution and dynamical evolution are assumed to be decoupled. This will be amended in future work. 

This paper is structured as follows. We discuss our methodology in \S~\ref{sect:meth}. In \S~\ref{sect:first}, we present the results pertaining to the first-generation mergers. We focus on second-generation mergers in \S~\ref{sect:second}. We discuss our results in \S~\ref{sect:discussion}, and conclude in \S~\ref{sect:conclusions}.

\section{Methods}
\label{sect:meth}
Our methodology is split into three stages: first, we model the relatively short-term (order 10-100 Myr) evolution of the two inner binaries of quadruple systems with massive components using a rapid binary population synthesis code, and track the systems in which two (bound) binaries are formed, each consisting of NSs and/or BHs. Subsequently, we integrate the long-term evolution (order 10 Gyr) using a secular dynamics approach. Lastly, for the systems that experience a merger in one of the two inner binaries, we take into account the GW merger recoil, and integrate the subsequent evolution using a direct $N$-body code. 

\subsection{Stellar and binary evolution to binary compact objects}
\label{sect:meth:bse}

\begin{table}
\begin{tabular}{lcccc}
\toprule
Model & $\sigma$ (km s$^{-1}$) & $Z$ (Z$_\odot$) & $f(a)$ & $f(q)$ \\
\midrule
1 (ref.) & 100 & 1     & $1/a$     & const\\
2 (low $\sigma$) & 10  & 1     & $1/a$     & const\\
3 (high $\sigma$) & 250 & 1     & $1/a$     & const\\
4 (Sana) & 100 & 1     & Sana+ 2012& const\\
5 (low $Z$) & 100 & 0.01  & $1/a$     & const\\
6 (ind. $m$) & 100 & 1     & $1/a$     & -\\
\bottomrule
\end{tabular}
\caption{Model parameters: velocity dispersion in natal kick distribution ($\sigma$), metallicity ($Z$), distribution of semi-major axis ($f(a)$), and distribution of mass ratios ($f(q)$).}
\label{table:models}
\end{table}

The stellar quadruples in our simulations are initialized as described in what follows. In total, we consider six different models (see Table~\ref{table:models}).

In all our models, we sample the initial mass $m_{1,\,\init}$ of the most massive star in the first binary from a \citet{kroupa2001} initial mass function,
\begin{equation}
\frac{\mathrm{d}N}{\mathrm{d}m} \propto m^{-2.3},
\label{eqn:massfunc}
\end{equation}
in the mass range $8\,\msun$-$150\,\msun$, reflecting the progenitor of a NS or a BH. We adopt a flat mass ratio distribution to determine the initial secondary mass ($m_{2,\,\init}$) in the inner binary, and the initial total mass ($m_{3,\,\init}+m_{4,\,\init}$) of the second binary with respect to the total mass of the first binary. This is consistent with observations of massive binary stars, which suggest a nearly flat distribution of the mass ratio \citep{sana12,duch2013,sana2017}. In Model 6, we sample the masses of the four stars independently from each other, directly from \eq~(\ref{eqn:massfunc}).

The distribution of the inner and outer semi-major axes is assumed to be flat in log-space (\"{O}pik's law; \citealt{1924PTarO..25f...1O}), roughly consistent with the results of \citet{kob2014}. Since any common-envelope phase of evolution will likely shrink the inner orbit so that the ZLK mechanism will become strongly suppressed by the relativistic precession, we impose a minimum orbital separation of $10$\,AU as in \citet{2017ApJ...841...77A} to make sure that the periapsis distance of the inner binary is large enough that likely no common-envelope phase or mass transfer occurs. As a maximum separation, we adopt $10^4$\,AU. Different maximum separations are likely not to affect our results \citep[see, e.g.,][]{2017ApJ...841...77A,FragioneKocsis2020}. In Model 4, we sample the semimajor axis from the distribution of \citet{sana12}. For the orbital eccentricities of the inner binary and outer binary, we assume a thermal distribution \citep{1919MNRAS..79..408J}. The latter distributions are uncertain, especially for massive quadruple systems for which orbital statistics are poorly constrained. We remark that observations of massive MS binaries do indicate that eccentricities are on average less eccentric than thermal \citep[e.g.,][]{sana12,2013ARA&A..51..269D}; our choice for a thermal distribution is mainly motivated by simplicity. With fewer initially eccentric systems, the typical secular timescales of the quadruple system would be longer, potentially producing fewer compact object mergers within a Hubble time, and potentially reducing the number of highly eccentric merger systems (cf. \S~\ref{sect:first:eLIGO}). However, a we do not consider a quantitative investigation here, also in light of other large uncertainties such as natal kicks (see below).

The system may receive a natal kick due to recoil from an asymmetric supernova explosion when stars evolve to form a NS or BH. We assume that the velocity kick is drawn from a Maxwellian distribution,
\begin{equation}
p(v_\mathrm{kick})\propto v_\mathrm{kick}^2 \exp \left ( -\frac{v_\mathrm{kick}^2}{2\sigma^2} \right ),
\label{eqn:vkick}
\end{equation}
with a velocity dispersion $\sigma$. The distribution of natal kick velocities of BHs and NSs is unknown. We implement momentum-conserving kicks, in which we assume that the momentum imparted to a BH is the same as the momentum given to a NS. As a consequence, the kick velocities for the BHs will be reduced with respect to those of NSs by a factor of $m_{\rm BH}/m_{\rm NS}$. We consider a non-zero natal kick velocity for the newly formed BHs and NSs, by adopting \eq~(\ref{eqn:vkick}) with $\sigma=100$ km s$^{-1}$, consistent with the distribution of natal kicks found by \citet{arz2002}. We run Model 3 where we adopt $\sigma=250$ km s$^{-1}$, which is consistent with the distribution deduced by \citet{hobbs2005}. Finally, we consider Model 2 where $\sigma=10$ km s$^{-1}$ is imparted during formation. For NSs, this would be consistent with the electron-capture supernova process, where the newly-formed NSs would receive no kick at birth or only a very small one due to an asymmetry in neutrino emissions \citep{pod2004}.

Given the above set of initial conditions, we evolve the two binaries using the stellar evolution code \textsc{BSE} \citep{hurley2000,2002MNRAS.329..897H}. We use the updated version of \textsc{BSE} from \citet{banerjee2020}, with the most up-to-date prescriptions for stellar winds and remnant formation; it produces remnant populations similar to those from \textsc{StarTrack} \citep{belc2008}. Importantly, the current version includes prescriptions for the most recent theoretical results on pulsational pair instability that limit the maximum BH mass to $\sim 50\,\msun$ \citep{bel2016b}. We run BSE models until a binary (BH-BH, BH-NS, NS-NS) of compact objects is formed.

In the different models (Table~\ref{table:models}), we vary natal kick properties, metallicity, and the semimajor axis and mass ratio distributions. We remark that properties of compact object binaries in \textsc{BSE} are also strongly affected by other (uncertain) parameters such as the CE efficiency, and wind mass loss in massive stars. For computational reasons, we restrict to the parameters described in Table~\ref{table:models}.

\subsection{Secular dynamical evolution}
\label{sect:meth:sec}
The binary population synthesis calculations yield the remnant masses of the NSs and BHs ($m_1$, $m_2$, $m_3$, and $m_4$), as well as their final orbital properties (semimajor axes $a_1$ and $a_2$, and eccentricities $e_1$ and $e_2$). Subsequently, we assign an outer orbit to the inner two binaries. Here, we assume that the outer orbit has a semimajor axis distribution which is flat in $\log a_3$ (with $10^{-3}\,\au<a_3<10^5\,\au$) and a thermal eccentricity distribution, $\mathrm{d} N/\mathrm{d} e_3 \propto e_3$ (with $0.01<e_3<0.99$). These distributions are the same as those assumed for the initial inner binaries. The upper limit on $a_3$ is motivated by the fact that orbits wider than $\sim 10^5\,\au$ are marginally bound in the Galaxy and are affected strongly by fly-bys and Galactic tides, both of which are not included in our simulations. We assume that all orbital orientations are random, i.e., we sample $i_j$ (inclination), $\omega_j$ (argument of periapsis), and $\Omega_j$ (longitude of the ascending node) for each orbit $j$ from distributions which are flat in $\cos(i_j)$, $\omega_j$, and $\Omega_j$, respectively. 

We reject a combination of $a_3$ and $e_3$ and orbital orientations $(i_j,\omega_j,\Omega_j)$ if the quadruple system would be dynamically unstable, or if the ZLK time-scale would be shorter than the one of the MS time-scales in the system. To assess stability, we use the stability criterion of \citet{2001MNRAS.321..398M}, i.e., 
\begin{align}
\label{eq:ma01}
\frac{a_3(1-e_3)}{a_\mathrm{in}} > 2.8 \, \left [ (1+q_\mathrm{out}) \frac{1+e_3}{\sqrt{1-e_3}} \right ]^{2/5} \, \left (1-0.3\, \frac{\Phi}{\pi} \right ),
\end{align}
for both both orbit pairs (1,3) and (2,3). Here, $a_\mathrm{in}$ is either $a_1$ or $a_2$, the mass ratio $q_\mathrm{out}$ is defined as $q_\mathrm{out}\equiv (m_3+m_4)/(m_1+m_2)$ for the (1,3) pair and $q_\mathrm{out}\equiv (m_1+m_2)/(m_3+m_4)$ for the (2,3) pair, and $\Phi$ is the mutual inclination angle between the inner and outer orbits (expressed in radians). To ensure that the lifetime of the stars is at least shorter than ZLK time-scale, we also require that
\begin{align}
t_\mathrm{MS} < t_\mathrm{ZLK},
\end{align}
where
\begin{align}
t_\mathrm{MS} = \mathrm{min}_i \left [ t_{\mathrm{MS},\,i} \right ] = \mathrm{min}_i \left [ 10^{10} \, \yr \, \left ( \frac{m_{i,\,\mathrm{init}}}{\msun} \right )^{-2.8} \right ],
\end{align}
and where the shortest ZLK time-scale is estimated as (e.g,. \citealt{2015MNRAS.452.3610A})
\begin{align}
\label{eq:tZLK}
\nonumber t_\mathrm{ZLK} &= \mathrm{min} \left [ t_\mathrm{ZLK,\,13}, t_\mathrm{ZLK,\,23} \right ] \\
& =  \left (1-e_3^2 \right )^{-3/2} \mathrm{min} \left [ \frac{P_3^2}{P_1} \frac{M_3}{M_2},\frac{P_3^2}{P_2} \frac{M_3}{M_1}  \right ].
\end{align}
Here, $M_1 \equiv m_1+m_2$, $M_2 \equiv m_3+m_4$ and $M_3\equiv M_1+M_2$, and $P_j = 2\pi \sqrt{a_j^3/(GM_j)}$ denotes the Keplerian period of orbit $j$. 

Our approach to assign an outer orbit {\it after} the formation of two compact object binaries is certainly a simplification. We do reject systems in which the {\it final} outer orbit would be relatively tight and could cause short-term ZLK oscillations and possibly a merger before the formation of two compact object binaries. However, in this approach, we neglect changes in the outer orbit during the {\it pre-compact object phase} due to stellar/binary evolution (most notably, mass loss due to stellar winds, CE evolution, and/or mass loss and kicks from SNe events). The latter processes, especially SNe kicks, will at least affect the distributions of $a_3$ and $e_3$ (and the orbital orientations as well). In the worst case, the outer orbit of the quadruple will become unbound due to the effects of SNe, implying an effectively lower formation rate of 2+2 quadruple systems with BHs and/or NSs. The motivation for our choice is that the {\it initial} distribution of $a_3$ is also uncertain, and the computational tools used here do not allow for a more self-consistent treatment in which stellar, binary, and dynamical evolution are all included simultaneously. We defer such a latter treatment to future work. 

\begin{figure}
\center
\includegraphics[scale = 0.45, trim = 5mm 25mm 0mm 30mm]{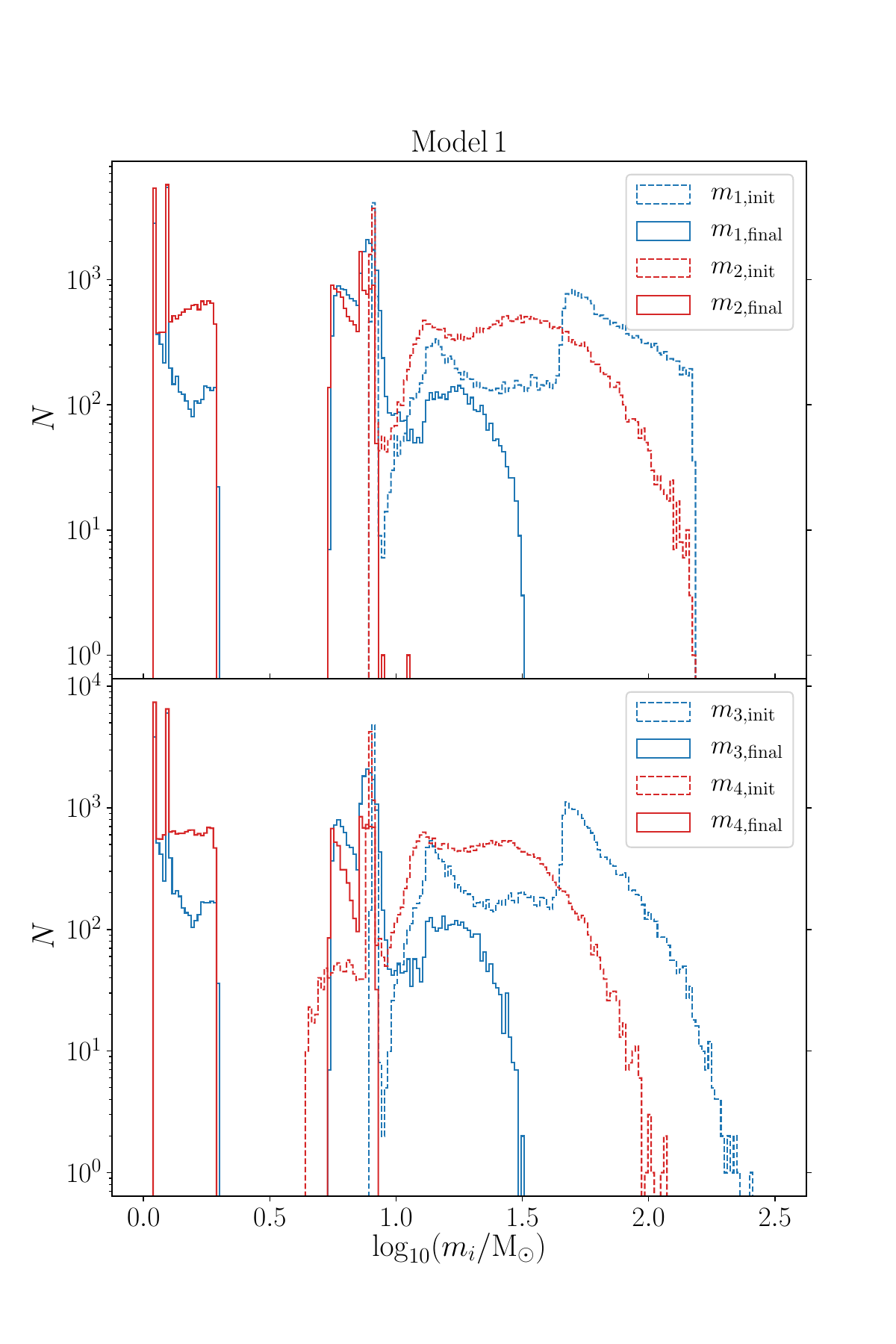}
\caption{ Distributions of the initial and final masses, where `final' refers to after evolution with \bse. The top and bottom panels correspond to the primary and secondary binary, respectively. These data apply to Model 1; data for other models are available in \AppICs. }
\label{fig:ics_masses_m1}
\end{figure}

\begin{figure}
\center
\includegraphics[scale = 0.45, trim = 5mm 25mm 0mm 30mm]{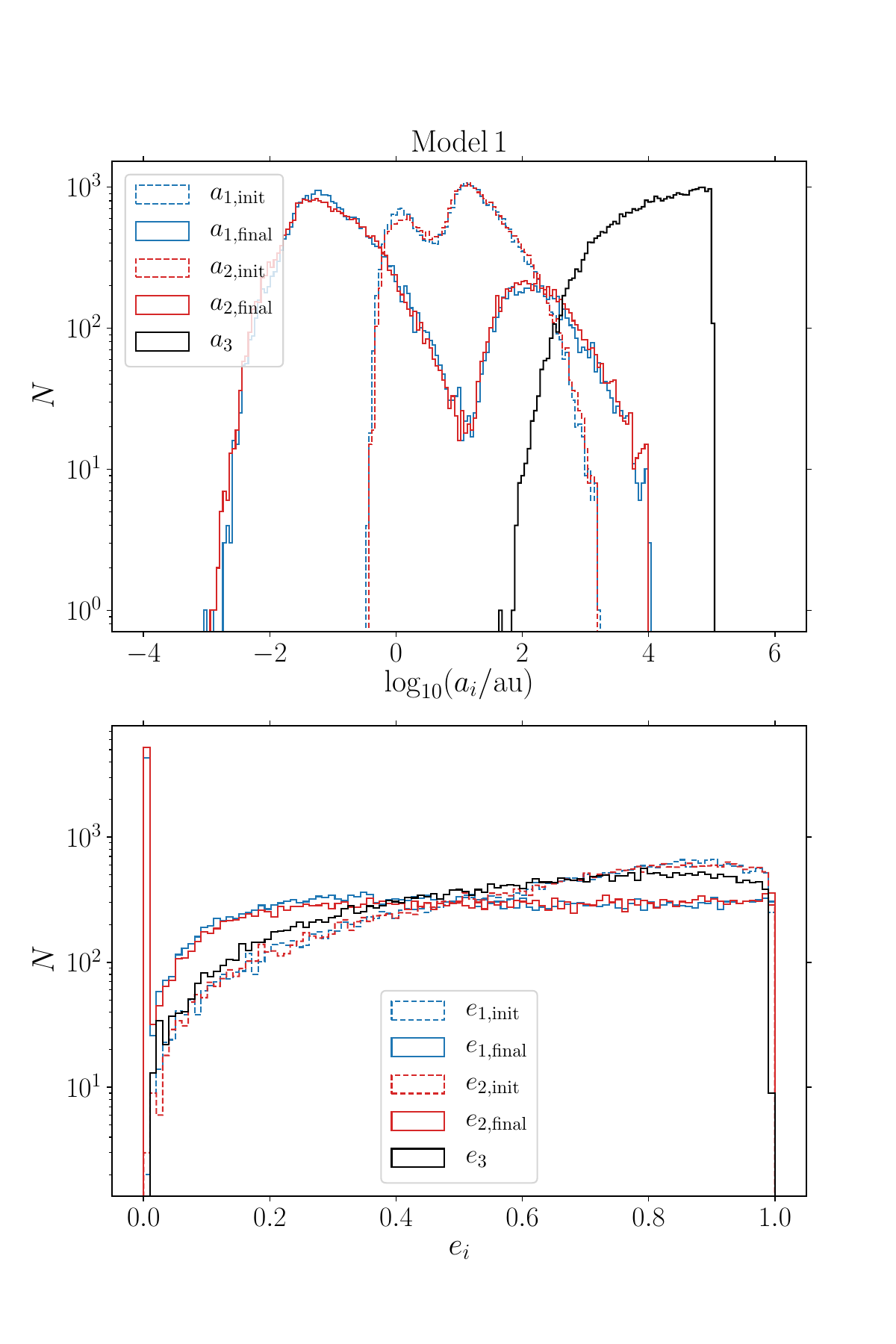}
\caption{ Distributions of the initial and final semimajor axes (top panel) and eccentricities (bottom panel), where `final' refers to after evolution with \bse. These data apply to Model 1; data for other models are available in \AppICs. Note that the peaks near $e_i=0$ in the bottom panel correspond to the final eccentricities; the initial eccentricities do not show such a peak.}
\label{fig:ics_orbits_m1}
\end{figure}

For the systems after compact object formation (as modelled with \bse), we show in \Fs~\ref{fig:ics_masses_m1} and \ref{fig:ics_orbits_m1} the distributions of the masses and orbital semimajor axes and eccentricities, respectively. Here, we present data from Model 1; data from other models are shown in \AppICs. In all of the models, the distributions of the final masses clearly show the `mass gap' between the most massive NSs ($\approx 2 \, \msun$ according to \bse), and the lowest-mass BHs ($\approx 4\,\msun$ according to \bse). The details of the final orbital distributions depend on the model. Generally speaking, however, $a_1$ and $a_2$ are broadly distributed between $\sim 10^{-3}$ and $10^4\,\au$ with some bimodality introduced by CE evolution (which significantly shrinks the orbit if it occurs). Furthermore, the outer orbit has a typical range of $10^2\,\au\lesssim a_3 \lesssim 10^5\,\au$ (the upper limit on $a_3$ is an artefact of our assumed range). The eccentricity distributions are generally reasonably close to the initially assumed distribution (thermal), but with a significant population of circular orbits for the inner orbits due to tidal evolution and/or CE evolution. The outer orbital eccentricity distribution shows a slight lack of very eccentric orbits due to dynamical stability requirements.

We integrate the subsequent dynamical evolution of the 2+2 compact object quadruple using the \textsc{SecularMultiple} code\footnote{The \textsc{SecularMultiple} code is freely available at \href{https://github.com/hamers/secularmultiple}{https://github.com/hamers/secularmultiple}.}  \citep{2016MNRAS.459.2827H,2018MNRAS.476.4139H,2020MNRAS.494.5492H}. This code (which applies to 2+2 quadruples but is also more general) is based on an expansion of the Hamiltonian in terms terms of ratios of adjacent orbital separations, and subsequent orbit averaging. It is significantly faster than direct summation $N$-body codes. However, there can be cases when the secular approximation breaks down; most notably, when the time-scale for angular-momentum change due to secular evolution is shorter than the orbital time-scale \citep{2012ApJ...757...27A,2014ApJ...781...45A,2016MNRAS.458.3060L,2018MNRAS.481.4907G,2018MNRAS.481.4602L,2019MNRAS.490.4756L}. Here, for computational reasons, we neglect this complication during the long-term evolution of the compact object quadruple. However, we take into account detailed $N$-body dynamics after a merger occurs in the system, possibly triggering a dynamical interaction with more subsequent mergers (see \S~\ref{sect:meth:nbody}). 

In \textsc{SecularMultiple}, we include expansion terms up to and including fifth order (dotriacontupole) for pairwise orbital interactions. We also include pairwise post-Newtonian (PN) terms of the 1PN and 2.5PN order (the former describe relativistic apsidal motion, and the latter describe the orbital effects of GW emission; \citealt{1964PhRv..136.1224P}).  We ignore PN `interaction' terms between different orbits (e.g., \citealt{2013ApJ...773..187N,2014CQGra..31x4001W,2020PhRvD.102f4033L}). Other included PN terms are the lowest-order spin-orbit terms \citep{1975PhRvD..12..329B},
\begin{align}
\label{eq:so}
    \frac{\mathrm{d}\unit{S}_i}{\mathrm{d} t } = \frac{2G }{c^2 a_j^3 \left(1-e_j^2\right)^{3/2}} \left ( 1 + \frac{3}{4} \frac{m_{i,\,\mathrm{comp}}}{m_i} \right ) \ve{L}_j \times \unit{S}_i.
\end{align}
Here, $\unit{S}_i$ is the spin-angular-momentum vector of body $i$, and $m_{i,\,\mathrm{comp}}$ is the mass of the companion object to body $i$ in orbit $j$. Furthermore, $L_j =\mu_j \sqrt{G M_j a_j (1-e_j^2)}$, where $\mu_j$ is the reduced mass, is the orbital momentum of orbit $j$. Note that the rate of precession of $\unit{S}$ does not depend on the magnitude of the spins, only on the spin-orbit misalignment angle (and orbital properties). Due to PN spin-orbit coupling, the orbit $\ve{L}_j$ also precesses around the spins; however, the latter effect is negligible if $S_i \ll L_j$, and this is satisfied in our case in which the bodies are of similar mass (see, e.g., \citealt{2020PhRvD.102b3020L} for a different situation with a high mass ratio). We ignore general relativistic spin-spin coupling, since this is generally only important during the last stages before inspiral. The initial spin directions $\unit{S}_i$ are chosen to lie within a $10^\circ$ cone of the orbital angular momenta $\ve{L}_j$, since -- in isolated binaries -- tidal evolution is generally expected to fully or nearly align the spins with the orbit (e.g., \citealt{1980A&A....92..167H}).

We integrate each system for a duration of $14\,\gyr$. Some of the systems are highly hierarchical and are therefore `decoupled', in the sense that the secular time-scale is much longer than the time-scale for apsidal motion due to general relativity. As is well known, the latter short-range force can reduce the efficiency of secular oscillations in triples and quadruples (e.g., \citealt{2003ApJ...589..605W,2007ApJ...669.1298F,2013ApJ...773..187N,2015MNRAS.447..747L,2017MNRAS.470.1657H}). If, for an inner orbit $j$, the time-scale for the lowest-order (1PN) relativistic apsidal motion, 
\begin{align}
\label{eq:tpn}
t_{\mathrm{1PN},\,j} = \frac{1}{3} P_j \frac{a_j c^2}{GM_j} \left (1-e_j^2 \right ),
\end{align}
is much shorter than the secular evolution time-scale of the system, then we consider the binary to be `decoupled'. The binary could still merge within a Hubble time due to isolated GW-driven evolution, but its coalescence will not be accelerated by secular evolution. 

To avoid unnecessary computation, we therefore do not integrate systems for which $t_{\mathrm{1PN},\,1} < f_\mathrm{PN} \, t_{\mathrm{ZLK},\,13}$ {\it and}  $t_{\mathrm{1PN},\,2} < \, f_\mathrm{PN}\, t_{\mathrm{ZLK},\,23}$ (i.e., both inner orbits must be decoupled). Here, $f_\mathrm{PN}=10^{-3}$ is a `safety' factor which ensures that the systems that we do not integrate are clearly decoupled. For these systems, we compute the merger time-scale due to GW emission as if the inner binaries were isolated \citep{1964PhRv..136.1224P}, i.e., for orbit an orbit $j$ with components labeled `$i$' and `$i,\,\mathrm{comp}$',
\begin{align}
\label{eq:tGW}
t_{\mathrm{GW},\,j} = \frac{12}{19} \frac{c_{0,\,j}^4}{\beta_j} \int_{0}^{e_j} \mathrm{d} e \, \frac{e^{29/19}}{\left (1-e^2 \right )^{3/2}} \left (1 + \frac{121}{304} e^2 \right )^{1181/2299},
\end{align}
where 
\begin{align}
\beta_j \equiv \frac{64}{5} \frac{G^3 m_i m_{i,\,\mathrm{comp}} M_j}{c^5},
\end{align}
and
\begin{align}
c_{0,\,j} \equiv \frac{a_j \left(1-e_j^2 \right )}{e_j^{12/19} \left (1+\frac{121}{304} e_j^2 \right )^{870/2299}}.
\end{align}
We label a merger as `decoupled' if $t_{\mathrm{1PN},\,1} < f_\mathrm{PN} \, t_{\mathrm{ZLK},\,13}$  and $t_{\mathrm{1PN},\,2} < \, f_\mathrm{PN}\, t_{\mathrm{ZLK},\,23}$, and $t_{\mathrm{GW},\,j} < 14\,\gyr$. 

For the systems that are integrated for $14\,\gyr$, we check during the integration for any mergers, i.e., if at any time $r_{\mathrm{p},\,j} \equiv a_j (1-e_j) < r_{\mathrm{col},\,j}$. Here, we set the `collision' distance to be a relatively large multiple of the binary's gravitational radius, i.e.,
\begin{align}
\label{eq:rcol}
r_{\mathrm{col},\,j} = 1000 \frac{G M_j}{c^2}.
\end{align}
Due to the factor 1000 in \eq~(\ref{eq:rcol}), the integration is terminated before actual merger. It is included for numerical reasons, since the 1PN apsidal motion rate (cf. \eq~\ref{eq:tpn}) diverges as $r_{\mathrm{p},\,j} \rightarrow 0$ and by this time the inner orbit can be considered to be completely decoupled from the secular evolution (also, note that the effective $\chi$ parameter, $\chi_\eff$, is no longer affected within our approximations, cf. \eq~\ref{eq:chi_eff}). We note, however, that the physical remaining merger time is very short. For example, if the orbit is circular with $a_j = \alpha \, G M_j/c^2$, the remaining merger time is \citep{1964PhRv..136.1224P}
\begin{align}
\nonumber t_{\mathrm{GW},\,j}  &= \frac{5}{256} \alpha^4 \frac{M_j^2}{m_i m_{i,\,\mathrm{comp}}} \frac{GM_j}{c^3} \\
&\simeq 167 \,\yr \, \left ( \frac{\alpha}{1000} \right )^4 \left ( \frac{M_j}{40\,\msun} \right )^3 \left ( \frac{m_i}{20 \, \msun} \right )^{-1} \left ( \frac{m_{i,\,\mathrm{comp}}}{20 \, \msun} \right )^{-1}.
\end{align}

Lastly, we terminate the secular integration of systems if the wall run time exceeds 18 hr (on Intel E5-2680v4 CPUs). The secular oscillation time-scale in some systems can be very short compared to the integration time of 14 Gyr, and these systems are computationally excessively expensive to integrate, even with the secular code. This creates an uncertainty in our outcome fractions, but we remark that only typically a few per cent of systems are affected (see Table~\ref{table:sec_fractions}).

\subsection{$N$-body evolution}
\label{sect:meth:nbody}

We assume that the merger remnant of NS and BH mergers is imparted a recoil kick $v_{\rm rec}$ from anisotropic emission of GWs \citep{lou12} and interacts with the components of the other binary. Let the most massive component of the merging binary be denoted as the primary (mass $m_\pd$), and the least massive as the secondary (mass $m_\sd$). The recoil kick depends on the asymmetric mass ratio $\eta=q/(1+q)^2$, where $q=m_\sd/m_\pd<1$, and on the dimensionless spin vectors, $\boldsymbol{\chi}_\pd$ and $\boldsymbol{\chi}_\sd$ (corresponding to $m_\pd$ and $m_\sd$). We model the recoil kick following \citet{lou12} as
\begin{equation}
\textbf{v}_{\mathrm{rec}}=v_m \hat{e}_{\perp,\pd}+v_{\perp}(\cos \xi \hat{e}_{\perp,1}+\sin \xi \hat{e}_{\perp,2})+v_{\parallel} \hat{e}_{\parallel},
\label{eqn:vkick2}
\end{equation}
where
\begin{eqnarray}
v_m&=&A\eta^2\sqrt{1-4\eta}(1+B\eta)\\
v_{\perp}&=&\frac{H\eta^2}{1+q}(\chi_{\sd,\parallel}-q\chi_{\pd,\parallel})\\
v_{\parallel}&=&\frac{16\eta^2}{1+q}[V_{1,1}+V_A \tilde{S}_{\parallel}+V_B \tilde{S}^2_{\parallel}+V_C \tilde{S}_{\parallel}^3]\times \nonumber\\
&\quad \times & |\boldsymbol{\chi}_{\sd,\perp}-q\boldsymbol{\chi}_{\pd,\perp}| \cos(\phi_{\Delta}-\phi_{1}).
\end{eqnarray}
The $\perp$ and $\parallel$ refer to the directions perpendicular and parallel to the orbital angular momentum, respectively, while $\hat{e}_{\perp,1}$ and $\hat{e}_{\perp,2}$ are orthogonal unit vectors in the orbital plane. We have also defined the vector
\begin{equation}
\tilde{\mathbf{S}}=2\frac{\boldsymbol{\chi}_{\sd,\perp}+q^2\boldsymbol{\chi}_{\pd,\perp}}{(1+q)^2},
\end{equation}
$\phi_{1}$ as the phase angle of the binary, and $\phi_{\Delta}$ as the angle between the in-plane component of the vector
\begin{equation}
\mathbf{\Delta}=M^2\frac{\boldsymbol{\chi}_{\sd}-q\boldsymbol{\chi}_{\pd}}{1+q}
\end{equation}
and the infall direction at merger. Finally, we adopt $A=1.2\times 10^4$ km s$^{-1}$, $H=6.9\times 10^3$ km s$^{-1}$, $B=-0.93$, $\xi=145^{\circ}$ \citep{gon07,lou08}, and $V_{1,1}=3678$ km s$^{-1}$, $V_A=2481$ km s$^{-1}$, $V_B=1793$ km s$^{-1}$, $V_C=1507$ km s$^{-1}$ \citep{lou12}. The final total spin of the merger product and its mass are computed following \citet{rezzolla2008}. For other accurate models for the GW recoil kick, see \citet{VarmaGerosa2019,VarmaField2019}. \footnote{We remark that the fits described here were not always applied entirely correctly because $q>1$ could occur in some systems, whereas the fits from \citet{lou12} assumed $q<1$. However, the absolute fractional differences in recoil speed, remnant spin parameter, and remnant mass between the calculations that were carried out and the correct calculations, are typically on the order of $10^{-7}$ and no larger than $10^{-4}$ for all mergers across all six models. We therefore do not believe that this significantly impacts our results on the second-generation mergers. }  In evaluating these expressions, we adopt the spin directions at the moment when our merger criterion was reached (cf. \Eq~\ref{eq:rcol}). We note that the recoil kick magnitude in NS binary mergers could be smaller than BH recoil kicks since the encounter takes place at a larger gravitational radius and hydrodynamic effects could become important. Nevertheless, the main contribution to NS recoil kicks could still be expected to be due to asymmetry in the emission of GWs \citep[e.g.,][]{BernuzziNagar2014,ZappaBernuzzi2018,ZappaBernuzzi2019}.

We use the \textsc{FEWBODY} numerical toolkit for computing these 1+2 close encounters \citep{fregeau04}. We include PN2.5 corrections, while we do not account for lower-order PN effects. In the scattering experiments, there is typically no statistically significant change due to the inclusion of lower-order non-dissipative PN terms \citep[see, e.g.,][]{ZevinSamsing2019}. We take into account the results of the previous section and different recoil kick velocities, assuming flat distributions for the dimensionless spin parameter $\chi_i$ with either $0<\chi_i<0.1$ or $0<\chi_i<1$. The impact parameter is drawn from a distribution
\begin{equation}
f(b)=\frac{b}{2b_{\rm max}^2}\,.
\end{equation}
Here, $b_{\rm max}$ is the maximum impact parameter of the scattering experiment, defined by the scattering cross-section,
\begin{equation}
\Sigma = \pi b_{\rm max}^2  = \pi a_{2}^2 \left(1+\frac{2Gm_{\rm tot}}{a_{2}v_{\rm rec}^2} \right),
\label{eqn:focus}
\end{equation}
where $m_{\rm tot}$ is the total quadruple mass, and sets the size of the fractional solid angle within which the recoil kick velocity vectors has to lie to have a scattering that could lead to an exchange \citep{2020ApJ...895L..15F} \begin{equation}
\mathcal{F}\sim \left(\frac{b_{\rm max}}{a_3}\right)^2\,.
\label{eqn:solidangle}
\end{equation}
We use as a target the binary that has not merged during the 2+2 evolution and as a projectile the merger remnant of the first merger. We assign to the merger remnant a velocity computed using \eq~(\ref{eqn:vkick2}) and perform the scattering event. We collect the results of the scattering experiments, which can result in an exchange, where the new binary contains the merger remnant of the first merger, or in a fly-by. We check if the binary merges using \eq~(\ref{eq:tGW}).

\section{First-generation mergers}
\label{sect:first}
In this section, we discuss first-generation mergers that occurred during the secular dynamical simulations (cf. \S~\ref{sect:meth:sec}). 

\begin{table}
\begin{tabular}{lcccccc}
\toprule
Model & \multicolumn{5}{c}{$f\,(\%)$} \\
& \multicolumn{2}{c}{No interaction} &  \multicolumn{2}{c}{Merger} & Dyn. & Time \\
& Coupled & Dec. & Coupled & Dec. & inst. & exc.\\
\midrule
1 (ref.) & 19.5 & 25.5 & 2.9 & 50.0 & 0.2 & 1.9 \\
2 (low $\sigma$) & 50.2 & 27.4 & 4.8 & 12.1 & 0.6 & 4.9 \\
3 (high $\sigma$) & 52.1 & 7.1 & 3.7 & 34.4 & 0.7 & 2.0 \\
4 (Sana) & 10.2 & 23.6 & 5.6 & 58.1 & 0.1 & 2.4 \\
5 (low $Z$) & 5.3 & 25.1 & 5.5 & 62.6 & 0.0 & 1.5 \\
6 (ind. $m$) & 5.4 & 31.5 & 2.5 & 57.2 & 0.1 & 3.3 \\
\bottomrule
\end{tabular}
\caption{Outcome fractions in the secular simulations (in per cent) for all six models. Fractions are defined with respect to the systems evolved with \textsc{BSE} that resulted in two individually bound binaries with BHs and/or NSs. Mergers include those in either of the inner orbits of the 2+2 quadruple system. For non-interacting and merging systems, we make a distinction between systems that were coupled and decoupled (`Dec.') as defined in \S~\ref{sect:meth:sec}. We also include fractions for dynamical instability (`Dyn. inst.'), and systems in which the maximum allowed run time of 18 hr was exceeded (`Time exc.'). }
\label{table:sec_fractions}
\end{table}

\subsection{Merger fractions}
\label{sect:first:frac}
In Table~\ref{table:sec_fractions}, we show the outcome fractions in the secular simulations for all six models. These fractions are defined with respect to the systems evolved with \textsc{BSE} that resulted in two individually bound binaries with BHs and/or NSs. For non-interacting and merging systems, we make a distinction between systems that were coupled and decoupled (`Dec.') as defined in \S~\ref{sect:meth:sec}. We also include fractions for dynamical instability (`Dyn. inst.'), and systems in which the maximum allowed run time of 18 hr was exceeded (`Time exc.'). 

The majority of systems either do not interact within 14 Gyr, or they merge. For the merging systems, the majority are decoupled, meaning that the outer orbit is relatively wide and secular dynamics are unimportant. Roughly 5 to 30 per cent of merging systems were identified as coupled systems. However, as we will discuss below, because of the conservative nature of our criterion to define the boundary between coupled and decoupled systems, secular evolution was unimportant in a significant fraction of systems originally marked as `coupled'.

Only few systems undergo dynamical instability (fractions less than $1\,\%$). Dynamical instability is mostly affected by changes in the outer orbital eccentricity ($e_3$; cf. \Eq~\ref{eq:ma01}) which are typically small in 2+2 quadruples, and those systems that become unstable tend to be initially only marginally stable. The fraction of systems in which the maximum CPU wall time of 18 hr was exceeded is typically a few per cent, and no larger than $5\,\%$. 

\begin{figure}
\center
\includegraphics[scale = 0.45, trim = 5mm 10mm 0mm 10mm]{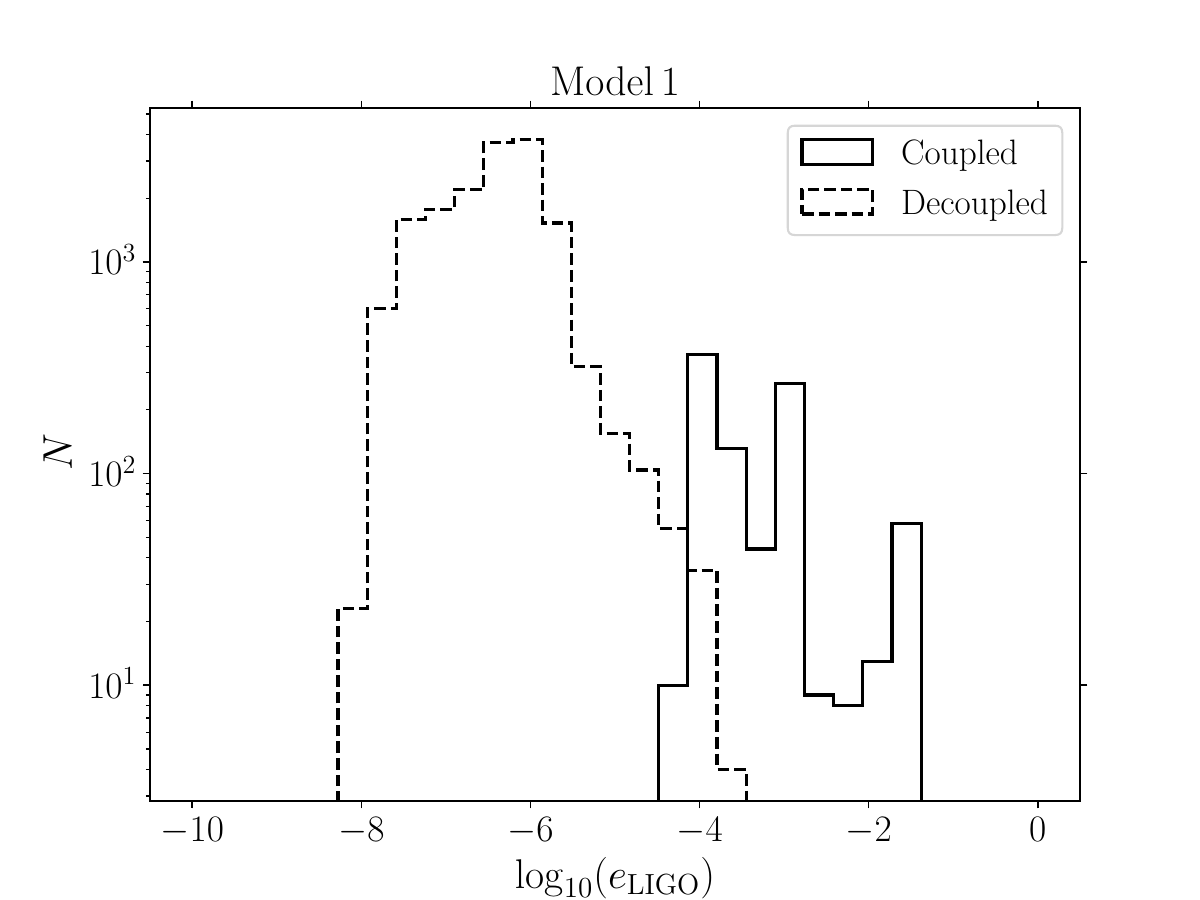}
\caption{ Distributions of the eccentricities in the LIGO band ($f_\ligo = 10\, \hz$) for the first-generation mergers in the secular simulations. Black solid (dashed) lines show distributions for coupled (decoupled) systems. These data apply to Model 1; data for other models are available in \AppGenOne. }
\label{fig:e_LIGO_m1}
\end{figure}

\subsection{Eccentricities in the LIGO band}
\label{sect:first:eLIGO}
For first-generation merging binaries, we use the orbit-averaged equations of \citet{1964PhRv..136.1224P} combined with an estimate for the GW peak frequency \citep{2003ApJ...598..419W} to calculate the eccentricity when reaching the LIGO band, $f_\ligo = 10\, \hz$. The resulting distributions of $e_\ligo$ are shown for Model 1 in \F~\ref{fig:e_LIGO_m1} (data for other models are available in \AppGenOne). We make a distinction between coupled and decoupled systems as defined in \S~\ref{sect:meth:sec}. The distribution is peaked around $10^{-6}$ for the decoupled systems, which can be understood from the fact that tidal evolution tends to circularise the orbit in isolated binaries prior to mass transfer and/or CE evolution \citep{2002MNRAS.329..897H}, and the latter interactions are usually necessary for the compact object binary to merge within a Hubble time. 

The coupled systems show a distribution of $e_\ligo$ which is in part a continuation of the high-$e_\ligo$ end of the decoupled systems, but also contains a smaller population of systems with significantly higher $e_\ligo$, up to $\sim 10^{-2}$. The higher eccentricities in the latter case can be attributed to secular evolution in 2+2 quadruple systems. We note that the results in \F~\ref{fig:e_LIGO_m1} are based on secular integrations, which can break down in some cases as discussed in \S~\ref{sect:meth:sec}. Non-secular effects tend to produce even higher eccentricities in the LIGO band, up to $\sim 10^0$ (e.g., \citealt{2016ApJ...816...65A}), so our distributions of $e_\ligo$ should be considered to be lower limits. 

We remark that eccentricities at $10\,\hz$ below 0.02 cannot be distinguished from zero observationally with the LIGO/VIRGO/KAGRA network \citep{2018PhRvD..98h3028L,2018PhRvD..97b4031H,2019ApJ...871..178G}. However, the eccentricity may be significant for systems in the LISA band. LISA may have the potential to identify GW sources with a quadruple origin either statistically based on the eccentricity distribution or by directly observing the ZLK oscillations \citep[e.g.,][]{2019arXiv190208604R,2021ApJ...914...75R}.

\begin{figure}
\center
\includegraphics[scale = 0.45, trim = 5mm 10mm 0mm 10mm]{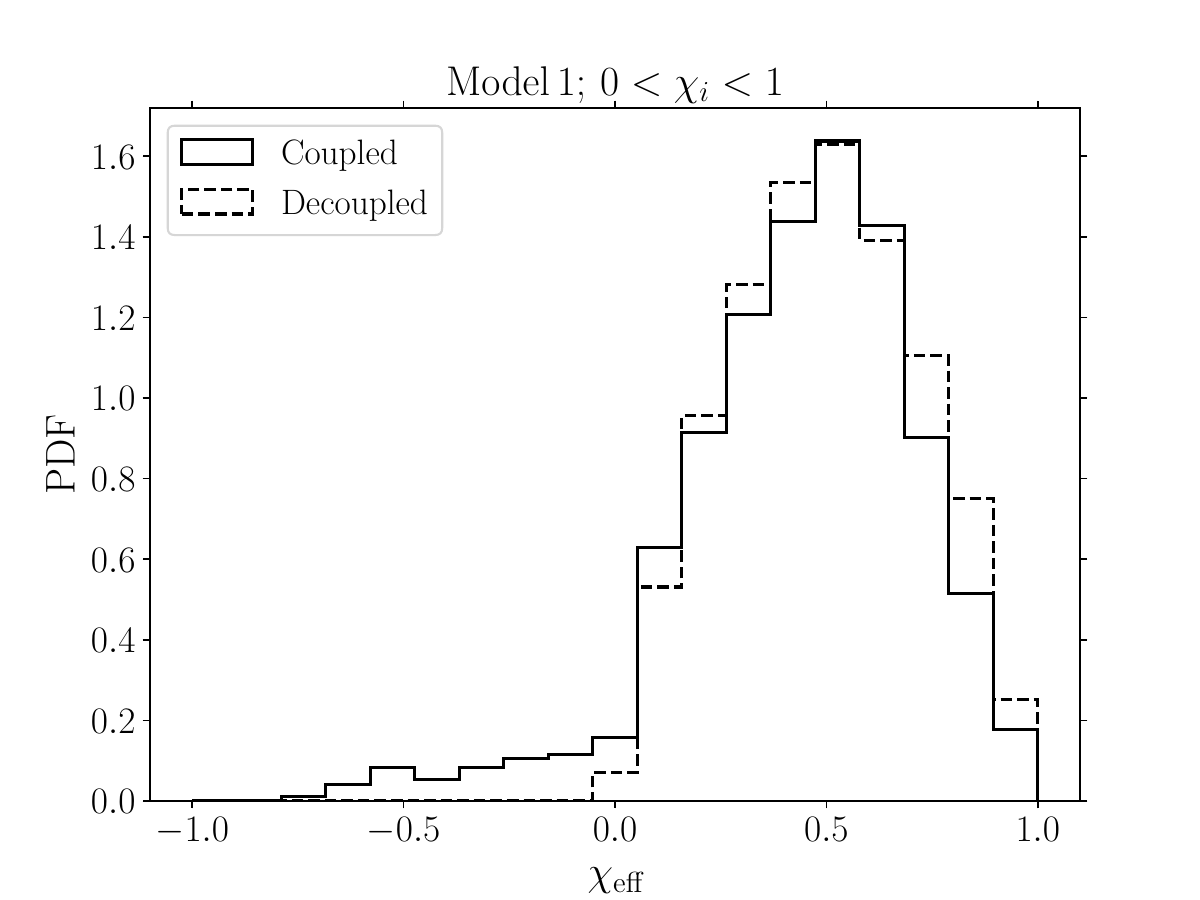}
\caption{ Distributions of $\chi_\eff$ (cf. \Eq~\ref{eq:chi_eff}) for the first-generation mergers in the secular simulations. Black solid (dashed) lines show distributions for coupled (decoupled) systems. These data apply to Model 1 and $0<\chi_i<1$; data for other models are available in \AppGenOne. }
\label{fig:chi_eff_m1}
\end{figure}

\subsection{Distributions of $\chi_\eff$}
\label{sect:first:chi}
In \F~\ref{fig:chi_eff_m1}, we show the distributions of $\chi_\eff$ for the first-generation mergers in Model 1 (data for other models are available in \AppGenOne). For two merging compact objects $i$ and $j$ in an orbit $k$, $\chi_\eff$ is defined according to
\begin{align}
\label{eq:chi_eff}
\chi_\eff \equiv \frac{m_i \chi_i \, \left ( \unit{S}_i \cdot \unit{L}_k \right ) + m_j \chi_j \, \left ( \unit{S}_j \cdot \unit{L}_k \right )}{m_i+m_j},
\end{align}
where $\chi_i$ is the dimensionless spin parameter of compact object $i$. We extract the spins and $\unit{L}_k$ when the stopping condition \eq~(\ref{eq:rcol}) was met, and assume that $\chi_\eff$ does not change significantly until actual merger (consistent with \eq~\ref{eq:so}; note also that $\chi_\eff$ is conserved to 2PN order, \citealt{2008PhRvD..78d4021R}). We assume flat distributions for $\chi_i$ with either $0<\chi_i<0.1$, or $0<\chi_i<1$. In \F~\ref{fig:chi_eff_m1}, we take $0<\chi_i<1$ and note that the choice $0<\chi_i<0.1$ only gives a re-scaling of the horizontal axis, while the shapes of the distributions remain qualitatively the same. 

The decoupled systems show an approximately symmetric distribution of $\chi_\eff$ centered around $\chi_\eff=0.5$ (assuming $0<\chi_i<1$). In these systems, the orbital angular momentum is largely unaffected by secular evolution, and $\chi_\eff$ reflects the initially assumed distributions of the spin-orbit misalignment angles ($10^\circ$; see \S~\ref{sect:meth:sec}). No retrograde orientations ($\chi_\eff<0$) occur for the decoupled systems. 

Coupled systems show a largely very similar distribution of $\chi_\eff$ as the decoupled systems. This is a reflection of the conservative nature of our definition of the boundary between coupled and decoupled systems. There exists a small population, however, of systems with negative $\chi_\eff$. This population can be attributed to secular evolution, since the latter tends to randomise the orbital angular momentum orientation, allowing for also retrograde spins-orbit angles (e.g., \citealt{2018MNRAS.480L..58A,2019ApJ...881...41L}).

\begin{figure}
\center
\includegraphics[scale = 0.45, trim = 5mm 35mm 0mm 48mm]{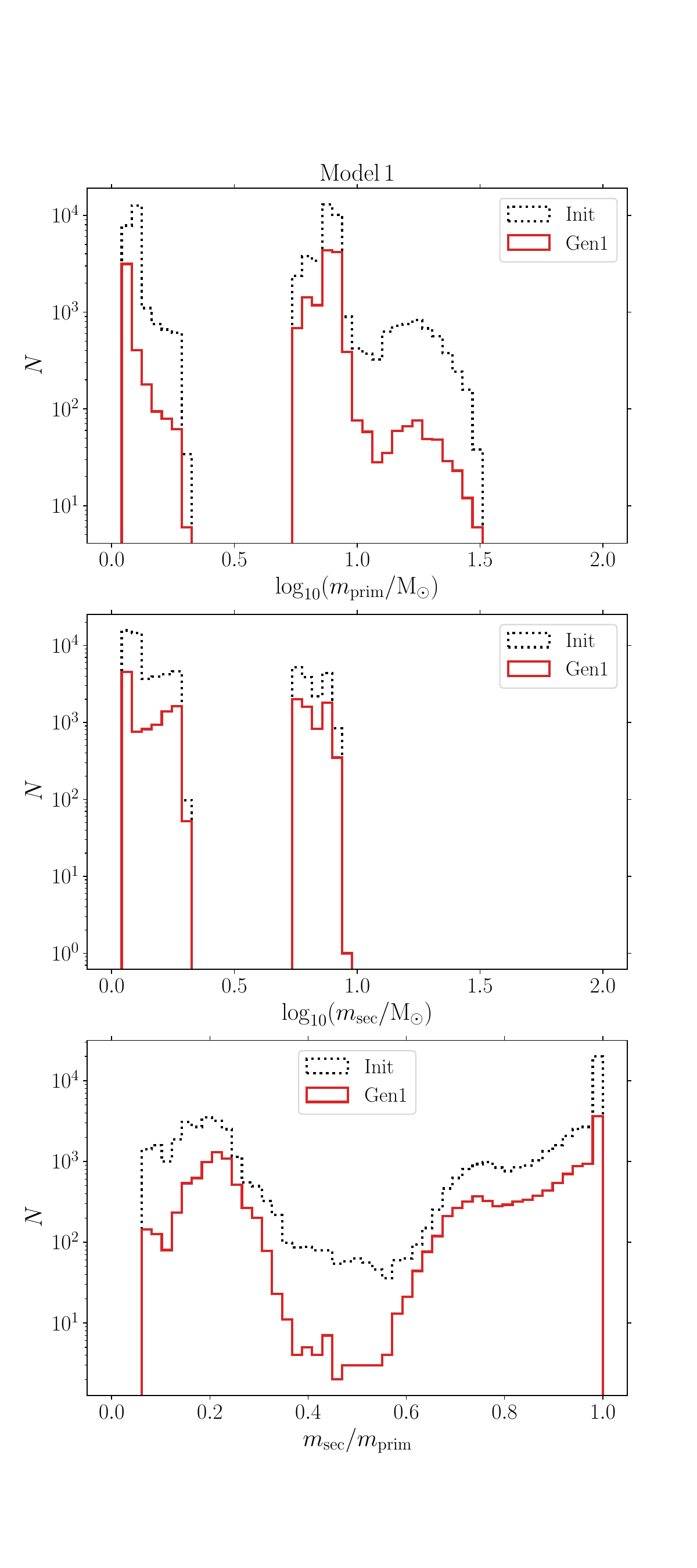}
\caption{ Distributions of the primary masses (top panel), secondary masses (second panel), and mass ratios (third panel) for first-generation mergers in the secular simulations. Here, we define the primary to be the more massive component of the binary. Solid red lines correspond to the first-generation mergers, whereas black dotted lines show distributions of the masses for all systems after compact object formation (i.e., after evolution with \bse). These data apply to Model 1; data for other models are available in \AppGenOne. }
\label{fig:masses_m1}
\end{figure}

\subsection{Masses}
\label{sect:first:m}
In \F~\ref{fig:masses_m1}, we consider the mass distributions for first-generation mergers in Model 1 (data for other models are available in \AppGenOne). Here, we define the primary (secondary) to be the most (least) massive component of the merging binary. The primary and secondary masses of merging systems show very little distinction compared to the initial systems following compact object formation. Merging systems do tend to disfavour mass ratios around 0.5, however, as shown in the bottom panel of \F~\ref{fig:masses_m1}. This can be mainly attributed to isolated binary evolution and the low-end mass gap for BHs between $\sim2$-$5\,\msun$. The majority of merging primary BH masses are around $8\,\msun$ (see the top panel of \F~\ref{fig:masses_m1}), whereas secondaries do not exist in the range $\sim2$-$5\,\msun$ (see the middle panel of  \F~\ref{fig:masses_m1}). This implies that mass ratios of $\sim 0.5$ are disfavoured.

\begin{figure}
\center
\includegraphics[scale = 0.45, trim = 5mm 25mm 0mm 30mm]{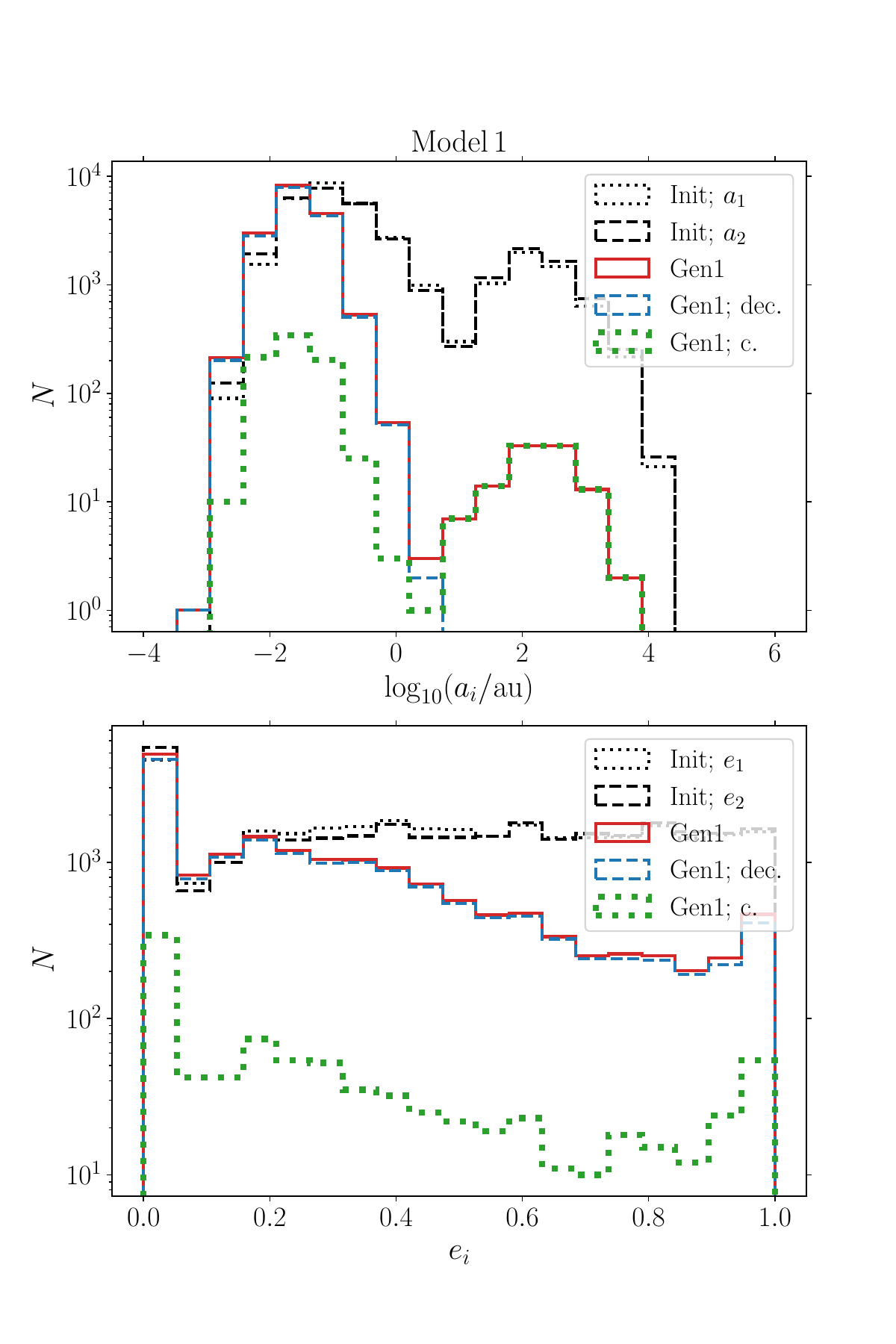}
\caption{ Distributions of the semimajor axes (top panel) and eccentricities (bottom panel) for first-generation mergers in the secular simulations. Black dotted and dashed lines show the initial distributions for orbits 1 and 2, respectively. All first-generation mergers are shown with solid red lines; among these, blue dashed and green dotted lines separate decoupled and coupled mergers, respectively.  These data apply to Model 1; data for other models are available in \AppGenOne. }
\label{fig:orbits_m1}
\end{figure}

\subsection{Orbits}
\label{sect:first:orbit}
Distributions of the semimajor axes and eccentricities of first-generation merging binaries (when the stopping condition \eq~\ref{eq:rcol} was reached) are shown in \F~\ref{fig:orbits_m1} (Model 1; other models are included in \AppGenOne). As expected, decoupled systems tend to have small semimajor axes (peaked around $10^{-2}\,\au$). Coupled mergers, on the other hand, show a bimodal distribution in their semimajor axes. One population has semimajor axes similar to the decoupled systems (peaking near $10^{-2}\,\au$), reflecting our conservative criterion for coupling/decoupling. However, another population exists with semimajor axes peaking around $10^2\,\au$. The latter are systems in which secular evolution is more effective since the secular time-scales are shorter (cf. \Eq~\ref{eq:tZLK}). Given their significantly wider orbits, eccentricity excitation is necessary for these systems to merge within a Hubble time. There are no significant differences between the coupled and decoupled systems in terms of the initial eccentricity (i.e., the eccentricity after compact formation, see the bottom panel of \F~\ref{fig:orbits_m1}). This can be understood by noting that secular eccentricity excitation of the inner orbits is relatively insensitive to the initial eccentricity.

\begin{figure}
\center
\includegraphics[scale = 0.45, trim = 5mm 10mm 0mm 10mm]{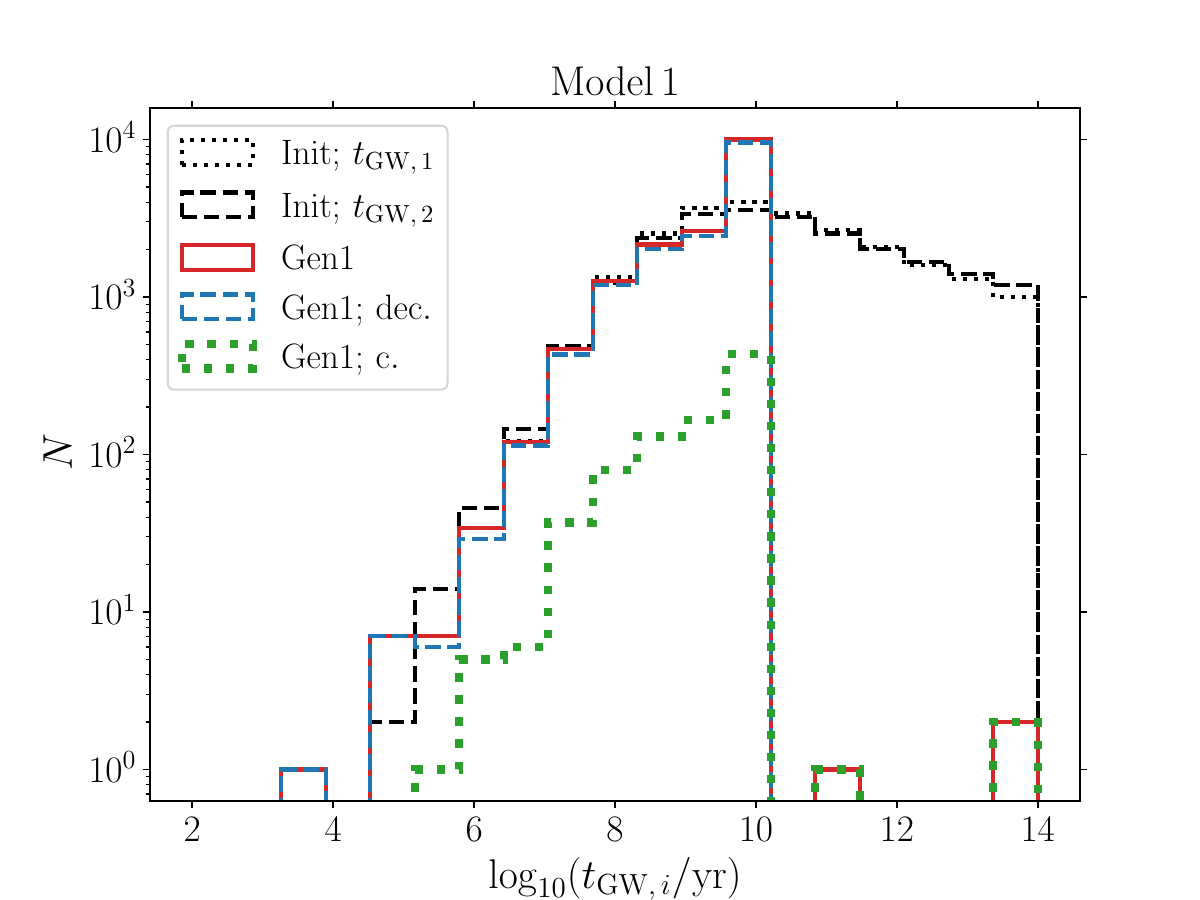}
\caption{ Distributions of the GW merger time-scales for first-generation mergers in the secular simulations, based on the orbital elements just after compact object formation. Black dotted and dashed lines show the merger time-scale distributions for orbits 1 and 2, respectively. All first-generation mergers are shown with solid red lines; among these, blue dashed and green dotted lines separate decoupled and coupled mergers, respectively.  These data apply to Model 1; data for other models are available in \AppGenOne. }
\label{fig:t_GW_m1}
\end{figure}

\subsection{Initial GW merger times}
\label{sect:first:time}
Lastly, \F~\ref{fig:t_GW_m1} shows the distributions of the first-generation GW merger times (cf. \Eq~\ref{eq:tGW}; we ignore spins when computing merger times). These times are calculated from the orbital elements right after binary compact object formation. Therefore, any isolated binary would only merge in our simulations if $t_{\mathrm{GW},\,i} < 14 \,\gyr$. Indeed, decoupled systems show a sharp cutoff in $t_{\mathrm{GW},\,i}$ near $14\,\gyr$. Most coupled systems also have $t_{\mathrm{GW},\,i}<14\,\gyr$ (showing that they would have merged also in isolation and indicating that our coupling/decoupling criterion is conservative). However, a small fraction of coupled systems has an initial merger time-scale greatly exceeding $14\,\gyr$. In these systems, the merger has been accelerated by secular eccentricity excitation.

\section{Second-generation mergers}
\label{sect:second}

\begin{figure}
\center
\includegraphics[scale=0.55]{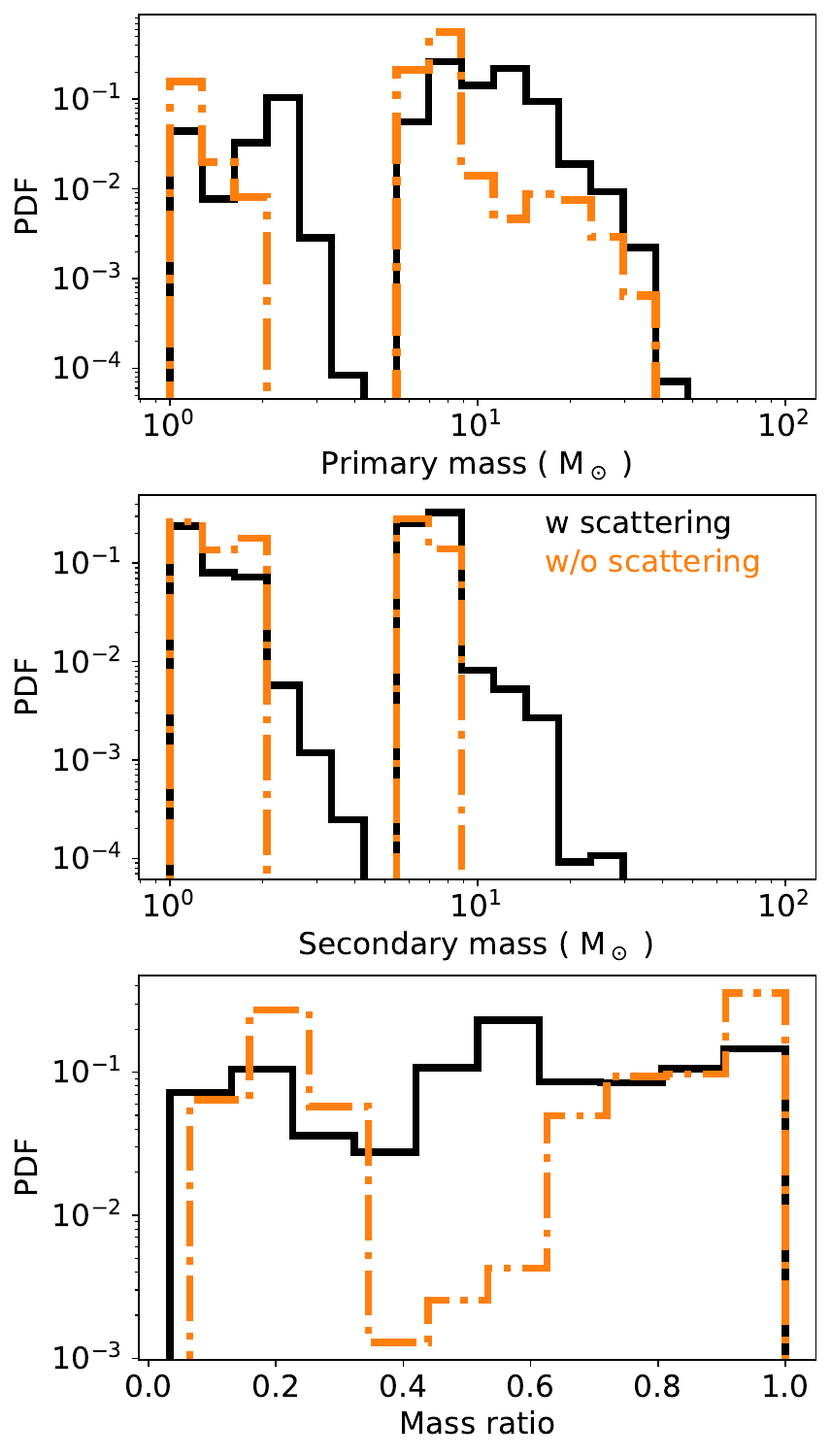}
\caption{ Distributions of the primary mass (top), secondary mass (center), and mass ratio (bottom) in second-generation mergers. Black and orange lines show the distributions with and without scattering. These data apply to Model 1 and $0<\chi_i<1$; data for other models are available in \AppGenTwo. }
\label{fig:2g_mod11_mass}
\end{figure}

\begin{figure}
\center
\includegraphics[scale=0.5]{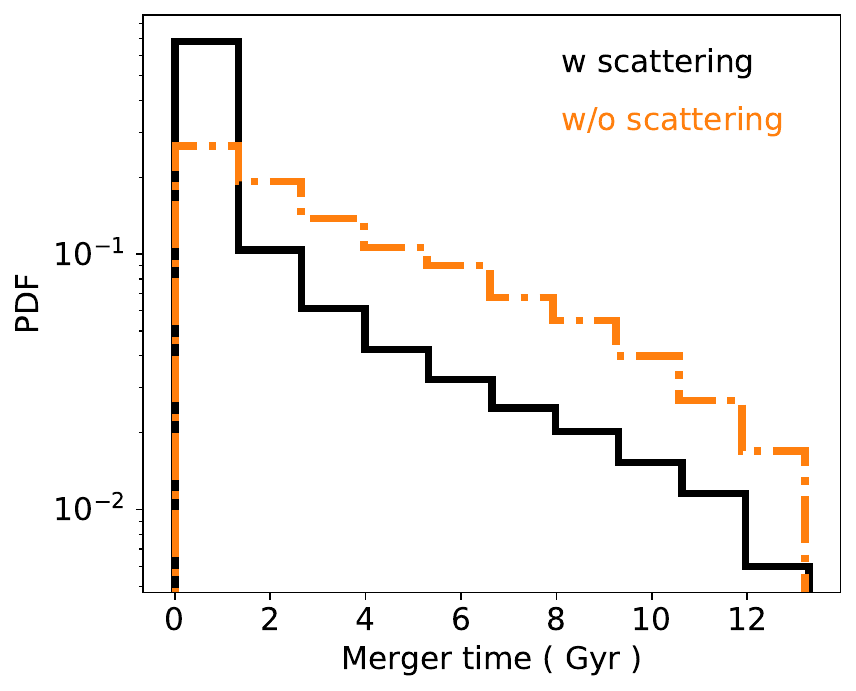}
\caption{ Distributions of the merger time in second-generation mergers. Black and orange lines show the distributions with and without scattering. These data apply to Model 1 and $0<\chi_i<1$; data for other models are available in \AppGenTwo. }
\label{fig:2g_mod11_time}
\end{figure}

\begin{figure}
\center
\includegraphics[scale=0.5]{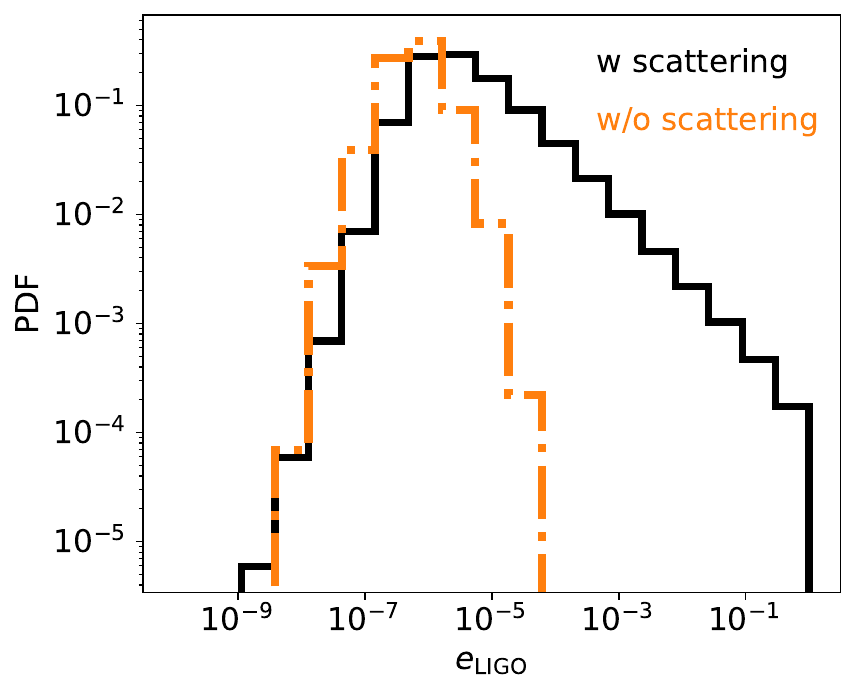}
\caption{ Distributions of the eccentricities in the LIGO band ($f_\ligo = 10\, \hz$) in second-generation mergers (cf. \F~\ref{fig:e_LIGO_m1} for first-generation mergers). Black and orange lines show the distributions with and without scattering. These data apply to Model 1 and $0<\chi_i<1$; data for other models are available in \AppGenTwo. }
\label{fig:2g_mod11_ecc}
\end{figure}

\begin{figure}
\center
\includegraphics[scale=0.5]{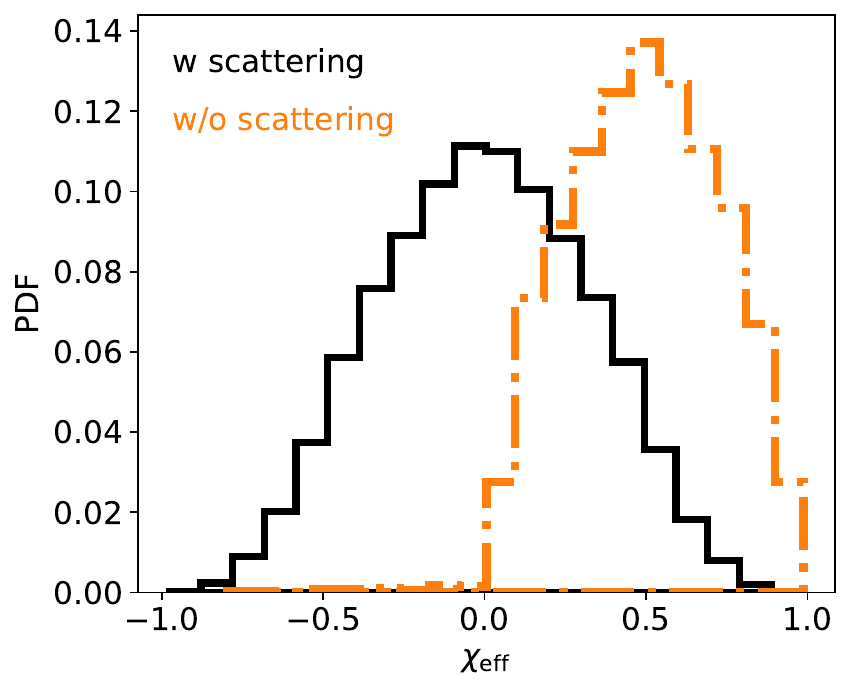}
\caption{ Distributions of the effective spin for the second-generation mergers. Black and orange lines show the distributions with and without scattering. These data apply to Model 1 and $0<\chi_i<1$; data for other models are available in \AppGenTwo. }
\label{fig:2g_mod11_spin}
\end{figure}

In this section, we discuss properties of second-generation compact objects formed after the remnant of the first merger (in one of the two inner binaries) merges with one of the components of the other binary after a scattering occurs (cf. \S~\ref{sect:meth:nbody}).

\begin{table}
\begin{tabular}{lcccc}
\toprule
Model & Max spin & \multicolumn{3}{c}{$f\,(\%)$} \\
& & `Scatt.' & `Scatt. No Exch.' & `No Scatt.' \\
\midrule
1 (ref.) & 1.0 & 40 & 12 & 28 \\
2 (low $\sigma$)  & 1.0 & 20 & 6.3 & 9.2 \\
3 (high $\sigma$) & 1.0 & 48 & 13 & 41 \\
4 (Sana)  & 1.0 & 43 & 12 & 31 \\
5 (low $Z$)  & 1.0 & 40 & 12 & 30 \\
6 (ind. $m$) & 1.0 & 37 & 12 & 25 \\

1 (ref.) & 0.1 & 42 & 11 & 28 \\
2 (low $\sigma$)  & 0.1 & 21 & 6.2 & 9.1 \\
3 (high $\sigma$) & 0.1 & 49 & 13 & 42 \\
4 (Sana)  & 0.1 & 45 & 11 & 31 \\
5 (low $Z$) & 0.1 & 42 & 12 & 30 \\
6 (ind. $m$) & 0.1 & 38 & 11 & 25 \\
\bottomrule
\end{tabular}
\caption{Second generation merger fractions in the $N$-body simulations (in per cent) with respect to the first-generation mergers for all six models. The actual merger fraction is $(b_\mathrm{max}/a_3)^{2}$ times the value shown in this table (cf. \eq~\ref{eqn:solidangle}), since the scattering experiments were carried out only for those small solid angles for which the first generation merger remnant points toward the companion binary. We include the fraction of systems that merge when the scattering is taking into account (`Scatt.'), the fraction of systems that merge in the scattering where none of the second binary components is replaced by the incoming remnant of the merger of the first binary (`Scatt. No Exch.'), and the fraction of systems that merge if there is no interaction (`No Scatt.').}
\label{table:nbody_fractions}
\end{table}

\subsection{Merger fractions}

In Table~\ref{table:nbody_fractions}, we show the outcome fractions in the $N$-body evolution. We report the fraction of systems that merge with respect to the first-generation mergers. For merging systems, we make a distinction between the number of systems that merge when the scattering is taken into account (`Scatt.') and the number of systems that merge if there is no interaction (`No Scatt.'). In the first category, we also report the fraction of systems that merge in the scattering where none of the second binary components is replaced by the incoming remnant of the merger of the first binary (`Scatt. No Exch.'). Since the recoil kick depends on the spins of the merging systems, we report the result in the case the maximum dimensionless spin parameter is either $0.1$ or $1.0$. We remark that the actual merger fraction is $(b_\mathrm{max}/a_3)^{2}$ times the value shown in Table \ref{table:nbody_fractions} (cf. \eq~\ref{eqn:solidangle}), since the scattering experiments were carried out only for those small solid angles for which the first generation merger remnant points toward the companion binary. 

The fraction of merging systems when the scattering between the merger remnant and the second binary is taken into account is $\ 20$-$49\%$. Independently of the model, about $75\%$ of the binaries that merge are exchanged binaries, with the remnant of the merger of the first binary as one of the components, while about $25\%$ are made up of the original members of the second binary. If the scattering is not taken into account, the fraction of mergers is $9$-$42\%$. The fraction of systems that merge when the scattering is considered is $\approx 1.5$ times larger. The merger fractions do not depend significantly on the assumed maximum spin.

\subsection{Masses}

In Fig.~\ref{fig:2g_mod11_mass}, we consider the mass and mass ratio distributions for second-generation mergers in Model 1 and $0<\chi_i<1$ (data for other models are available in \AppGenTwo). Here, we define the primary (secondary) to be the most (least) massive component of the merging binary. We show the results in the case the scattering is or is not considered. The primary and secondary masses of merging systems are larger for scattering cases than for non-scattering cases. The reason is that about $75\%$ of the merging systems contain the previous merger remnant, which is typically more massive than one or both the components of the second binary. Moreover, systems that merge in scatterings have almost uniform distribution of mass ratios, while there is a gap at $\sim 0.5$ (similar to the first-generation mergers) if scatterings are not taken into account.

\subsection{GW merger times}

Fig.~\ref{fig:2g_mod11_time} shows the distributions of the second-generation GW merger times (cf. \Eq~\ref{eq:tGW}) in Model 1 and $0<\chi_i<1$ (data for other models are available in \AppGenTwo). These times are calculated from the orbital elements right after binary formation in scattering or from the final orbital elements of the second binary when no scattering is considered. It is clear that in the former case the merger times are smaller than the latter case. The reason is twofold. First, about $75\%$ of the merging systems contain the previous merger remnant, which is typically more massive than one or both the components of the second binary, thus reducing the merger time. Second, the merger time is smaller since scatterings tend to increase the eccentricity of the binaries. 

\subsection{Eccentricity in the LIGO band}

For second-generation merging binaries, we use the orbit-averaged equations of \citet{1964PhRv..136.1224P} combined with an estimate for the GW peak frequency \citep{2003ApJ...598..419W} to calculate the eccentricity when reaching the LIGO band, $f_\ligo = 10\, \hz$ (similarly to first-generation mergers). The resulting distributions of $e_\ligo$ are shown for Model 1 and $0<\chi_i<1$ in Fig.~\ref{fig:2g_mod11_ecc} (data for other models are available in \AppGenTwo). We make a distinction between systems that merge when scattering is taken into account and systems that merge when there is no scattering. As for the first-generation mergers, the distribution is peaked around $10^{-6}$ in the case there is no scattering, since tidal evolution in the second binary tends to circularise the orbit \citep{2002MNRAS.329..897H}. For systems that merge in scattering simulations, the distribution is always peaked around $10^{-6}$, but with a tail up to $\sim 0.1$. The higher eccentricities in this case can be attributed to the fact that scatterings originate eccentric binaries, that later merge with eccentricity in the LIGO band larger than the second binary in isolation. The maximum eccentricity we find is about $0.9$.

\subsection{Distributions of $\chi_\eff$}

In Fig.~\ref{fig:2g_mod11_spin}, we show the distributions of $\chi_\eff$ (cf. \Eq~\ref{eq:chi_eff}) for the second-generation mergers in Model 1 and $0<\chi_i<1$ in Fig.~\ref{fig:2g_mod11_ecc} (data for other models are available in \AppGenTwo). We find that systems show a distribution of $\chi_\eff$ centered around $\chi_\eff=0.5$ in the case no scattering is considered. Here, as in first-generation mergers, $\chi_\eff$ reflects the initially assumed distributions of the spin-orbit misalignment angles ($10^\circ$; see \S~\ref{sect:meth:sec}). No retrograde orientations occur. For systems that merge when scatterings are considered, the $\chi_\eff$ distribution is peaked at $\sim 0$ and is consistent with isotropy, consistent with mergers that occur as a result of dynamical interactions. In the case $0<\chi_i<0.1$ (see Fig.~C4 in \AppGenTwo), systems show a distribution of $\chi_\eff$ centered around $\chi_\eff=0.05$ in the case no scattering is considered. When we consider scatterings, the distribution is approximately symmetric around $\chi_\eff=0$, again consistent with a dynamical merger origin.

\section{Discussion}
\label{sect:discussion}

\subsection{Limitations of the simulations}
\label{sect:discussion:lim}
As mentioned in \S~\ref{sect:meth}, in our methodology we greatly simplified the simulations by separating the stellar/binary and multi-body dynamical evolution, i.e., by considering the inner two binaries as initially isolated (taking into account their stellar and binary evolution only), and including multi-body dynamics only after compact object formation in a 2+2 quadruple system. Also, we assumed that the outer orbit of the 2+2 quadruple always remains bound, which is unlikely given the mass loss events and/or natal kicks that occur during stellar evolution. These simplifications introduce large uncertainties in the merger rates (\S~\ref{sect:discussion:rate}), although we expect the impact on eccentricity and spin distributions in the LIGO band to be less severe. Another caveat is the assumption of initial thermal eccentricity distributions, which may be more eccentric than observed \citep[e.g.,][]{sana12,2013ARA&A..51..269D}, and could lead to higher merger rates. In future work, we will consider these complicating factors with more sophisticated methods that take into account these processes simultaneously (e.g., \textsc{MSE}, \citealt{2021MNRAS.502.4479H}), and more realistic initial conditions.

\subsection{Merger rate estimates}
\label{sect:discussion:rate}
Here, we briefly estimate the rate of BH and NS mergers from the first and second-generation mergers in our simulations. We emphasize that these estimates are very approximate, and they are moreover affected by large uncertainties in the modelling given the limitations in our simulations (see \S~\ref{sect:discussion:lim}). 

We start our estimates by adopting a star formation rate of $R_\mathrm{SF} \approx 1.5 \times 10^{7} \, \msun \, \pgpc \, \pyr$ at redshift $z=0$ \citep{2014ARA&A..52..415M}. The total mass represented by our population synthesis calculations is $M_\mathrm{sim} \sim M_\mathrm{Kr} (\alpha_\mathrm{s}+ \frac{3}{2} \alpha_\mathrm{bin} ) N_\mathrm{sys}$, where $M_\mathrm{Kr} \simeq 0.50 \, \msun$ is the average mass in a Kroupa mass distribution (see, e.g., \citealt{2013MNRAS.430.2262H}), and $\alpha_\mathrm{s}$ and $\alpha_\mathrm{bin}$ are the single and binary star fractions, respectively. Here, we neglect the mass contributed to the total population by higher-order systems, and ignore the dependence of multiplicity on stellar mass (e.g., \citealt{2017ApJS..230...15M}). We motivate the neglect of higher-order systems in their contribution in the mass normalisation by noting that, although common among massive systems, the overall number of high-order multiplicity systems is small and the mass of any stellar population is dominated by low-mass stars. Furthermore, assuming a flat mass ratio distribution, the typical mass of a binary system is $3/2$ times the mass of a single-star system. For simplicity, we assume mass-independent $\alpha_\mathrm{s}=\alpha_\mathrm{bin}=0.5$ for the purposes of the total mass calculation.

The number of sampled systems in the simulations, $N_\mathrm{sample}$, can be expressed in terms of the number of all systems (of any multiplicity), $N_\mathrm{sys}$, according to $N_\mathrm{sample} = f_\mathrm{calc} \alpha_\mathrm{quad} N_\mathrm{sys}$, where $f_\mathrm{calc}$ is the fraction of sampled systems compared to the total population, and $\alpha_\mathrm{quad}$ is the quadruple fraction (in this work, we consider 2+2 quadruples exclusively and ignore the contribution from 3+1 quadruples). We adopt $\alpha_\mathrm{quad}=0.4$ \citep{2017ApJS..230...15M}. The calculated fraction $f_\mathrm{calc}$ is estimated by noting that we restricted to systems with primary masses $>8\,\msun$ and orbits with semimajor axes $a_i>a_\mathrm{min,\,sim} = 10\,\au$. The former restriction implies a contribution to $f_\mathrm{calc}$ of $\sim 2.6 \times 10^{-3}$ (assuming a Kroupa mass distribution; see, e.g., \citealt{2013MNRAS.430.2262H}). The latter restriction implies a contribution to $f_\mathrm{calc}$ of $\sim \log(a_\mathrm{min,\,sim}/a_\mathrm{max})/\log(a_\mathrm{min}/a_\mathrm{max}) = \log(10^1/10^4)/\log(10^{-2}/10^4) = \frac{1}{2}$ (assuming a flat distribution in $\log a_i$ with minimum separation $a_\mathrm{min}=10^{-2}\,\au$, and maximum separation $a_\mathrm{max} = 10^4 \, \au$). This gives $f_\mathrm{calc} \sim 1.3 \times 10^{-3}$. 

\begin{table}
\begin{tabular}{lcc}
\toprule
Model & \multicolumn{2}{c}{BH/NS merger rate ($\pgpc\,\pyr$)} \\
& First generation & Second generation \\
\midrule
1 (ref.) & 31 & $1.2\times 10^{-5}$\\
2 (low $\sigma$) & 61 & $1.2\times 10^{-5}$\\
3 (high $\sigma$) & 7.6 & $3.6\times 10^{-6}$\\
4 (Sana) & 80 & $3.4\times 10^{-5}$\\ 
5 (low $Z$) & 88 & $3.5\times 10^{-5}$\\
6 (ind. $m$) & 76 & $2.8\times 10^{-5}$\\
\bottomrule
\end{tabular}
\caption{Estimates (upper limits) of first and second-generation merger rates in our simulations. These rates include all combinations of BH and NS mergers; a compositional breakdown for the two merger generations and the different models is given in Table~\ref{table:comp_fractions} (the rates in this table can be multiplied with these fractions to obtain absolute rates for different subgroups of mergers).}
\label{table:rates}
\end{table}

\begin{table*}
\begin{tabular}{lcccccc}
Model & \multicolumn{6}{c}{Compositional fraction compared to all BH/NS mergers (\%)} \\
& \multicolumn{3}{c}{First generation} & \multicolumn{3}{c}{Second generation} \\
& BH-BH & NS-NS & BH-NS & BH-BH & NS-NS & BH-NS \\
\midrule
1 (ref.) & 39.5 & 23.7 & 36.8 & 60.1 & 20.8 & 19.2 \\
2 (low $\sigma$) & 41.8 & 28.7 & 29.5 & 72.3 & 19.3 & 8.4 \\
3 (high $\sigma$)  & 29.6 & 31.8 & 38.6 & 53.5 & 23.5 & 23.0 \\
4 (Sana) & 42.8 & 29.8 & 27.4 & 59.8 & 17.3 & 22.9 \\
5 (low $Z$)  & 63.4 & 22.8 & 13.8 & 64.4 & 13.7 & 21.9 \\
6 (ind. $m$)  & 34.2 & 13.5 & 52.3 & 54.5 & 41.5 & 4.0 \\
\bottomrule
\end{tabular}
\caption{Compositional fractions (in per cent) of first and second-generation individual BH-BH, NS-NS, and BH-NS mergers, compared to all BH/NS mergers. }
\label{table:comp_fractions}
\end{table*}

In Model 1, the number of first-generation mergers in the secular integrations was $N_\mathrm{merge} = 16,725$, whereas the number of initially sampled systems was $N_\mathrm{sample} = 6,668,191$. Combining the above numbers, we arrive at a merger rate in Model 1 of
\begin{align}
    \nonumber R_\mathrm{1G;\,M1} &\sim \frac{R_\mathrm{SF} N_\mathrm{merge}}{M_\mathrm{sim}} \sim R_\mathrm{SF} \frac{N_\mathrm{merge} f_\mathrm{calc} \alpha_\mathrm{quad}}{N_\mathrm{sample} \left (\alpha_\mathrm{s}+ \frac{3}{2} \alpha_\mathrm{bin} \right ) M_\mathrm{Kr}} \\
    \quad &\simeq 31 \, \pgpc \, \pyr.
\end{align}
Estimates for the other models are given in Table~\ref{table:rates}. Our rate estimates in the latter table include all combinations of BH and NS mergers; a compositional breakdown for the two merger generations and the different models is given in Table~\ref{table:comp_fractions} (the rates in Table~\ref{table:rates} can be multiplied with these fractions to obtain absolute rates for different subgroups of mergers).

For reference, our first-generation rate estimates (all BH and NS mergers) are on the order of $\sim 10^{1}$ and up to $\sim 10^2\,\pgpc\,\pyr$; previous theoretical estimates of BH merger rates in triples and quadruples are each up to $\sim 10^{0}-10^1\,\pgpc\,\pyr$ (e.g., \citealt{2017ApJ...836...39S,2017ApJ...841...77A,2019MNRAS.486.4781F}). 

Rates inferred from LIGO/Virgo are $23.9^{+14.9}_{-8.6} \, \pgpc\,\pyr$ for BH-BH mergers and $320^{+490}_{-240}\,\pgpc\,\pyr$ for NS-NS mergers \citep{LIGO_O3_Catalog}. Based on two recently-announced GW sources from merging BH-NS systems, \citet{2021arXiv210615163T} report a BH-NS merger rate of $45^{+75}_{-33}\, \pgpc\,\pyr$ when assuming the sources are representative of the NS BH population, or $130^{+112}_{-69} \, \pgpc\,\pyr$ under the assumption of a broader distribution of component masses. For BH-BH mergers (cf. Table~\ref{table:comp_fractions}), our rates are compatible with LIGO/Virgo, whereas our NS-NS rates fall short of the LIGO/Virgo rate by about an order of magnitude. Our first-generation BH-NS rates are $\sim$ 3-40 $ \pgpc\,\pyr$ depending on the model, which are on the lower end of, but compatible with the recent LIGO rates. Again, we stress that our estimated rates are highly uncertain, and are likely overestimates given that we did not self-consistently take into account the effects of mass loss on the outer orbit (the latter will be addressed in future work). 

To estimate the rates of second-generation mergers, we have to account for the fact that the velocity kick vector has to lie in the fractional solid angle $\sim \left(a_{\rm 2b}/a_{\rm 3}\right)^2$, where $a_{\rm 2b}$ and $a_{\rm 3}$ are the semi-major axis of the second binary and the outer semi-major axis, respectively \citep{2020ApJ...895L..15F}. For our models, we estimate it to be $\left(a_{\rm 2b}/a_{\rm 3}\right)^2 \sim 10^{-6}$. From Table~\ref{table:nbody_fractions}, we find that the merger fraction for Model 1 is $0.4$. Combining these two numbers, we find that the rate of second-generation mergers in Model 1 is about

\begin{equation}
R_\mathrm{2G;\,M1} \sim 1.2\times 10^{-5}\, \pgpc \, \pyr,
\end{equation}
and estimates for the other models are included in Table~\ref{table:rates}. These estimates are consistent with the results of \citet{2020ApJ...895L..15F}. Similarly to the first-generation mergers, these estimates are highly uncertain; a more detailed modelling of the 2+2 systems is highly desirable in future studies. Nevertheless, we can compare our results of the relative rates of second generation+first-generation mergers compared to first generation+first generation mergers. Based on the second LIGO-Virgo Gravitational Wave Transient Catalog, \citet{2020arXiv201105332K} infer a relative rate of $5.3 \times 10^{-3}$ of second generation+first-generation mergers compared to first generation+first generation mergers, with a $99\%$ upper limit of $0.04$. This is significantly larger than in our simulations, where we find that the same corresponding relative rate is $\sim 10^{-7}$.

\subsection{LISA sources}
\label{sect:discussion:lisa}
In this paper we focused on properties for ground-based GW detectors such as LIGO and VIRGO. Here, we briefly comment on the potential for detecting GWs from compact objects in our channel with space-based detectors such as LISA ($0.1\,\mathrm{mHz} \lesssim f_\mathrm{GW} \lesssim 10\,\mathrm{mHz}$). In our simulations, we considered objects to have merged when the periapsis distance $r_\mathrm{p} = \alpha \, G M/c^2$, where $M$ is the total binary mass, and where we set $\alpha=1000$ (see \S~\ref{sect:meth:sec}). If the orbit at this point is highly eccentric, then the GW frequency can be estimated as
\begin{align}
f_\mathrm{GW} \sim \frac{v_\mathrm{p}}{r_\mathrm{p}} = \frac{c^3}{G M} \alpha^{-3/2} \simeq 0.11\,\mathrm{Hz} \, \left ( \frac{M}{60\,\msun} \right )^{-1} \left (\frac{\alpha}{1000} \right )^{-3/2},
\end{align}
where we set the periapsis speed to $v_\mathrm{p} \sim \sqrt{GM/r_\mathrm{p}}$. This frequency is significantly above the LISA band, hence the system may have emitted GWs in the LISA band before reaching the periapsis distance $r_\mathrm{p} = \alpha \, G M/c^2$, and effects associated with a third or fourth body in the system might be detectable. In addition, the typical eccentricity in the LISA band ($10^{-4} \, \mathrm{Hz} \lesssim f_\lisa \lesssim 10^{-3} \, \mathrm{Hz}$) can be roughly estimated by noting that $f_\mathrm{GW} \propto a^{-3/2} \propto e^{-18/19} \sim e^{-1}$ in the limit of small eccentricity \citep{1964PhRv..136.1224P}. Therefore, $e_\lisa \sim e_\ligo \, (f_\ligo/f_\lisa)$, so $e_\lisa \sim 0.01-0.1$ for the dominant population of first-generation mergers with $e_\ligo \sim 10^{-6}$ (cf. \F~\ref{fig:e_LIGO_m1}), which may be measurable with LISA.  We leave a more detailed investigation fur future work.

\section{Conclusions}
\label{sect:conclusions}
Hierarchical quadruple systems are potential sources of BH and NS mergers, and in particular they might produce compact object mergers in the low and high-end mass gap regimes (with masses between $\sim 2$-$5\,\msun$ and exceeding $\sim 50\,\msun$, respectively) if successive mergers occurred. Here, we considered repeated mergers of NSs and BHs in 2+2 quadruples, taking into stellar/binary evolution and dynamical evolution (treating them separately). Secular evolution could accelerate the merger of one of the inner binaries after compact object formation. Subsequently, the merger remnant could interact with the companion binary, yielding a second-generation merger event of compact objects. Our main conclusions are given below. 

\medskip \noindent 1. We treated the evolution of the 2+2 quadruple in a simplified manner in which two massive binaries are first evolved as if in isolation, using a rapid binary population synthesis code. To the surviving BH/NS binaries, we then assigned an outer orbit in a stable configuration. Of the subsequent 2+2 quadruple systems, a significant fraction ($\sim 15$-$70\%$ depending on the model assumptions, see Table~\ref{table:sec_fractions}) will undergo a merger in one of the two inner orbits within a Hubble time. Although secular chaotic evolution in 2+2 quadruple systems can generally lead to strong eccentricity excitation (e.g., \citealt{2013MNRAS.435..943P,2015MNRAS.449.4221H,2016MNRAS.461.3964V,2017MNRAS.470.1657H,2018MNRAS.476.4234F,2018MNRAS.474.3547G,2019MNRAS.483.4060L,2019MNRAS.486.4781F}), we find here that the majority of merging compact object binaries can be considered to be dynamically decoupled from the outer orbit. In other words, they are sufficiently close that they would merge within a Hubble time (as a result of binary interactions prior to compact object formation), and no secular eccentricity excitation is needed to drive the merger. 

\medskip \noindent 2. The majority of first-generation mergers are decoupled and show a distribution of eccentricity in the LIGO band ($e_\ligo$; GW peak frequency reaching 10 Hz) which is peaked near $10^{-6}$ (\F~\ref{fig:e_LIGO_m1}), similarly to isolated binaries. In a small number of systems, secular evolution results in efficient eccentricity excitation, and significant eccentricity (with $e_\ligo$ up to $\sim10^{-2}$). The latter result is based on secular integrations and so likely represents a lower limit on $e_\ligo$ (e.g., \citealt{2016ApJ...816...65A}).

\medskip \noindent 3. We considered second-generation mergers by continuing the evolution of first-generation merger systems, assigning a GW recoil velocity to the merger remnant, and following the subsequent gravitational dynamical evolution with direct $N$-body integrations. In particular, the first-generation merger remnant can perturb the companion binary (either a weak perturbation, i.e., a fly-by, or a strong interaction possibly involving exchange interactions), potentially accelerating the companion binary's merger, or causing direct mergers between the first-generation remnant and a component in the companion binary. The latter type constitutes a second-generation merger. The scattering effect of the first-generation merger remnant on the companion binary is significant: the companion binary has a $\sim 1.5$ larger probability of merging when scattering is taken into account.

\medskip \noindent 4. Although highly uncertain due to observational uncertainties and simplifications made in the numerical modelling, we made rough estimates of first and second-generation merger rates of BH and NS mergers in 2+2 quadruples. We stress that our estimated rates are likely overestimates given that we did not self-consistently take into account the effects of mass loss on the outer orbit (the latter will be addressed in future work). Depending on the model assumptions (see Table~\ref{table:rates}), the combined merger rate in the local Universe of BH-BH, NS-NS, and BH-NS mergers is on the order of $\sim 10^{1}$ and up to $\sim 10^2 \, \pgpc\,\pyr$ for first-generation mergers, and $\sim 10^{-5} \, \pgpc \, \pyr$ for second-generation mergers. These estimates are roughly consistent with previous studies (which were typically based on less comprehensive population synthesis calculations). Overall, we find that the rate of first-generation mergers dominates over that of second-generation mergers by a large factor of $\sim 10^7$.

\medskip \noindent 5. Second-generation mergers are able to populate most of the lower-end mass gap (between 2 and 5 $\msun$) when scattering is taken into account (see \F~\ref{fig:2g_mod11_mass}). This can be attributed to repeated mergers. However, we find that repeated mergers in 2+2 quadruples cannot explain the high-end mass gap (masses in excess of 50 $\msun$). 

\medskip \noindent 6. For first-generation mergers that are decoupled, the distributions of the effective spin parameter, $\chi_\eff$, reflect the assumed initial conditions (i.e., symmetric around 0.5 assuming initial individual dimensionless spin parameters of $0<\chi_i<1$, and small initial obliquities). Some imprint of secular eccentricity excitation exists in a smaller subset of systems, which show negative $\chi_\eff$ consistent with a dynamical origin.

\medskip \noindent 7. The mass distributions of first-generation mergers strongly reflect the initial distribution of masses after compact object formation. Secular evolution plays no major role in shaping these distributions. 

\medskip \noindent 8. When scattering is not taken into account, second-generation mergers show a distribution of $e_\ligo$ similar to that of isolated binaries (peaking around $10^{-6}$). Scattering processes, on the other hand, produce a tail in the distribution extending up to $e_\ligo\sim 0.1$, although with low probability.

\medskip \noindent 9. Second-generation mergers show a distribution of $\chi_\eff$ peaked around 0, consistent with a dynamical origin. Without scattering, their distribution of $\chi_\eff$ is fully consistent with that of isolated binaries.

\section*{Acknowledgements}
We thank the anonymous referees for critical and helpful reports. A.S.H. thanks the Max Planck Society for support through a Max Planck Research Group. GF acknowledges support from CIERA at Northwestern University. This work received founding from the European Research Council (ERC) under the European Union's Horizon 2020 Programme for Research and Innovation ERC-2014-STG under grant agreement No. 638435 (GalNUC) (to BK).

\section*{Data availability}
The data underlying this article are available in the article and in its online supplementary material.

\bibliographystyle{mnras}
\bibliography{literature}

\label{lastpage}
\end{document}


\title[Supplementary appendices]{Supplementary appendices to {\it First and second-generation black hole and neutron star mergers in 2+2 quadruples: population statistics}}
\author[Hamers et al.]{Adrian S. Hamers$^{1}$\thanks{E-mail: hamers@mpa-garching.mpg.de}, Giacomo Fragione$^{2,3}$, Patrick Neunteufel$^{1}$, and Bence Kocsis$^{4}$\\
$^{1}$Max-Planck-Institut f\"{u}r Astrophysik, Karl-Schwarzschild-Str. 1, 85741 Garching, Germany \\
$^2$Center for Interdisciplinary Exploration \& Research in Astrophysics (CIERA), Evanston, IL 60202, USA\\
$^3$Department of Physics \& Astronomy, Northwestern University, Evanston, IL 60202, USA\\
$^4$Rudolf Peierls Centre for Theoretical Physics, Clarendon Laboratory, Parks Road, Oxford OX1 3PU, UK}
\label{firstpage}
\pagerange{\pageref{firstpage}--\pageref{lastpage}}
\maketitle

\appendix

\section{Initial compact object systems}
\label{app:ICs}
\begin{figure}
\center
\includegraphics[scale = 0.35, trim = 5mm 10mm 0mm 0mm]{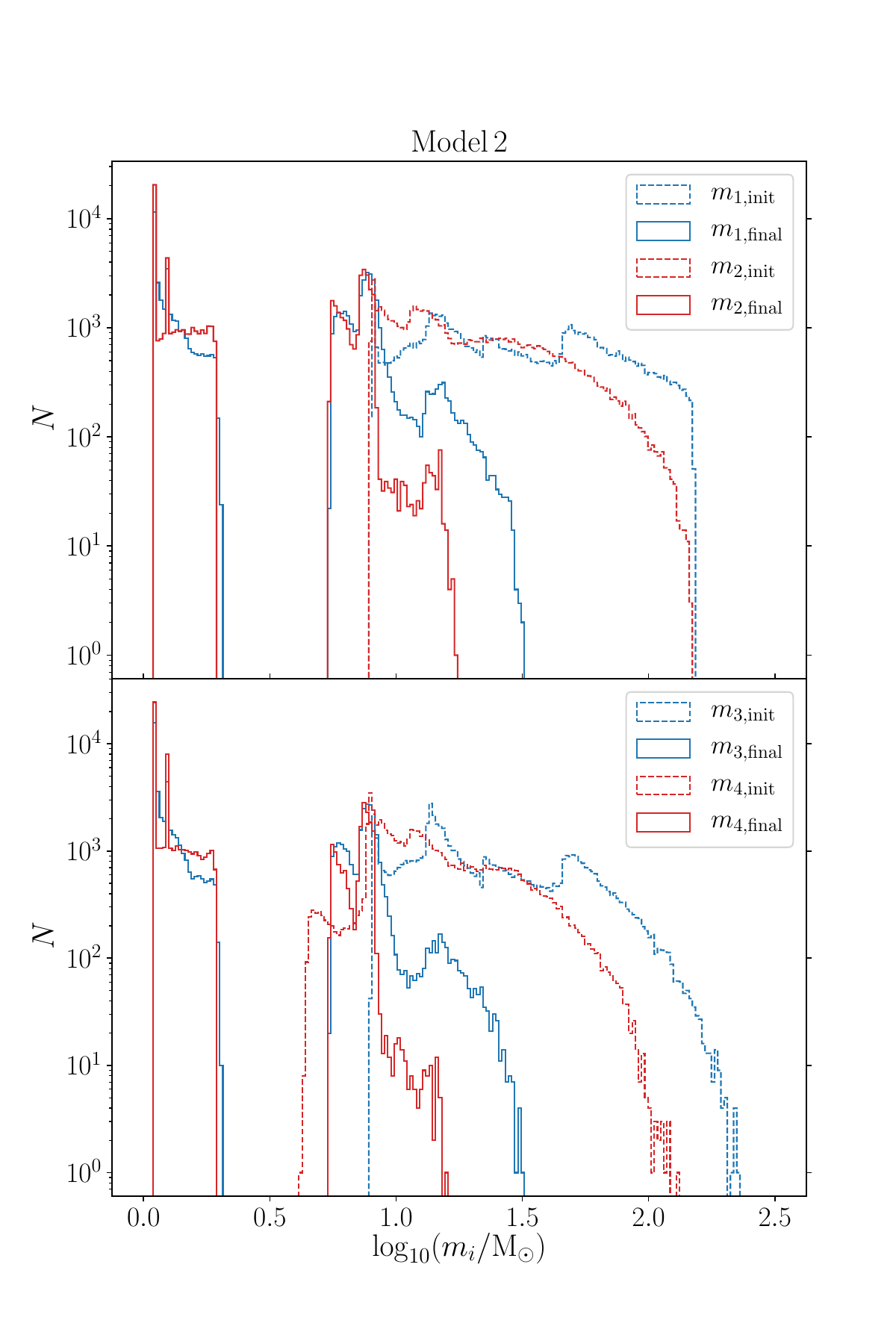}
\caption{ Similar to \ICsMassesMOne, here for Model 2. }
\label{fig:ics_masses_m2}
\end{figure}

\begin{figure}
\center
\includegraphics[scale = 0.35, trim = 5mm 10mm 0mm 0mm]{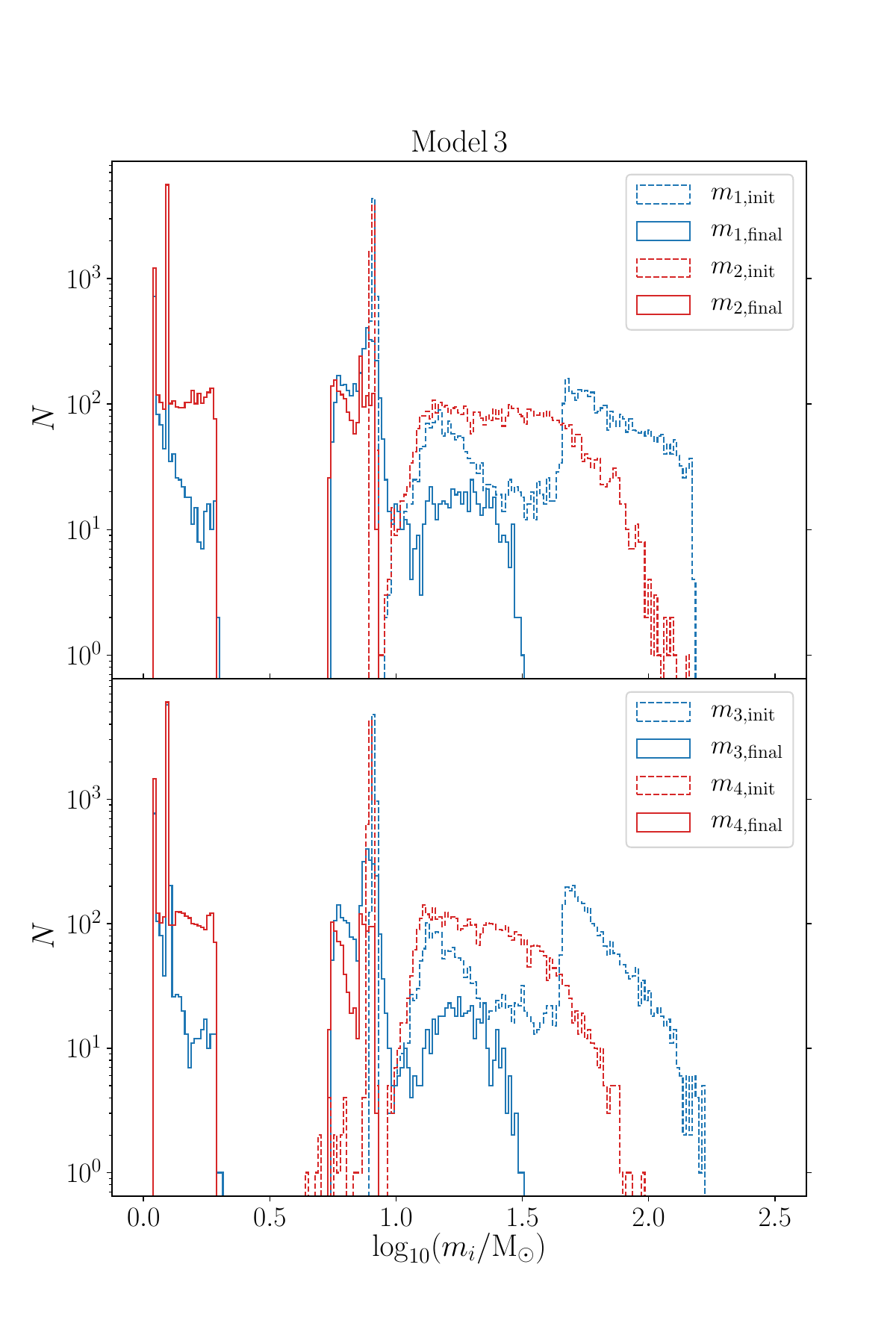}
\caption{ Similar to \ICsMassesMOne, here for Model 3. }
\label{fig:ics_masses_m3}
\end{figure}

\begin{figure}
\center
\includegraphics[scale = 0.35, trim = 5mm 10mm 0mm 0mm]{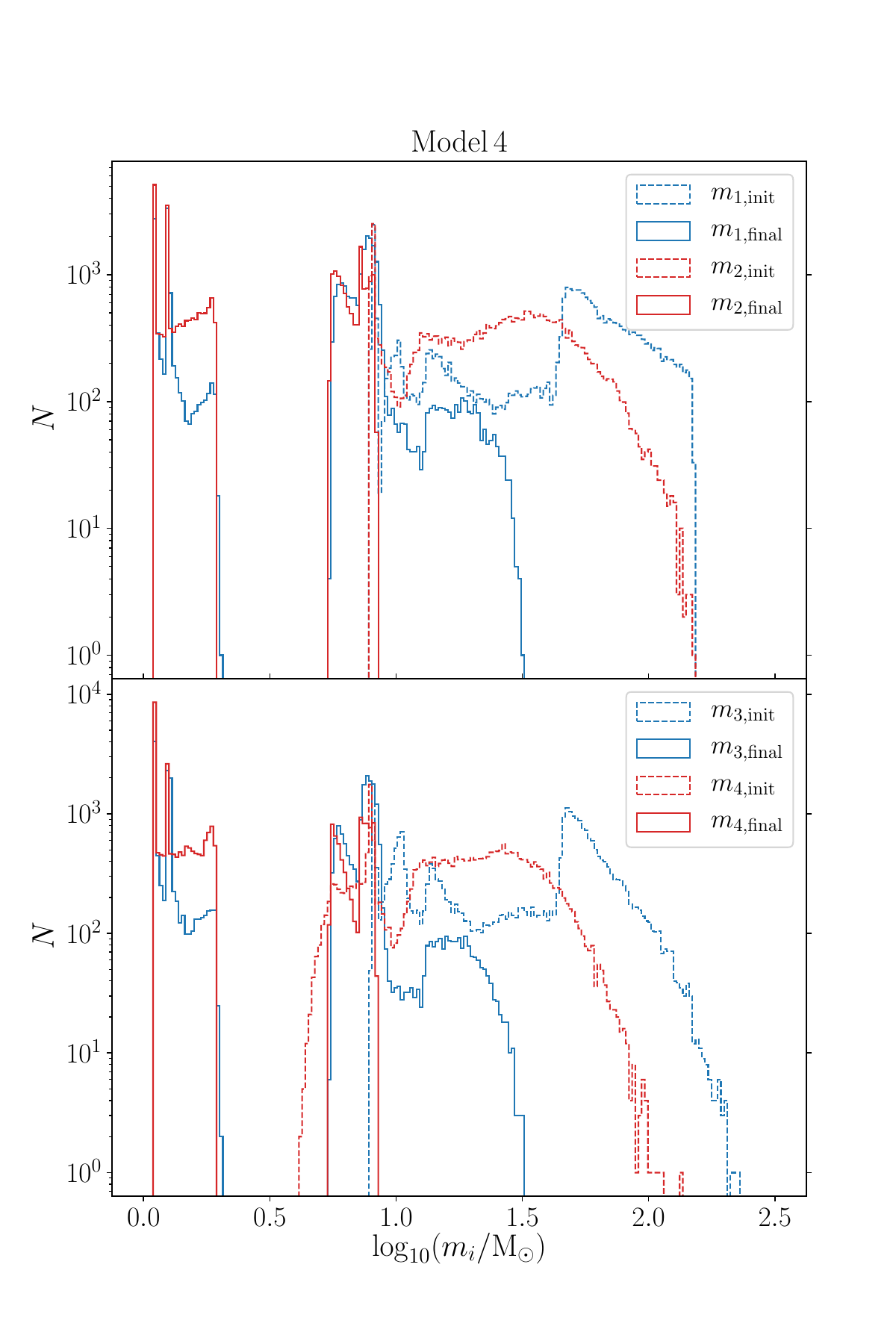}
\caption{ Similar to \ICsMassesMOne, here for Model 4. }
\label{fig:ics_masses_m4}
\end{figure}

\begin{figure}
\center
\includegraphics[scale = 0.35, trim = 5mm 10mm 0mm 0mm]{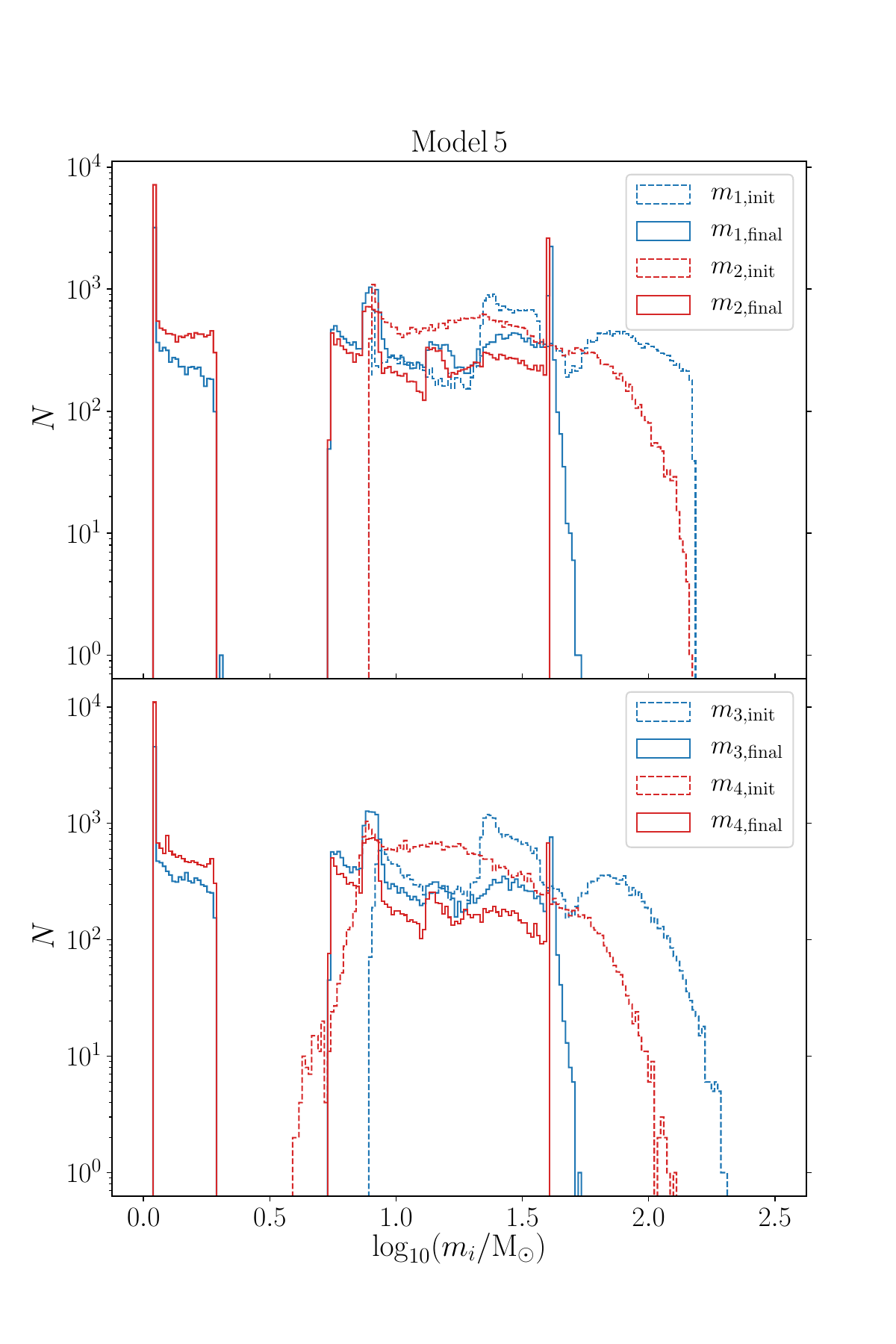}
\caption{ Similar to \ICsMassesMOne, here for Model 5. }
\label{fig:ics_masses_m5}
\end{figure}

\begin{figure}
\center
\includegraphics[scale = 0.35, trim = 5mm 10mm 0mm 0mm]{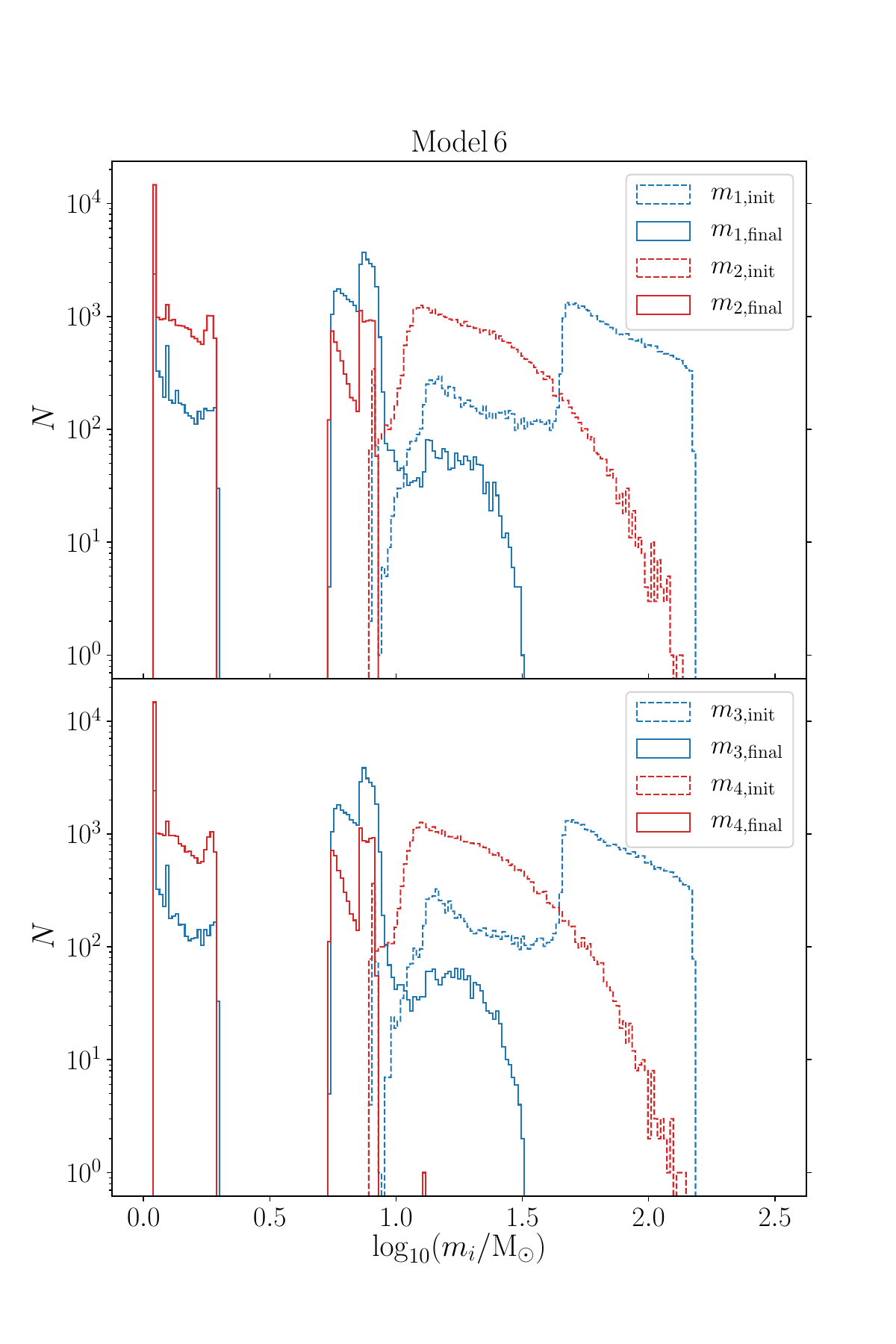}
\caption{ Similar to \ICsMassesMOne, here for Model 6. }
\label{fig:ics_masses_m6}
\end{figure}

\begin{figure}
\center
\includegraphics[scale = 0.35, trim = 5mm 10mm 0mm 0mm]{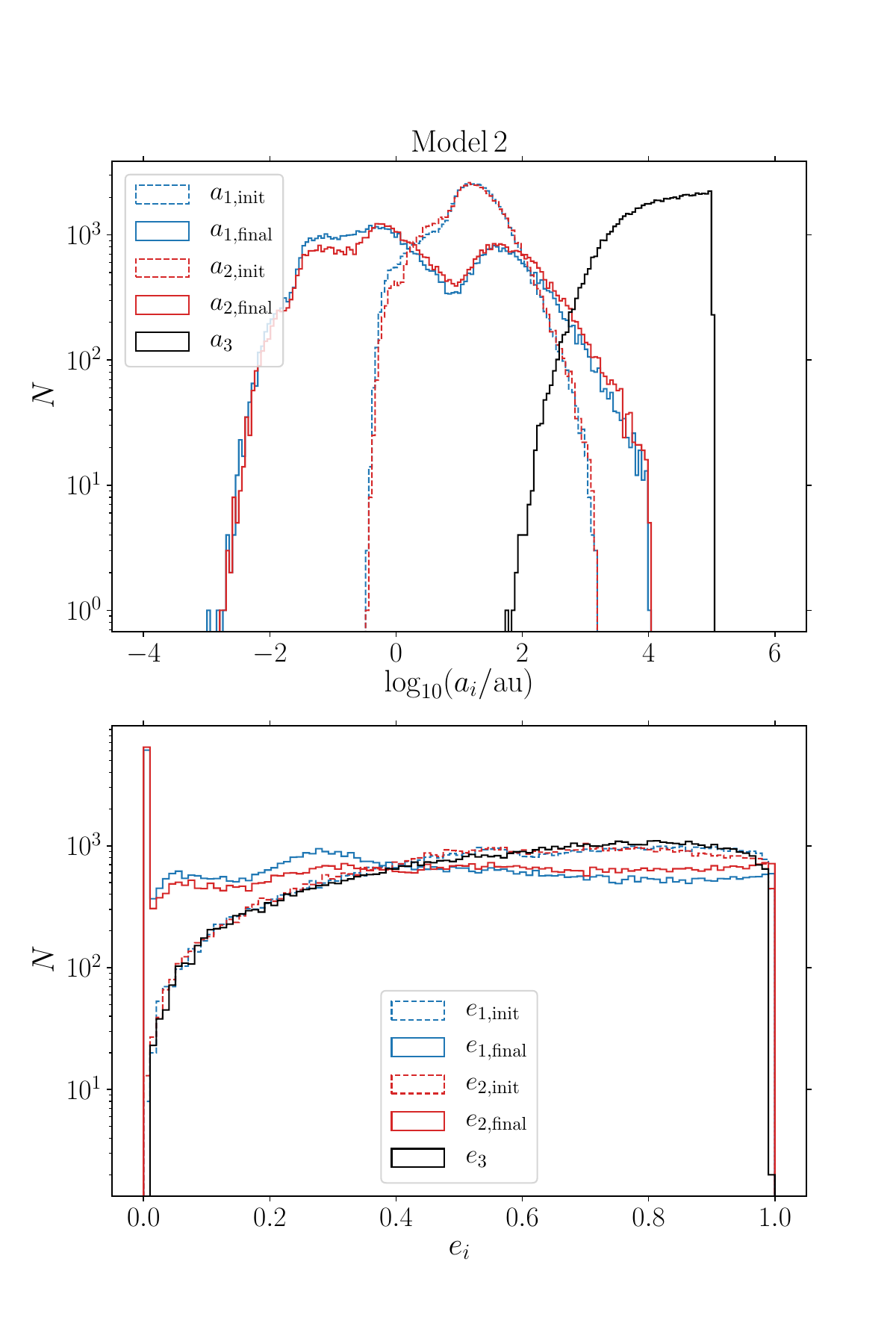}
\caption{ Similar to \ICsOrbitsMOne, here for Model 2. }
\label{fig:ics_orbits_m2}
\end{figure}

\begin{figure}
\center
\includegraphics[scale = 0.35, trim = 5mm 10mm 0mm 0mm]{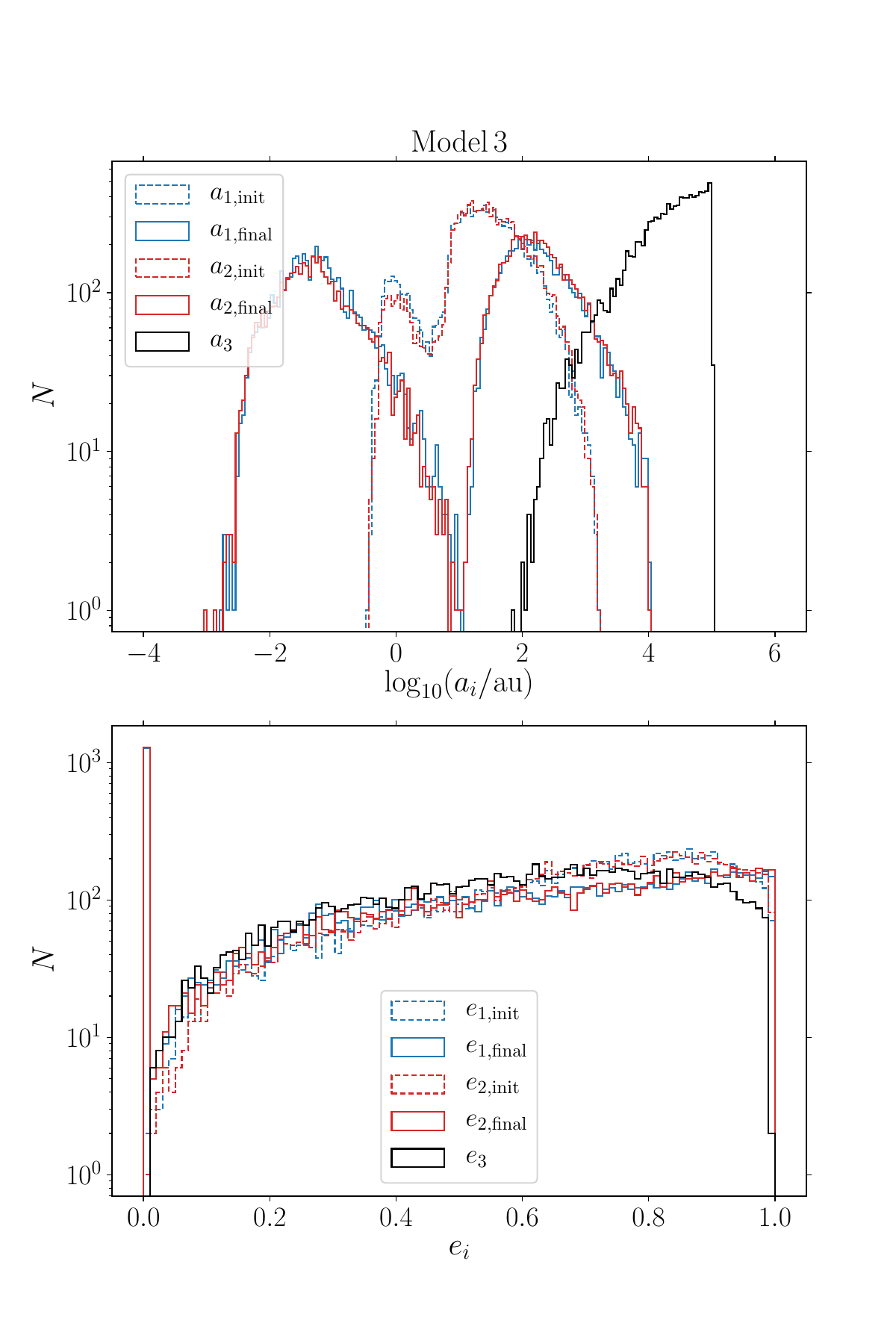}
\caption{ Similar to \ICsOrbitsMOne, here for Model 3. }
\label{fig:ics_orbits_m3}
\end{figure}

\begin{figure}
\center
\includegraphics[scale = 0.35, trim = 5mm 10mm 0mm 0mm]{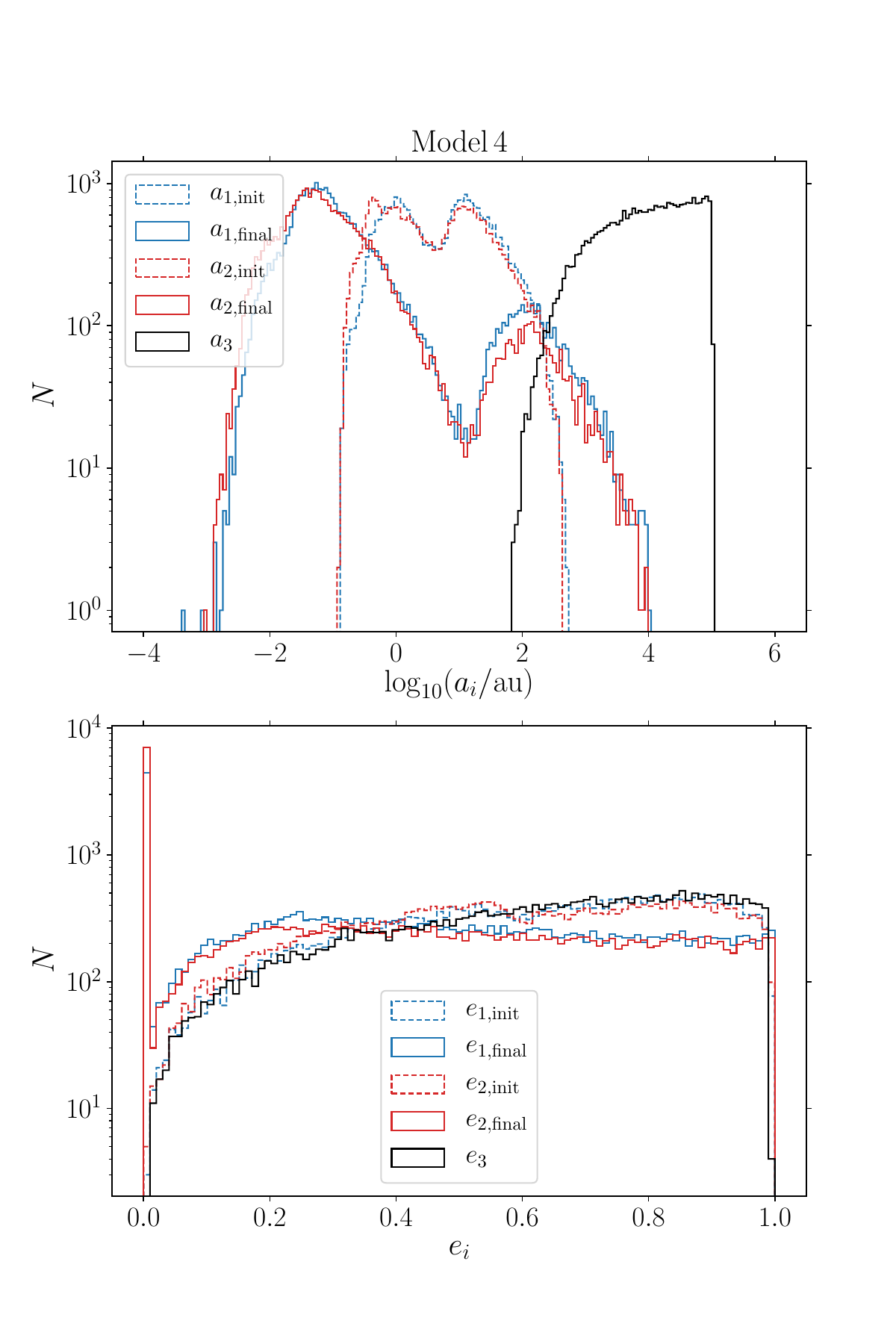}
\caption{ Similar to \ICsOrbitsMOne, here for Model 4. }
\label{fig:ics_orbits_m4}
\end{figure}

\begin{figure}
\center
\includegraphics[scale = 0.35, trim = 5mm 10mm 0mm 0mm]{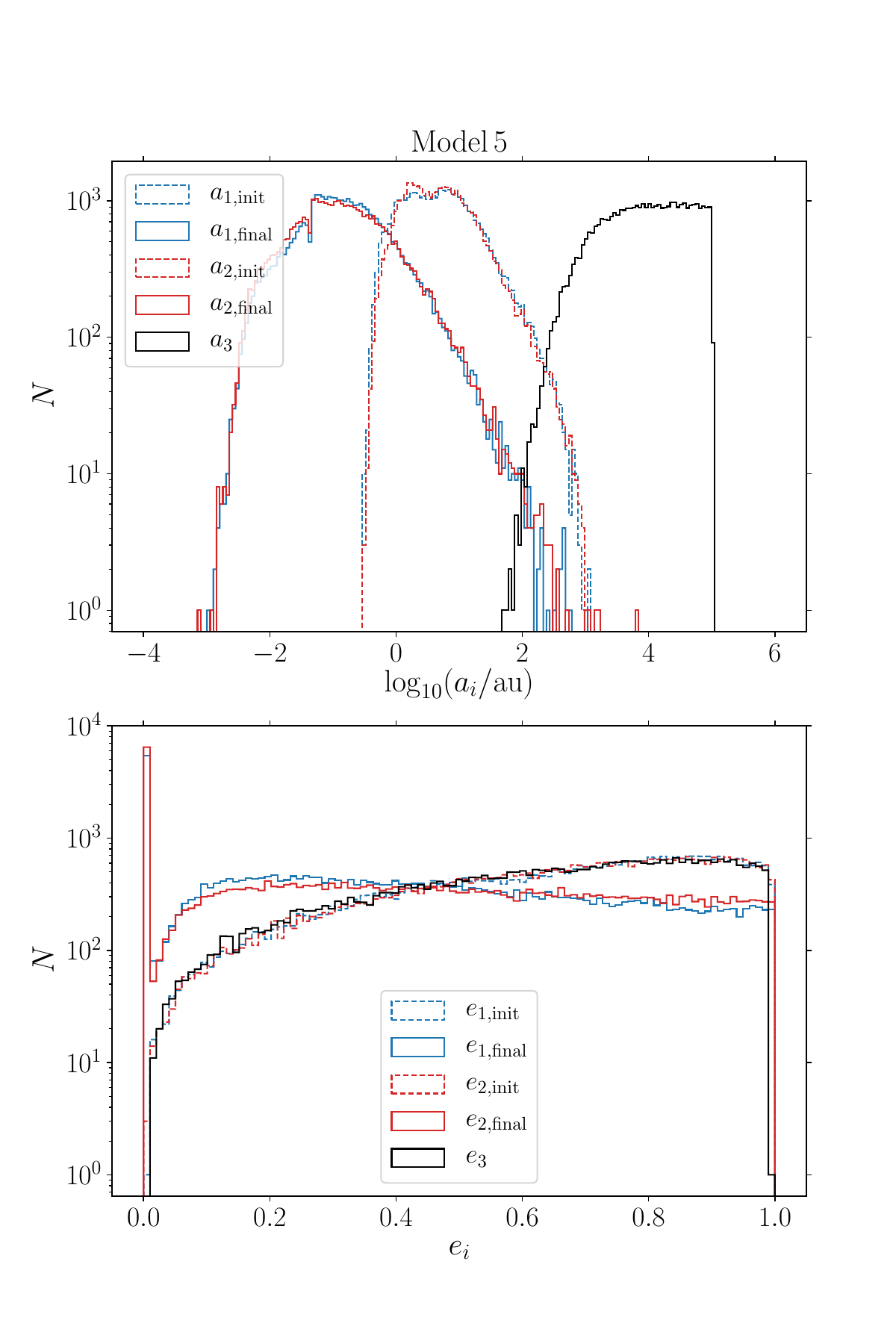}
\caption{ Similar to \ICsOrbitsMOne, here for Model 5. }
\label{fig:ics_orbits_m5}
\end{figure}

\begin{figure}
\center
\includegraphics[scale = 0.35, trim = 5mm 10mm 0mm 0mm]{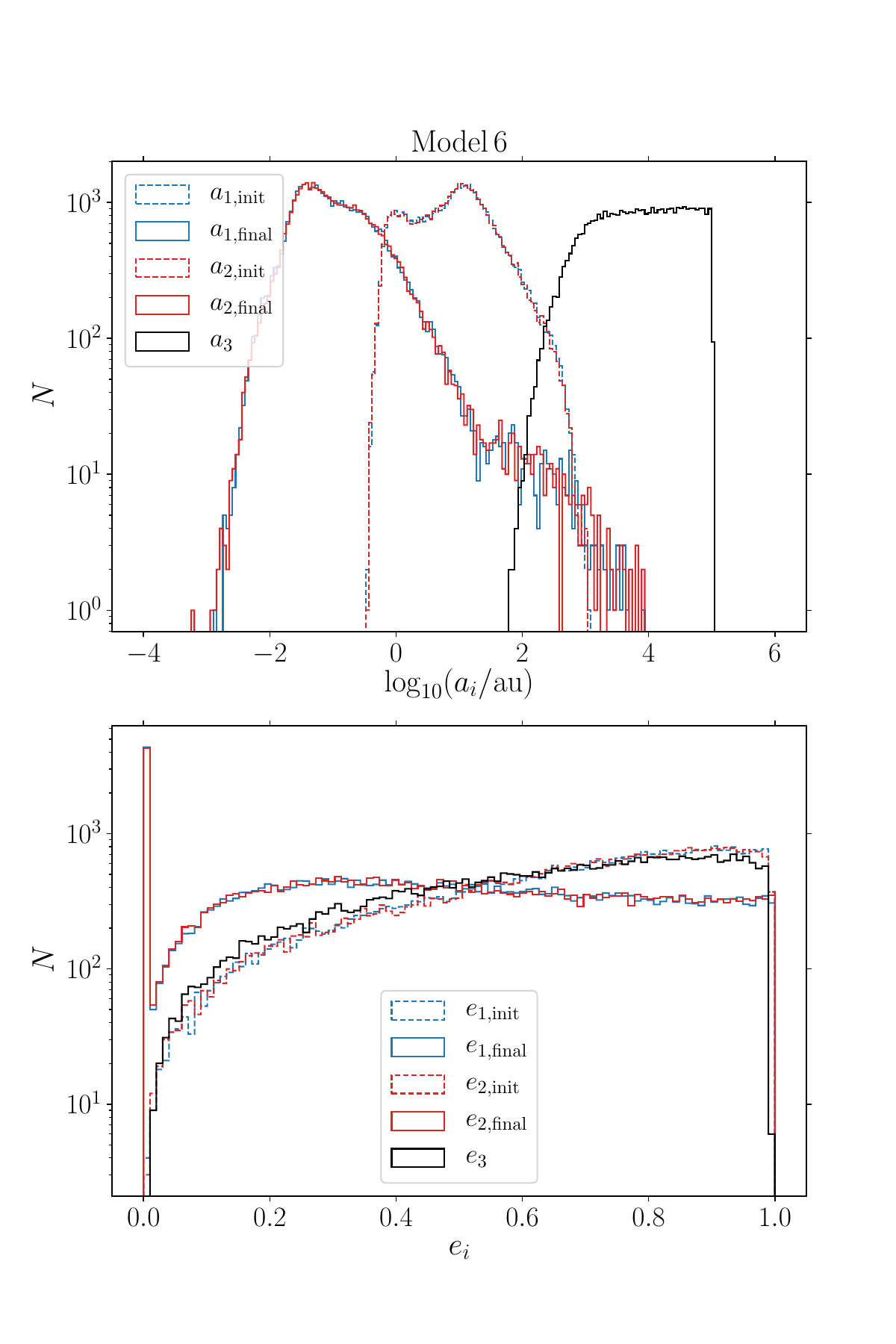}
\caption{ Similar to \ICsOrbitsMOne, here for Model 6. }
\label{fig:ics_orbits_m6}
\end{figure}

For the systems directly after compact object formation and for Models 2 through 6, we show in \Fs~\ref{fig:ics_masses_m2} through \ref{fig:ics_orbits_m6} the distributions of the masses, and orbital semimajor axes and eccentricities. 



\section{First-generation merger properties}
\label{app:gen1}
\begin{figure}
\center
\includegraphics[scale = 0.45, trim = 5mm 10mm 0mm 0mm]{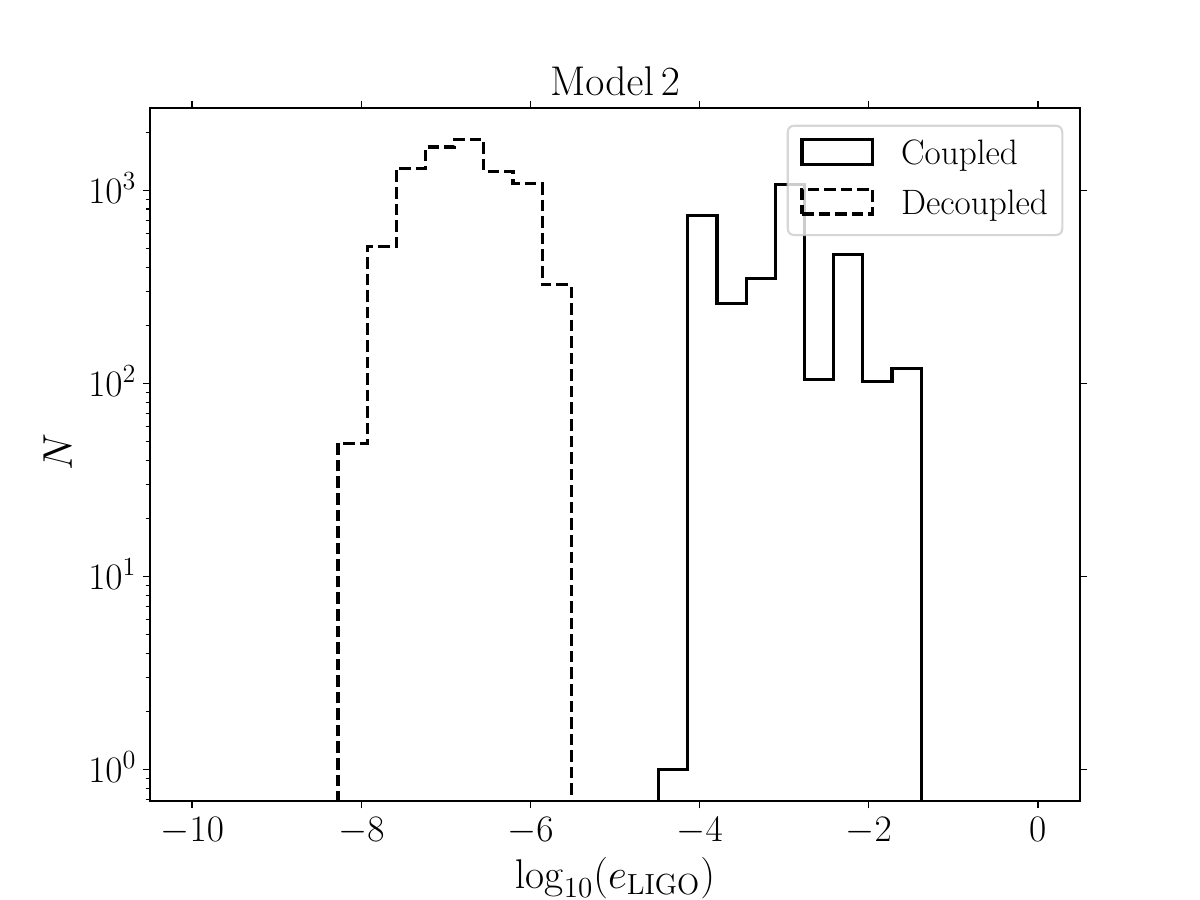}
\caption{ Similar to \eLIGOMOne, here for Model 2. }
\label{fig:e_LIGO_m2}
\end{figure}

\begin{figure}
\center
\includegraphics[scale = 0.45, trim = 5mm 10mm 0mm 0mm]{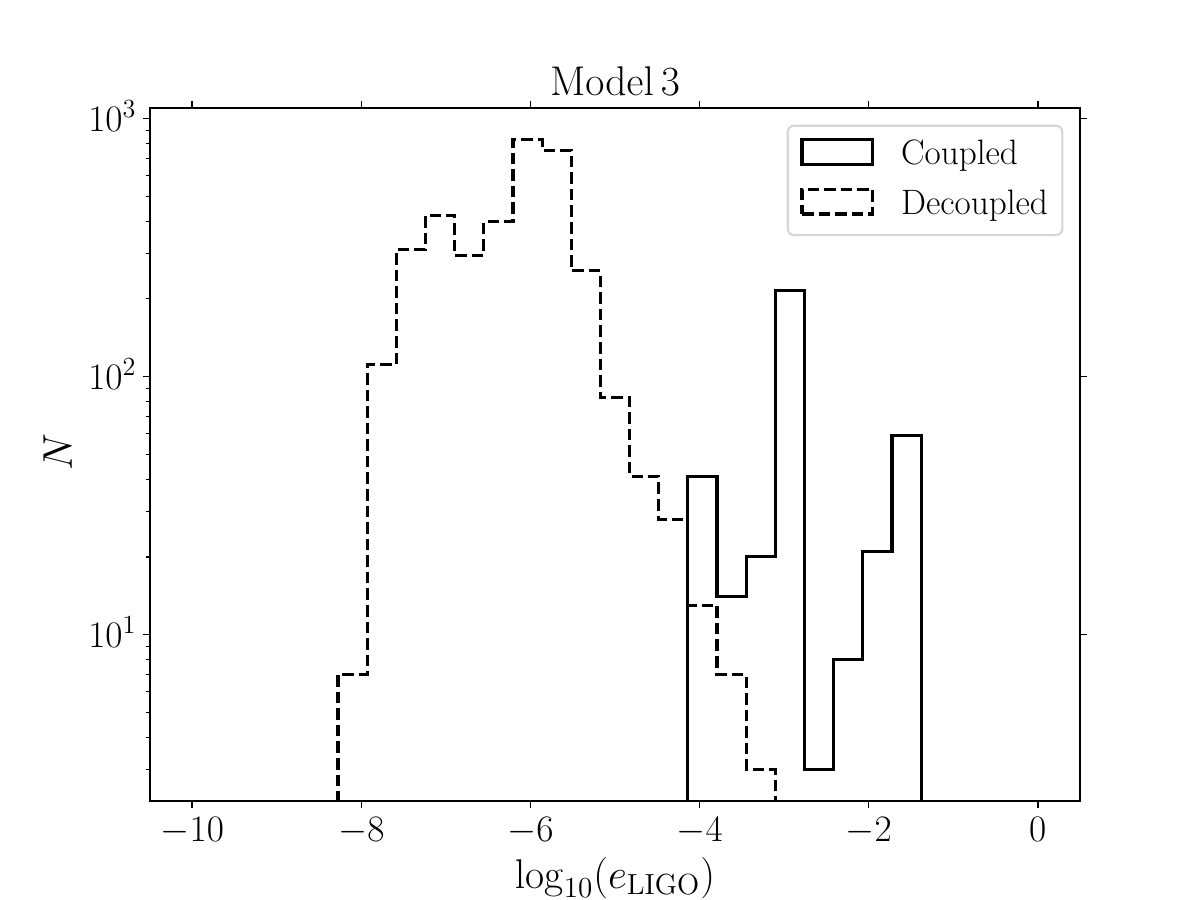}
\caption{ Similar to \eLIGOMOne, here for Model 3. }
\label{fig:e_LIGO_m3}
\end{figure}

\begin{figure}
\center
\includegraphics[scale = 0.45, trim = 5mm 10mm 0mm 0mm]{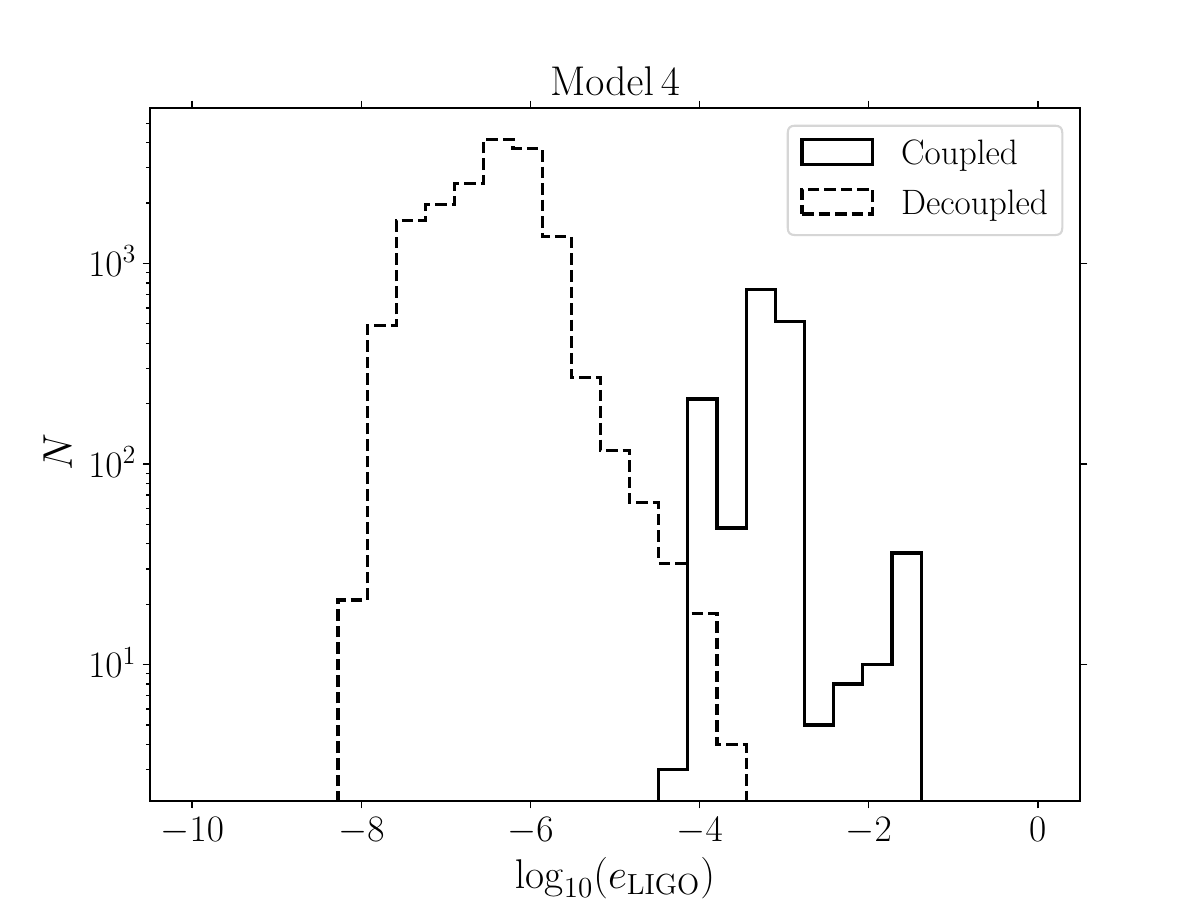}
\caption{ Similar to \eLIGOMOne, here for Model 4. }
\label{fig:e_LIGO_m4}
\end{figure}

\begin{figure}
\center
\includegraphics[scale = 0.45, trim = 5mm 10mm 0mm 0mm]{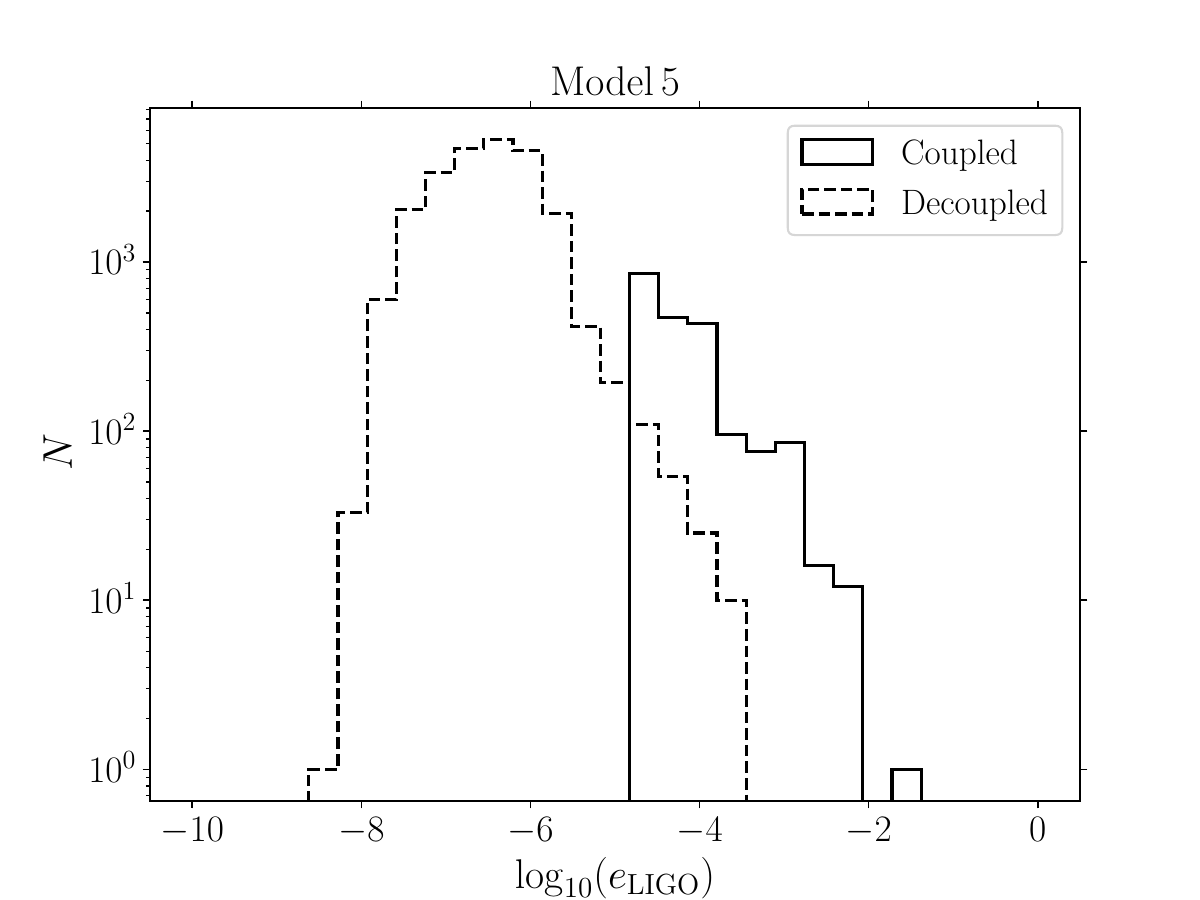}
\caption{ Similar to \eLIGOMOne, here for Model 5. }
\label{fig:e_LIGO_m5}
\end{figure}

\begin{figure}
\center
\includegraphics[scale = 0.45, trim = 5mm 10mm 0mm 0mm]{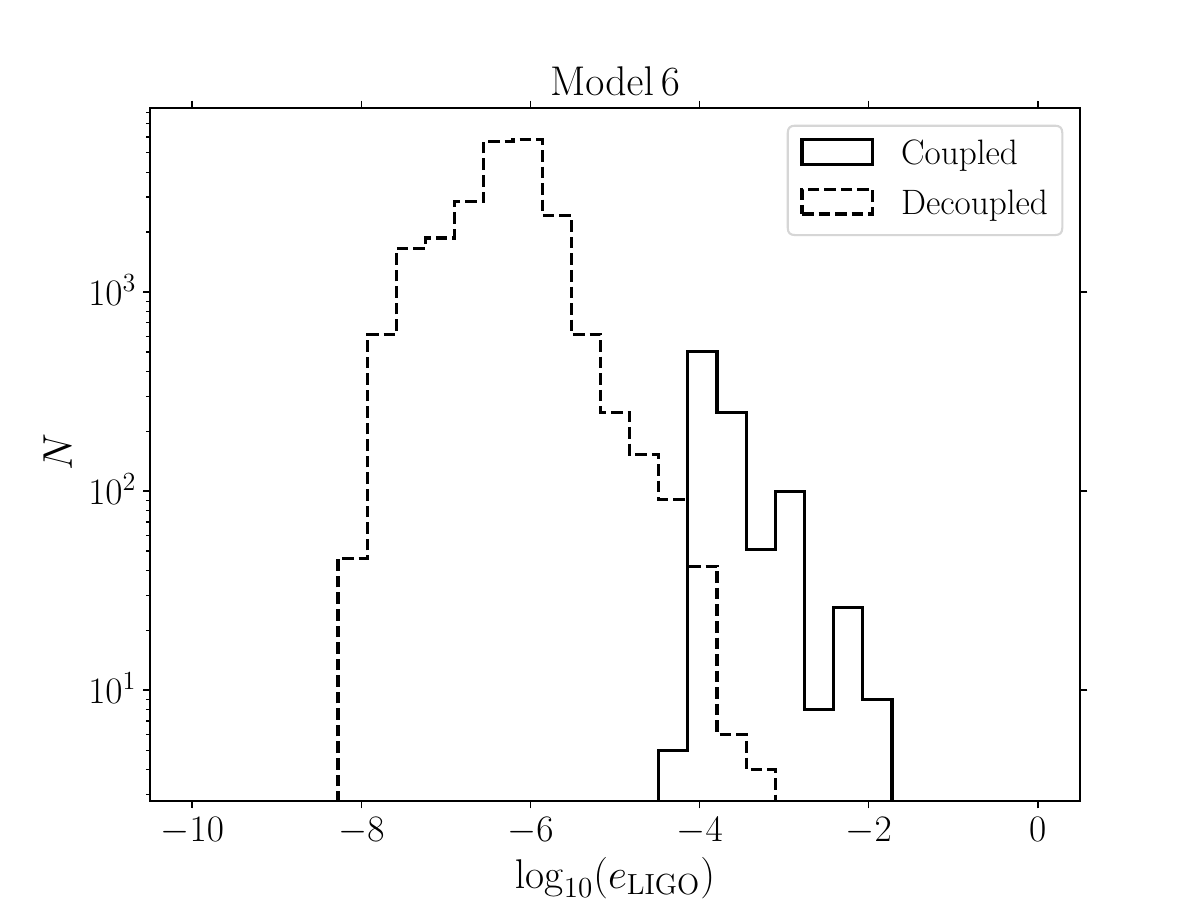}
\caption{ Similar to \eLIGOMOne, here for Model 6. }
\label{fig:e_LIGO_m6}
\end{figure}

\begin{figure}
\center
\includegraphics[scale = 0.45, trim = 5mm 10mm 0mm 0mm]{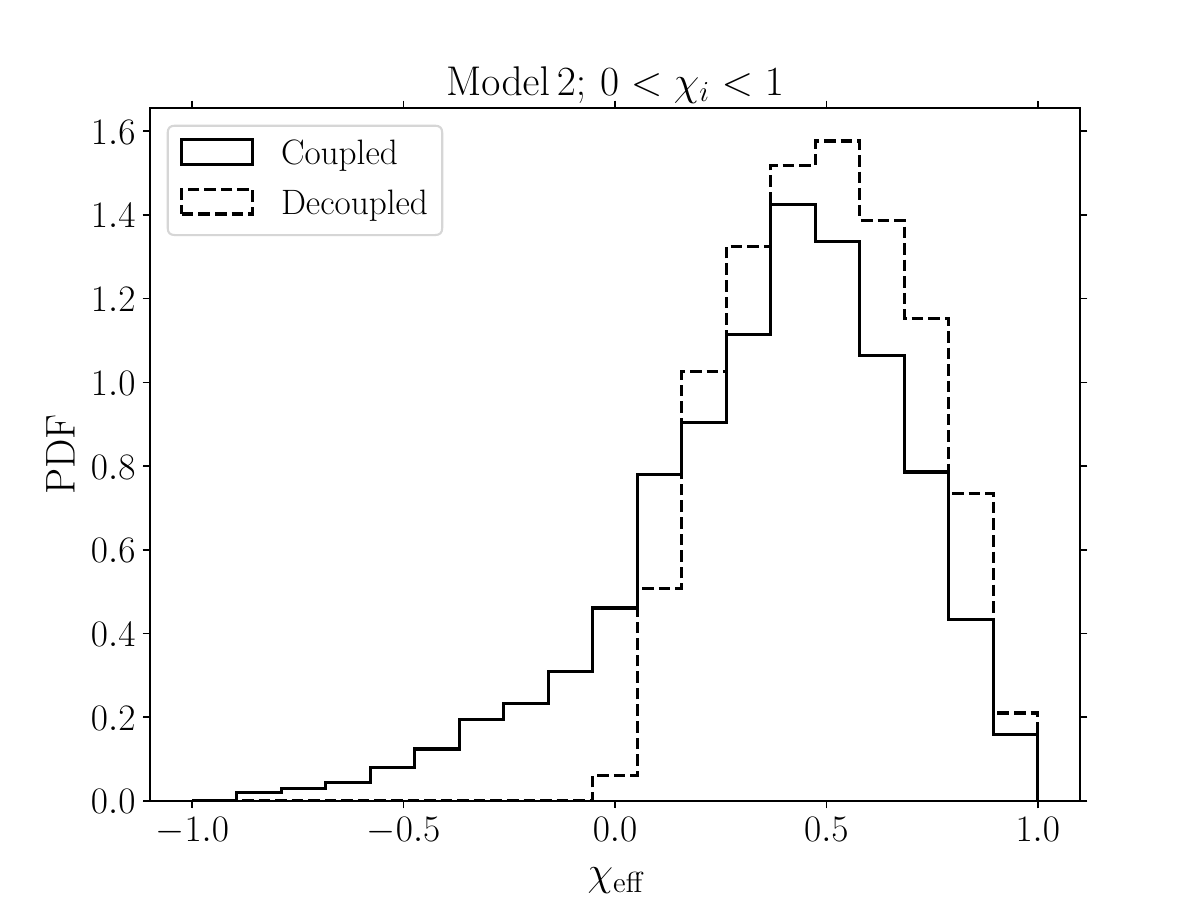}
\caption{ Similar to \ChiEffMOne, here for Model 2. }
\label{fig:chi_eff_m2}
\end{figure}

\begin{figure}
\center
\includegraphics[scale = 0.45, trim = 5mm 10mm 0mm 0mm]{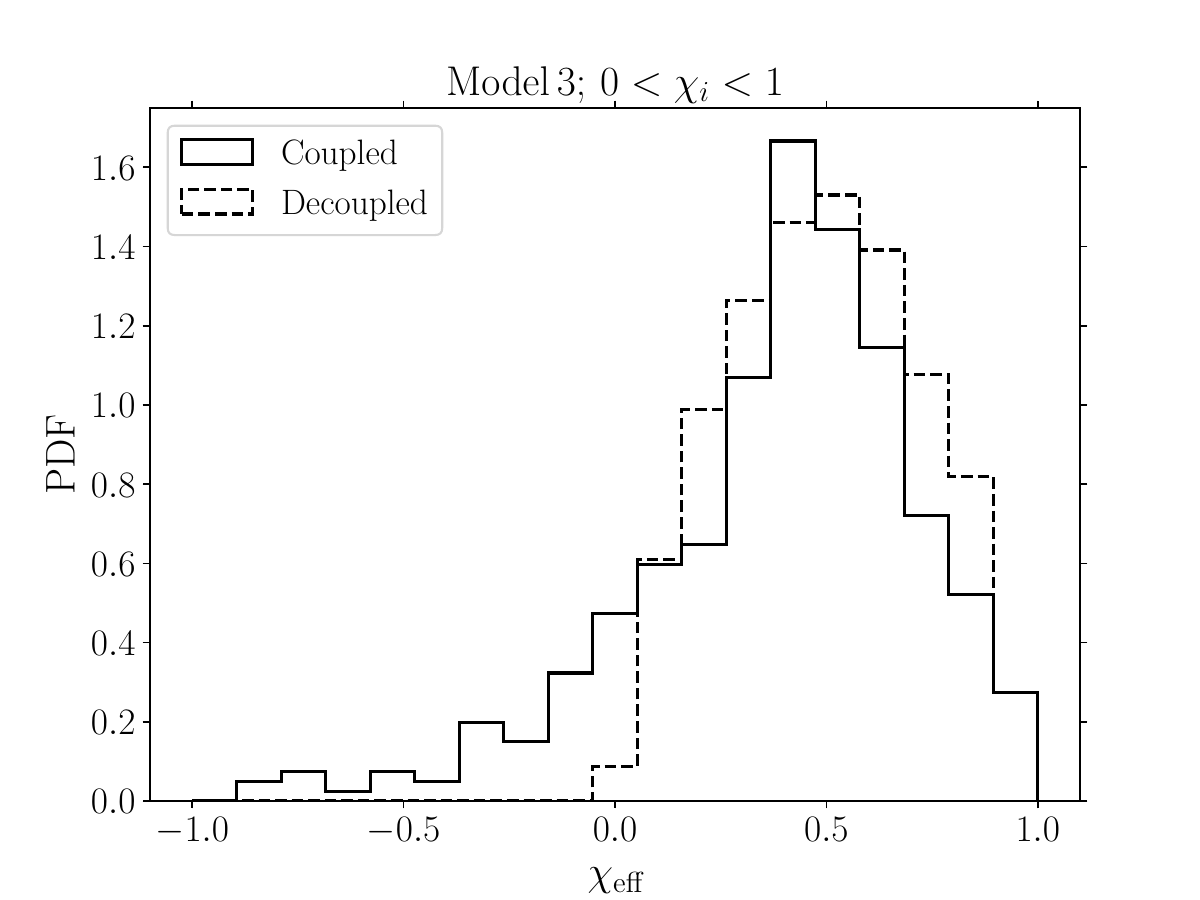}
\caption{ Similar to \ChiEffMOne, here for Model 3. }
\label{fig:chi_eff_m3}
\end{figure}

\begin{figure}
\center
\includegraphics[scale = 0.45, trim = 5mm 10mm 0mm 0mm]{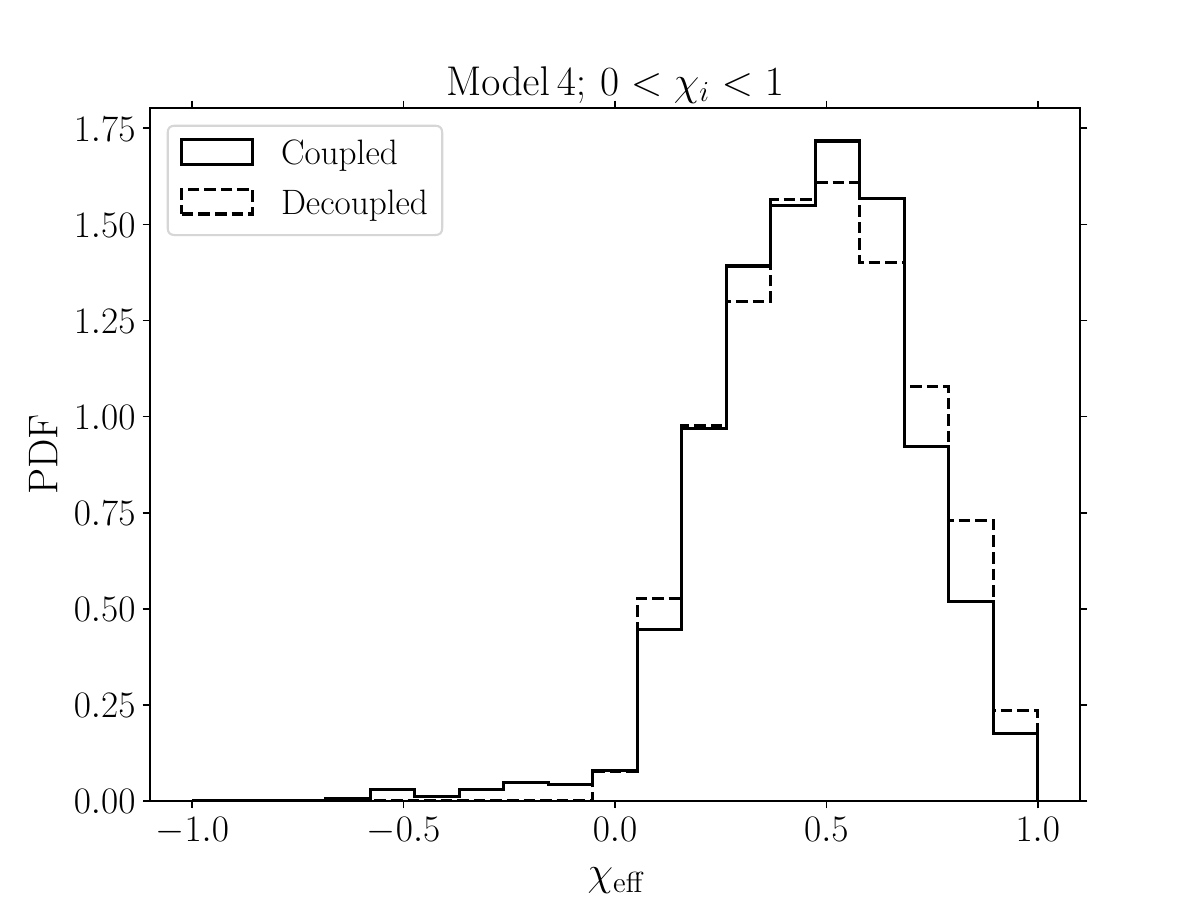}
\caption{ Similar to \ChiEffMOne, here for Model 4. }
\label{fig:chi_eff_m4}
\end{figure}

\begin{figure}
\center
\includegraphics[scale = 0.45, trim = 5mm 10mm 0mm 0mm]{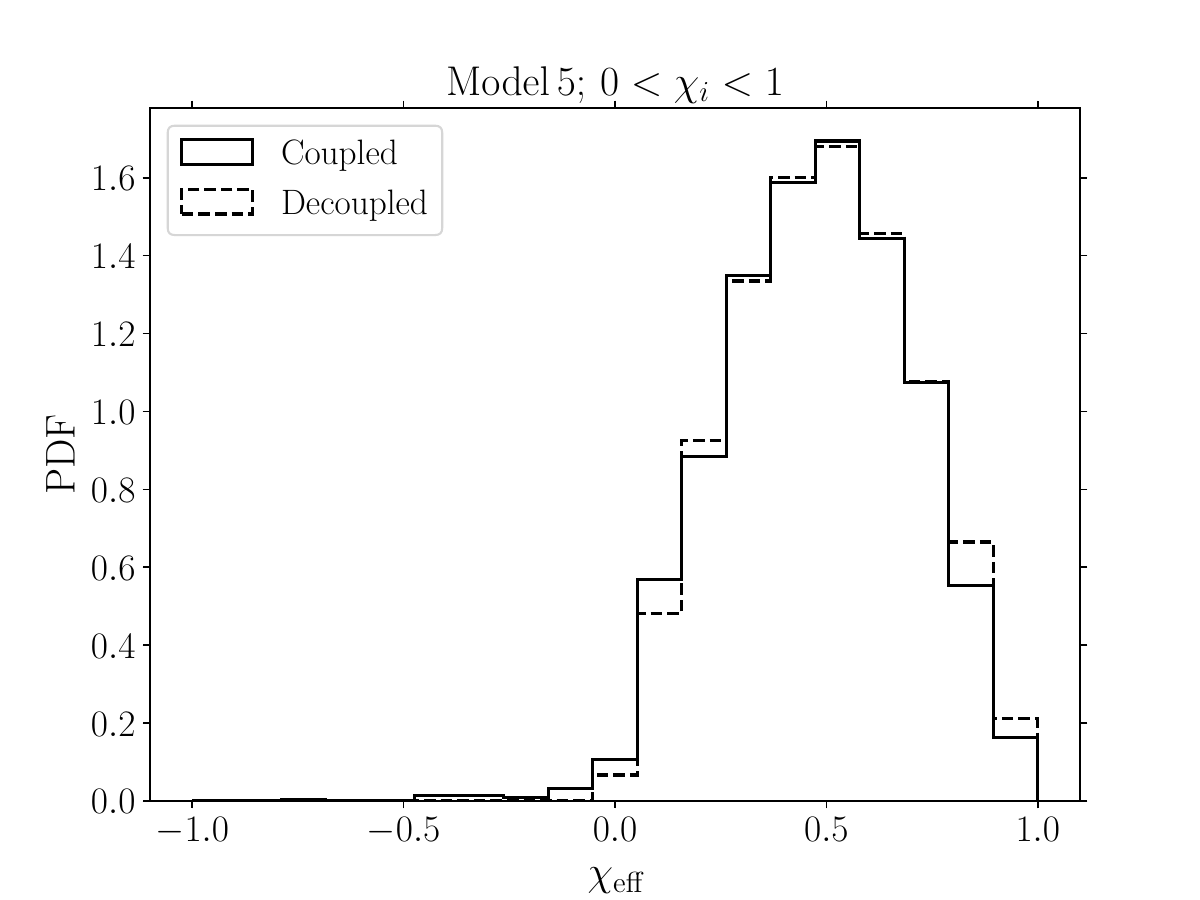}
\caption{ Similar to \ChiEffMOne, here for Model 5. }
\label{fig:chi_eff_m5}
\end{figure}

\begin{figure}
\center
\includegraphics[scale = 0.45, trim = 5mm 10mm 0mm 0mm]{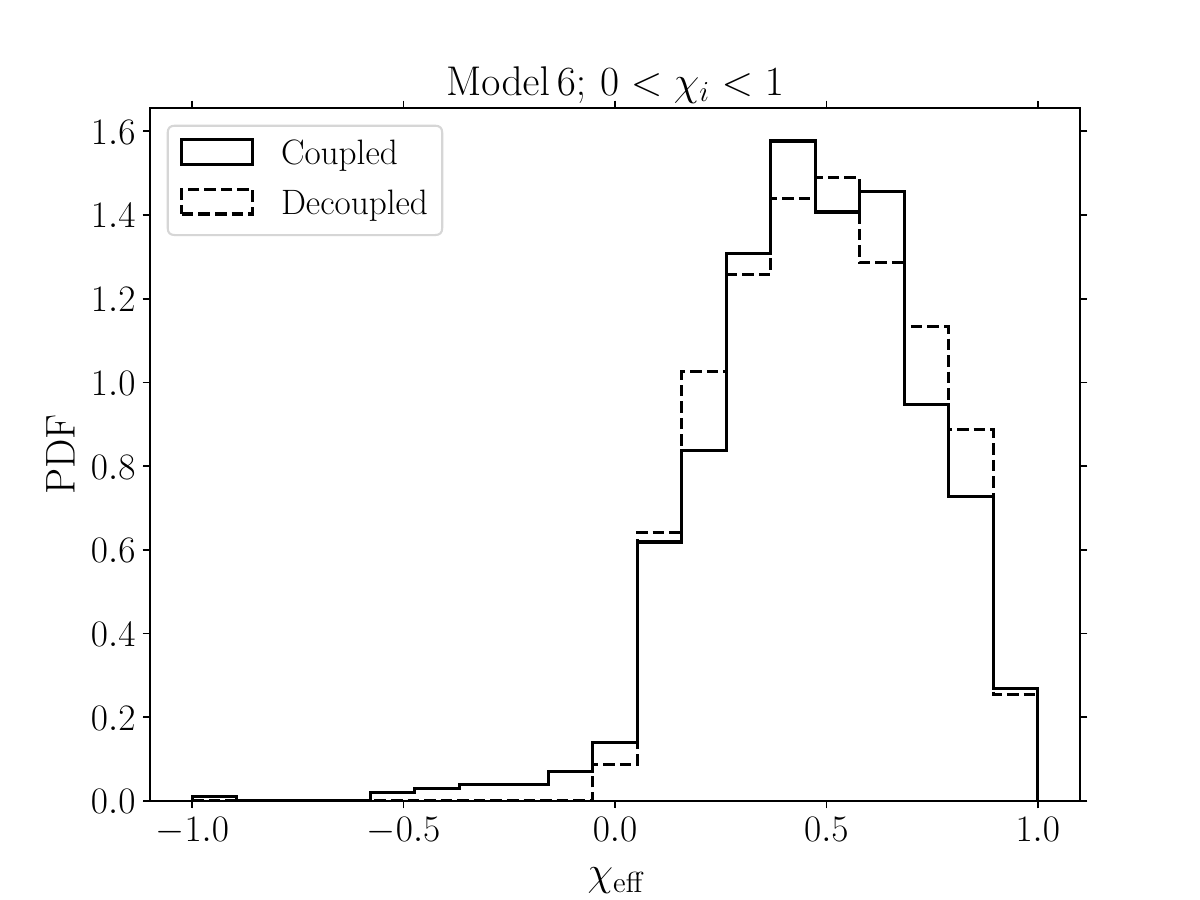}
\caption{ Similar to \ChiEffMOne, here for Model 6. }
\label{fig:chi_eff_m6}
\end{figure}

\begin{figure}
\center
\includegraphics[scale = 0.45, trim = 5mm 10mm 0mm 0mm]{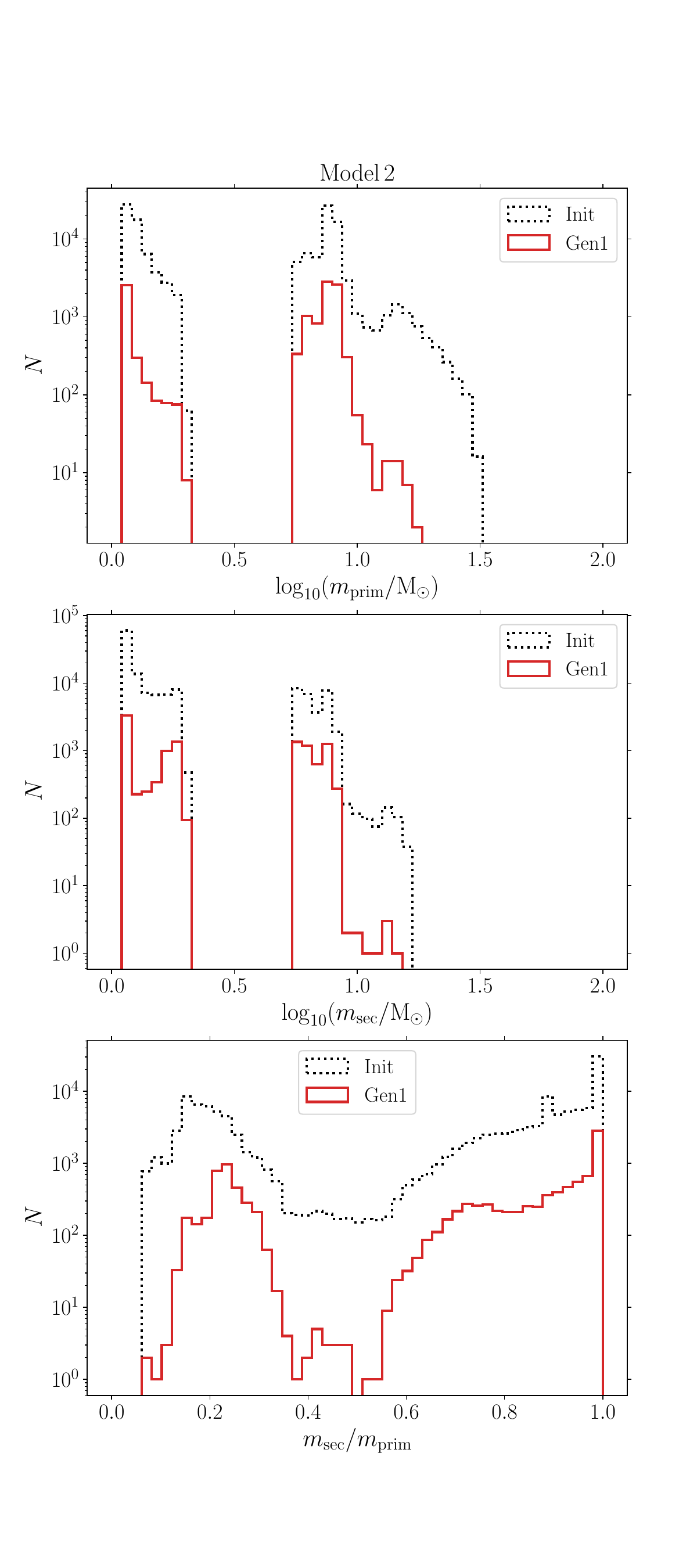}
\caption{ Similar to \MassesMOne, here for Model 2. }
\label{fig:masses_m2}
\end{figure}

\begin{figure}
\center
\includegraphics[scale = 0.45, trim = 5mm 10mm 0mm 0mm]{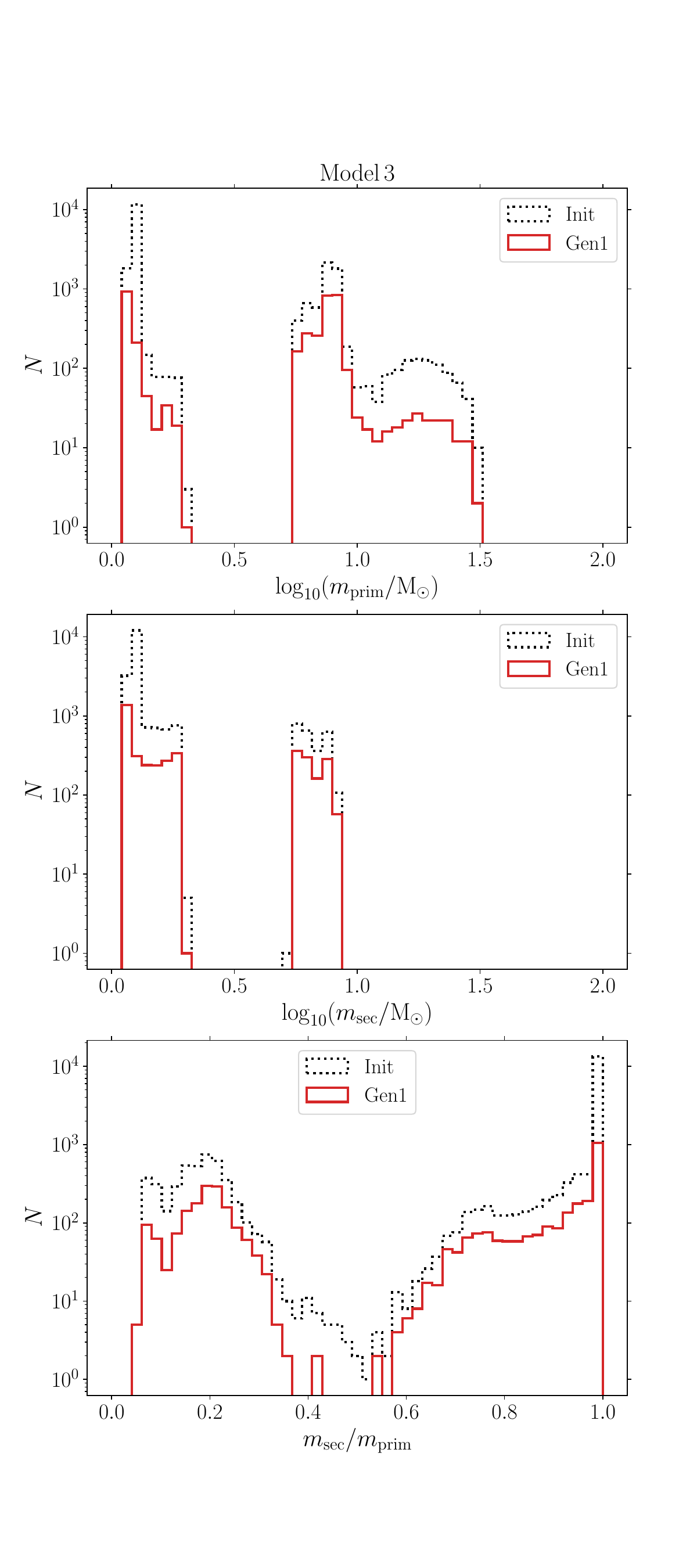}
\caption{ Similar to \MassesMOne, here for Model 3. }
\label{fig:masses_m3}
\end{figure}

\begin{figure}
\center
\includegraphics[scale = 0.45, trim = 5mm 10mm 0mm 0mm]{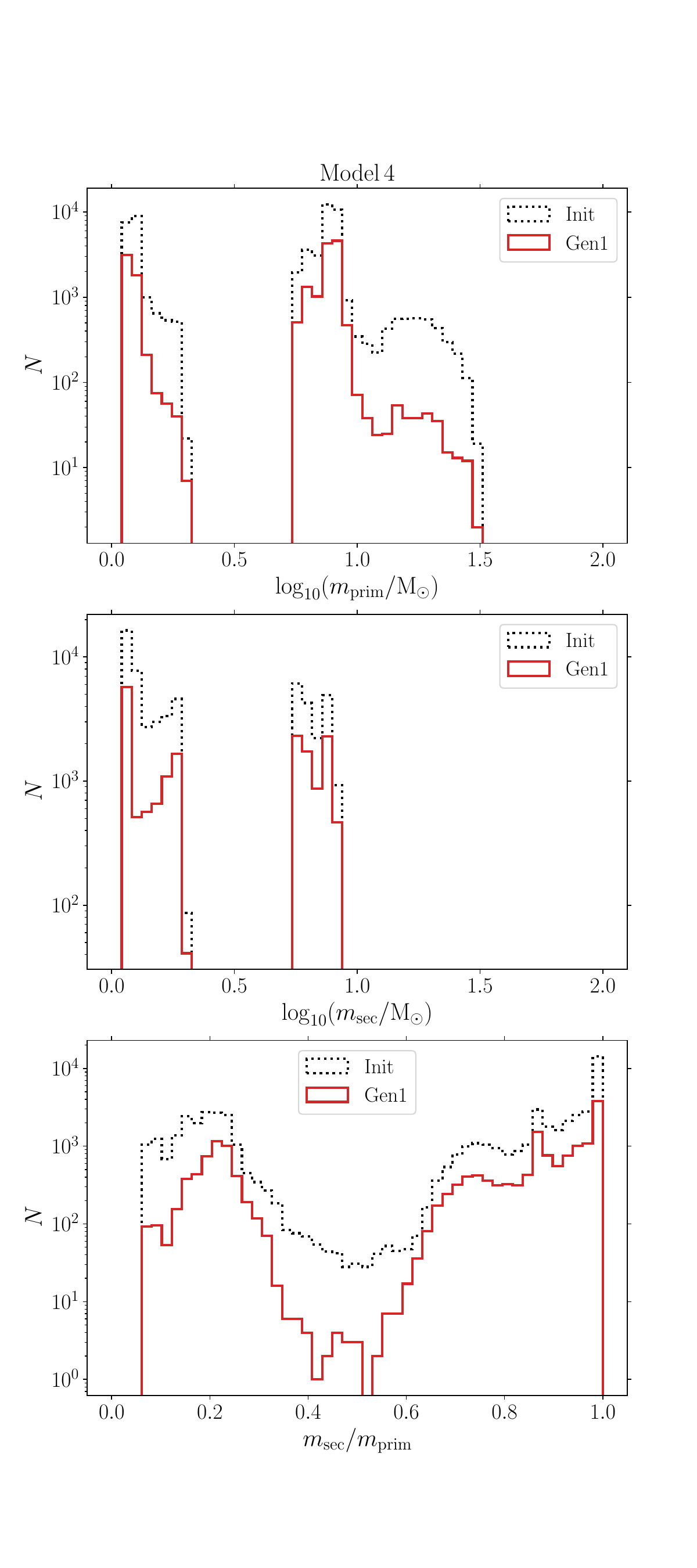}
\caption{ Similar to \MassesMOne, here for Model 4. }
\label{fig:masses_m4}
\end{figure}

\begin{figure}
\center
\includegraphics[scale = 0.45, trim = 5mm 10mm 0mm 0mm]{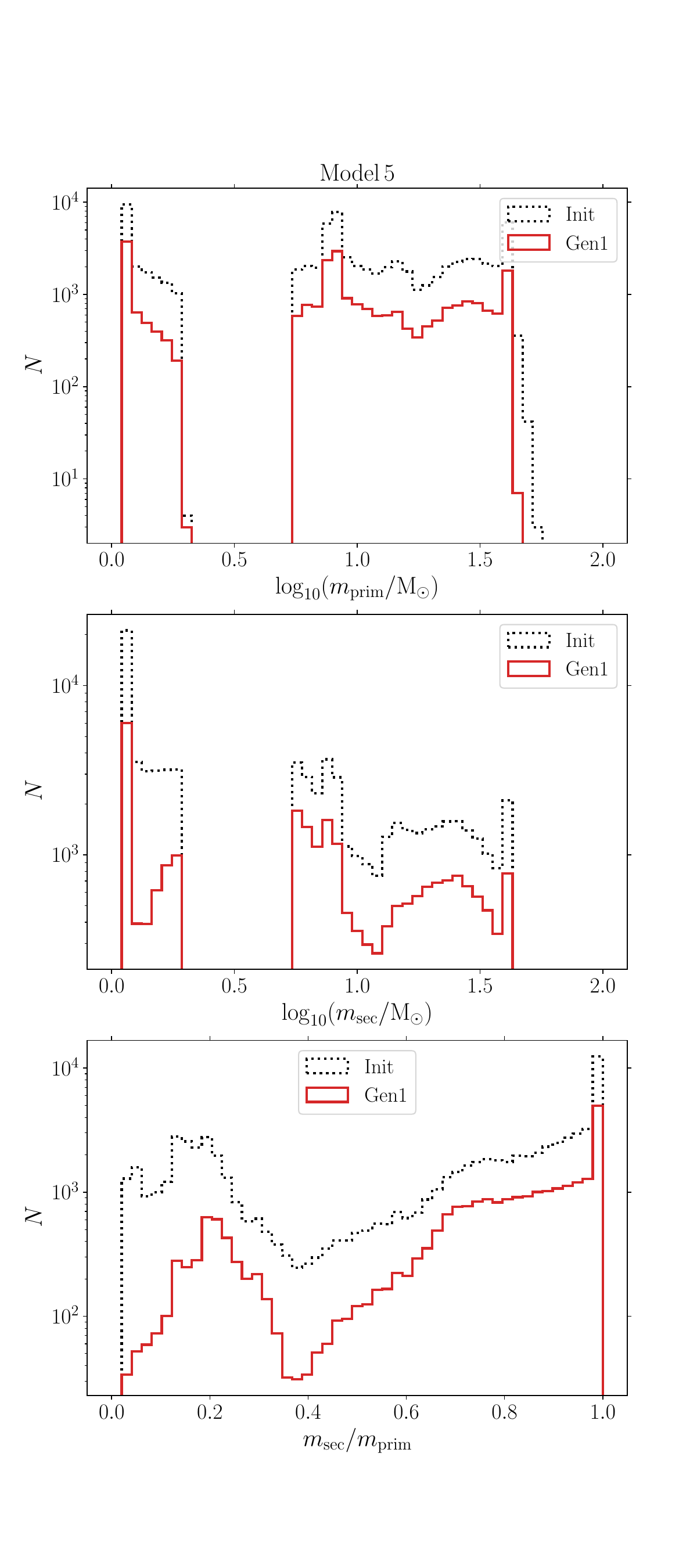}
\caption{ Similar to \MassesMOne, here for Model 5. }
\label{fig:masses_m5}
\end{figure}

\begin{figure}
\center
\includegraphics[scale = 0.45, trim = 5mm 10mm 0mm 0mm]{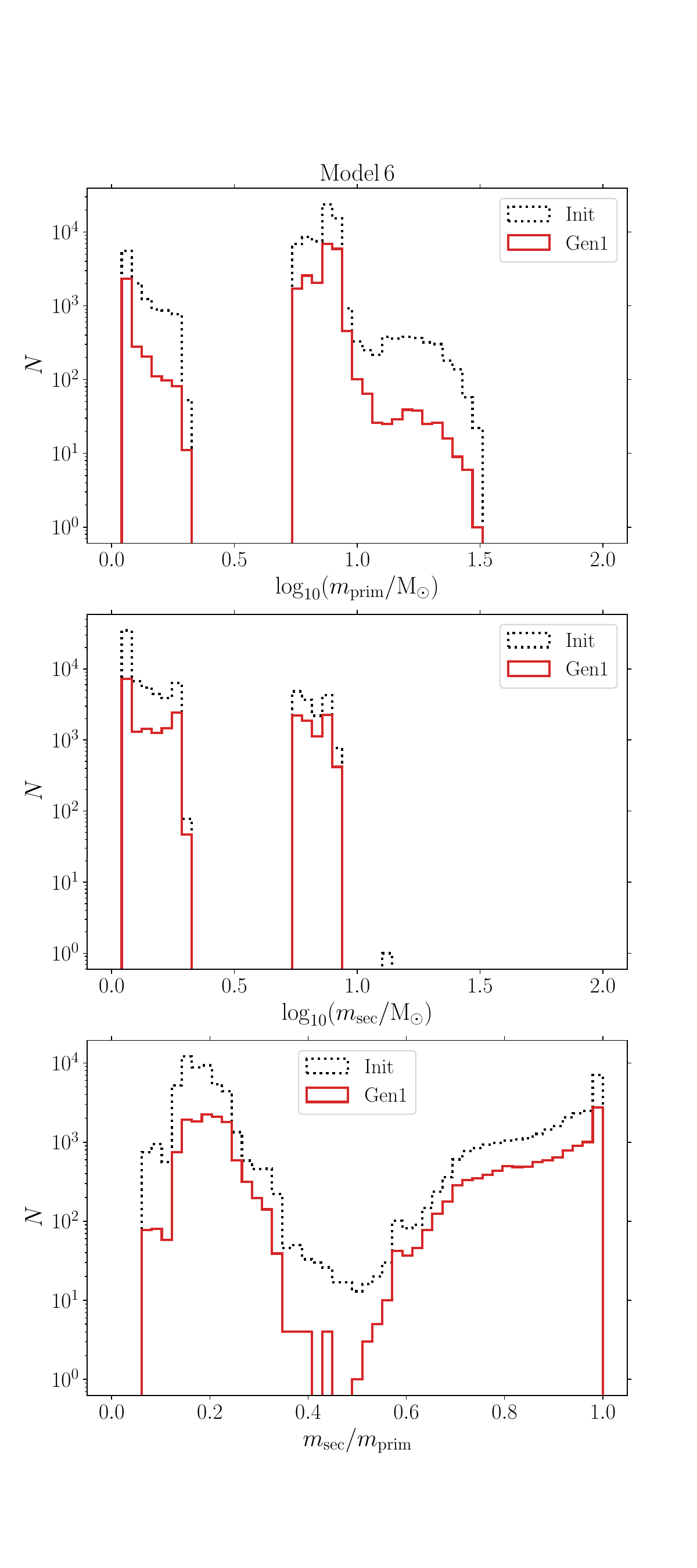}
\caption{ Similar to \MassesMOne, here for Model 6. }
\label{fig:masses_m6}
\end{figure}

\begin{figure}
\center
\includegraphics[scale = 0.45, trim = 5mm 10mm 0mm 0mm]{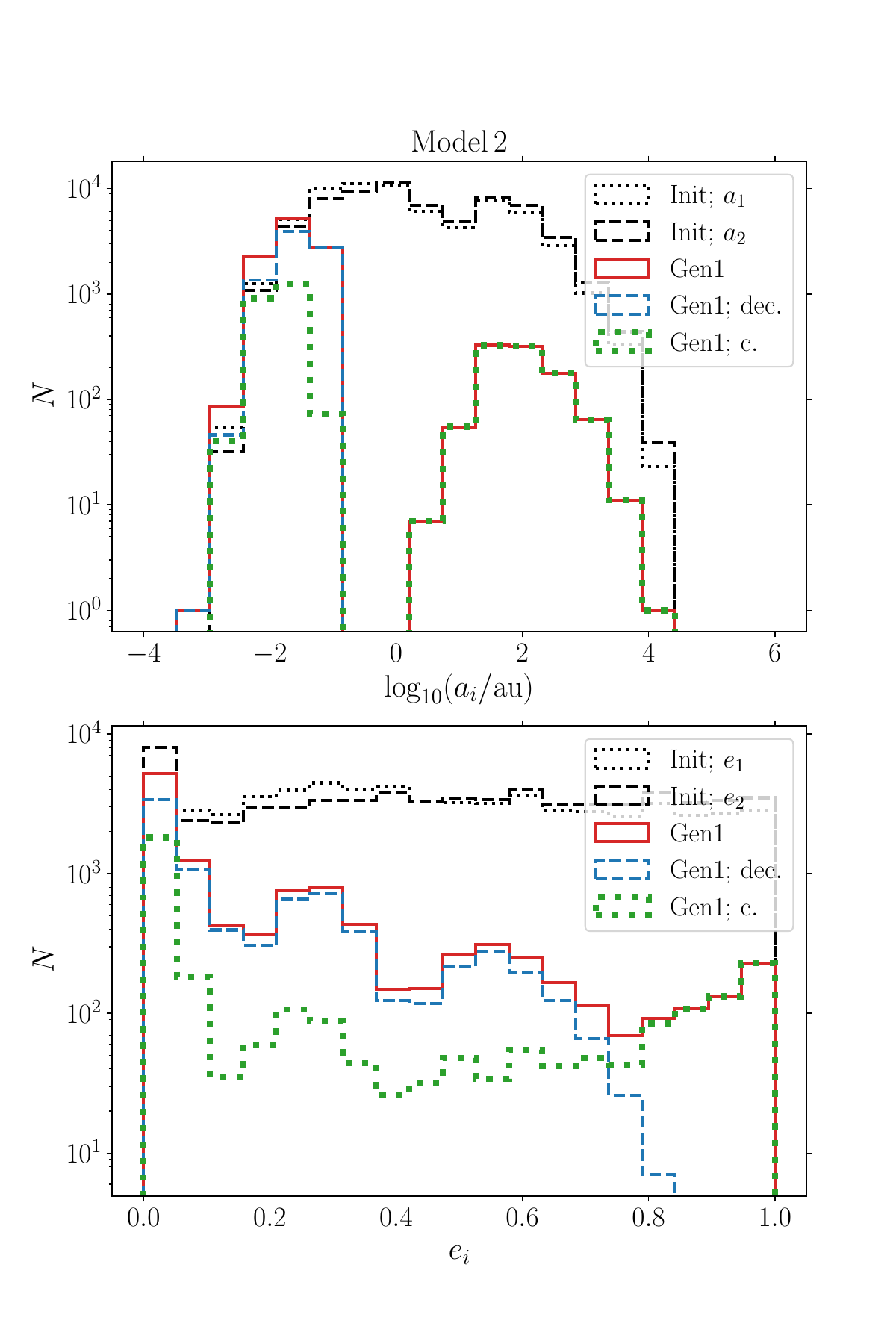}
\caption{ Similar to \OrbitsMOne, here for Model 2. }
\label{fig:orbits_m2}
\end{figure}

\begin{figure}
\center
\includegraphics[scale = 0.45, trim = 5mm 10mm 0mm 0mm]{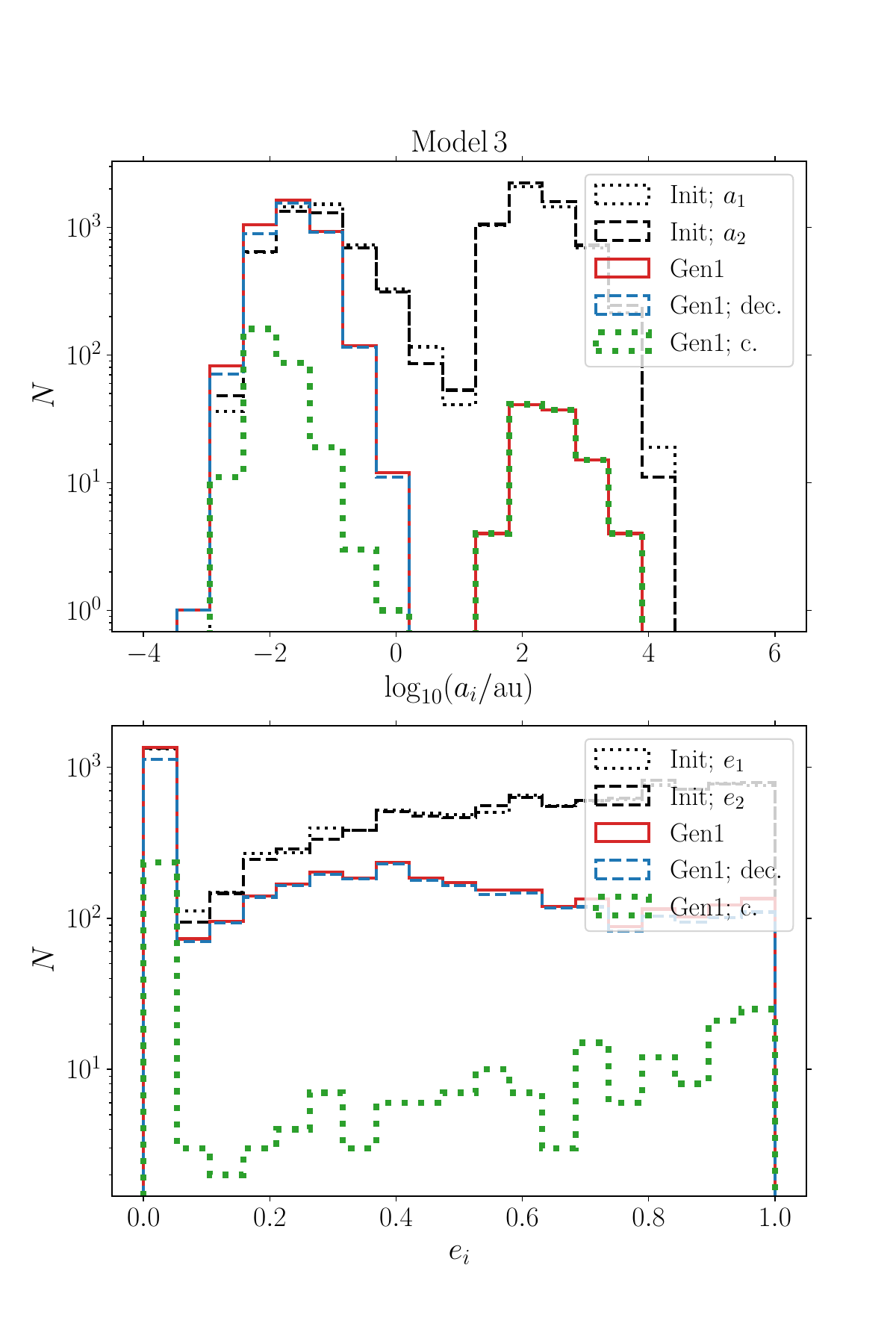}
\caption{ Similar to \OrbitsMOne, here for Model 3. }
\label{fig:orbits_m3}
\end{figure}

\begin{figure}
\center
\includegraphics[scale = 0.45, trim = 5mm 10mm 0mm 0mm]{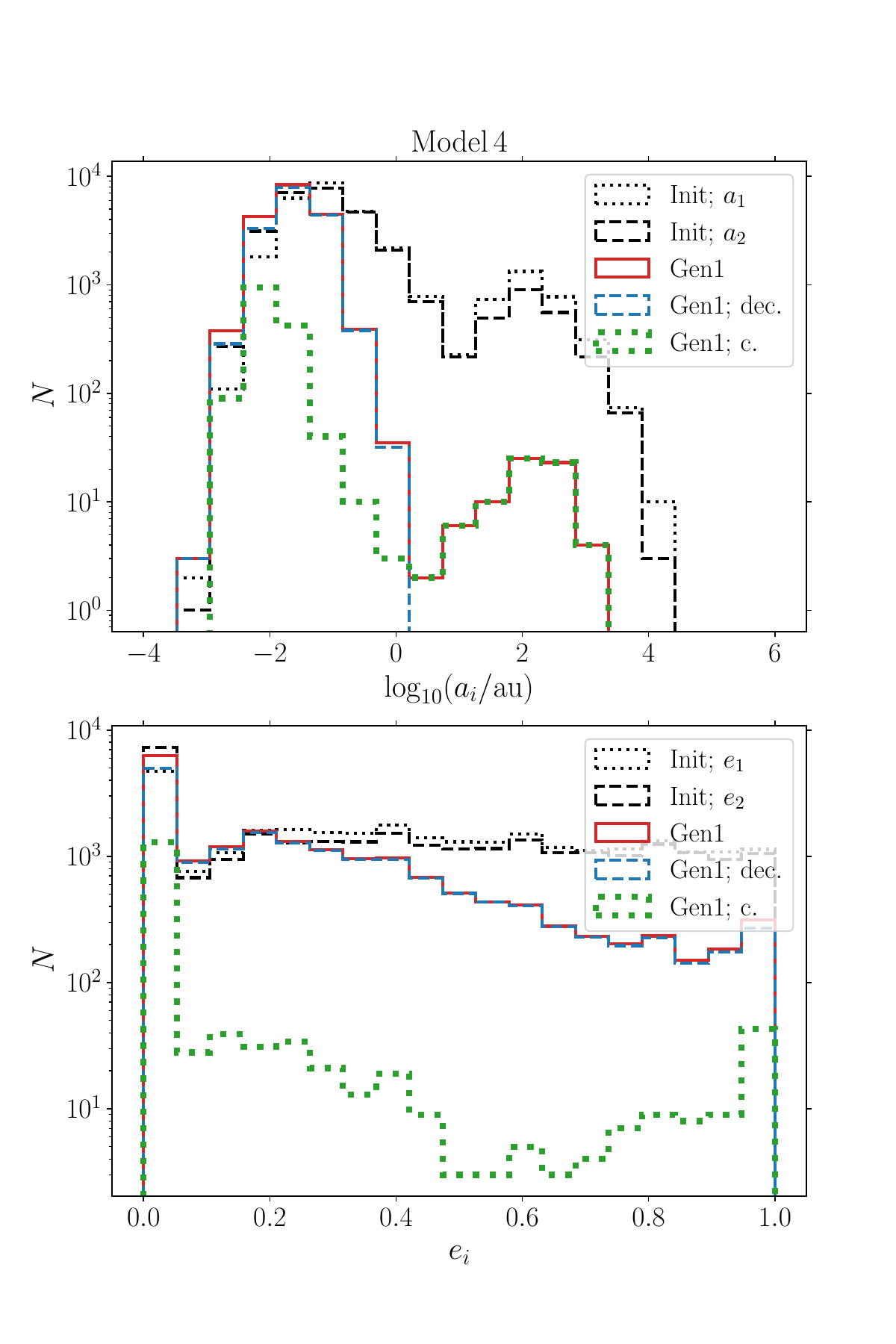}
\caption{ Similar to \OrbitsMOne, here for Model 4. }
\label{fig:orbits_m4}
\end{figure}

\begin{figure}
\center
\includegraphics[scale = 0.45, trim = 5mm 10mm 0mm 0mm]{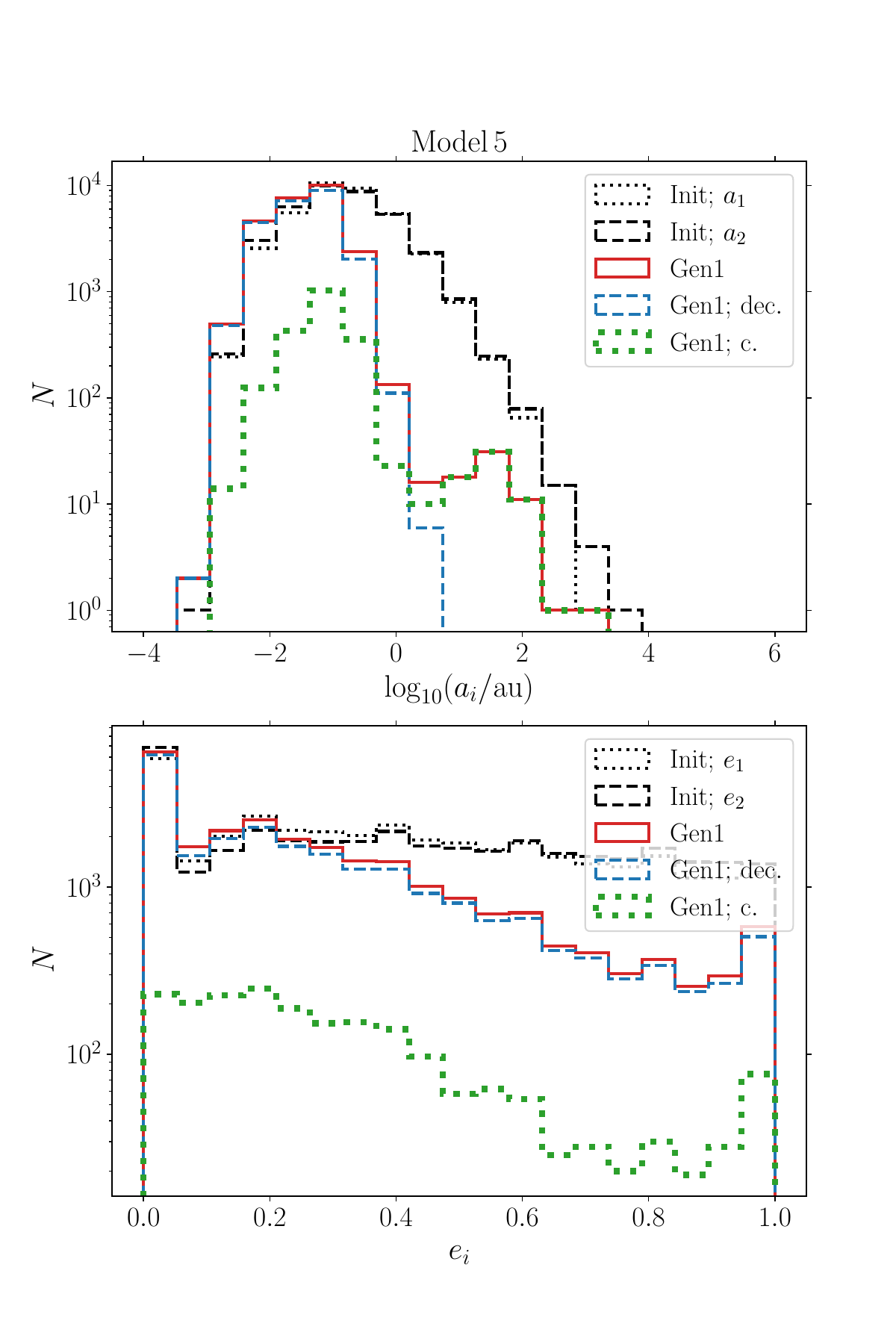}
\caption{ Similar to \OrbitsMOne, here for Model 5. }
\label{fig:orbits_m5}
\end{figure}

\begin{figure}
\center
\includegraphics[scale = 0.45, trim = 5mm 10mm 0mm 0mm]{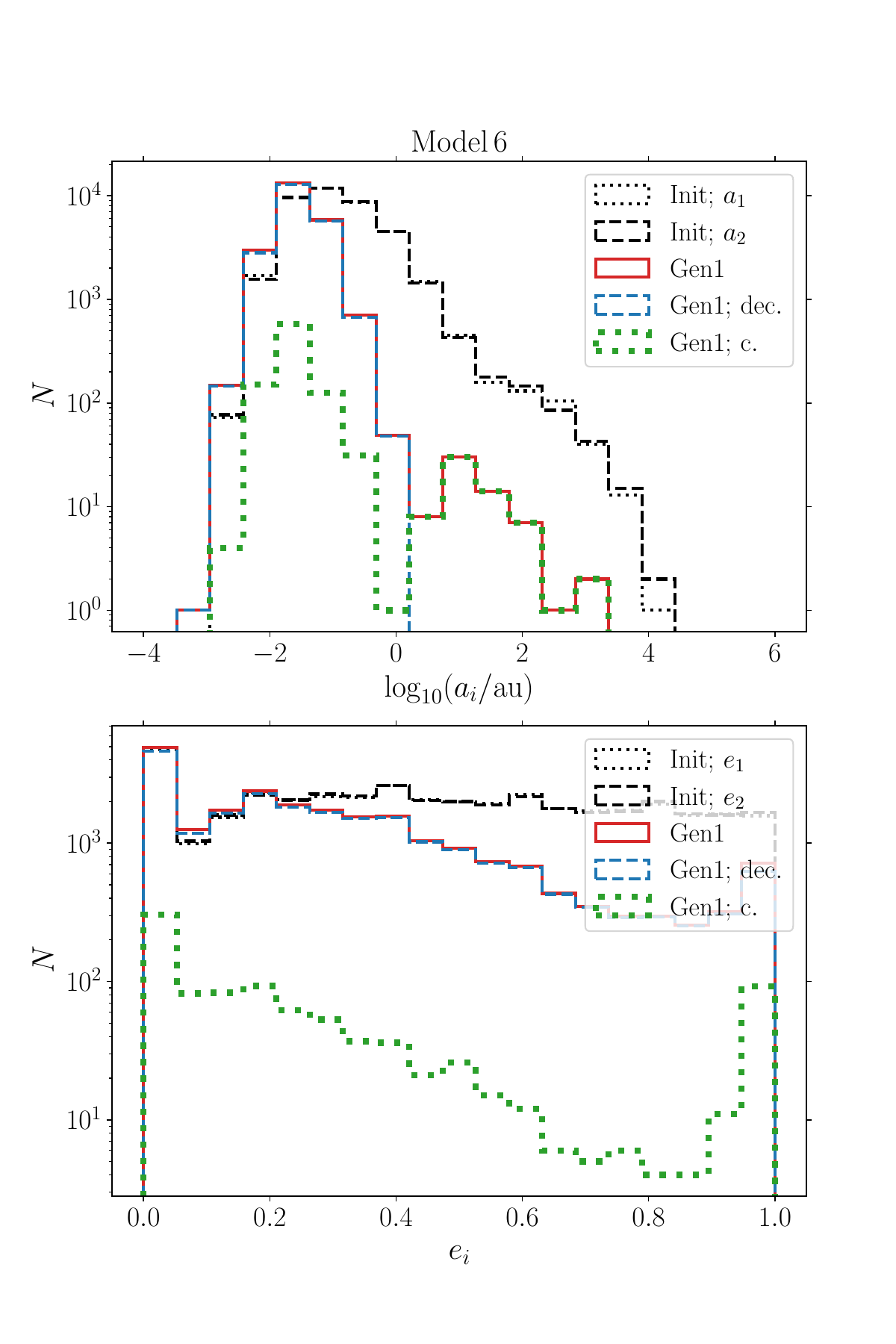}
\caption{ Similar to \OrbitsMOne, here for Model 6. }
\label{fig:orbits_m6}
\end{figure}

\begin{figure}
\center
\includegraphics[scale = 0.45, trim = 5mm 10mm 0mm 0mm]{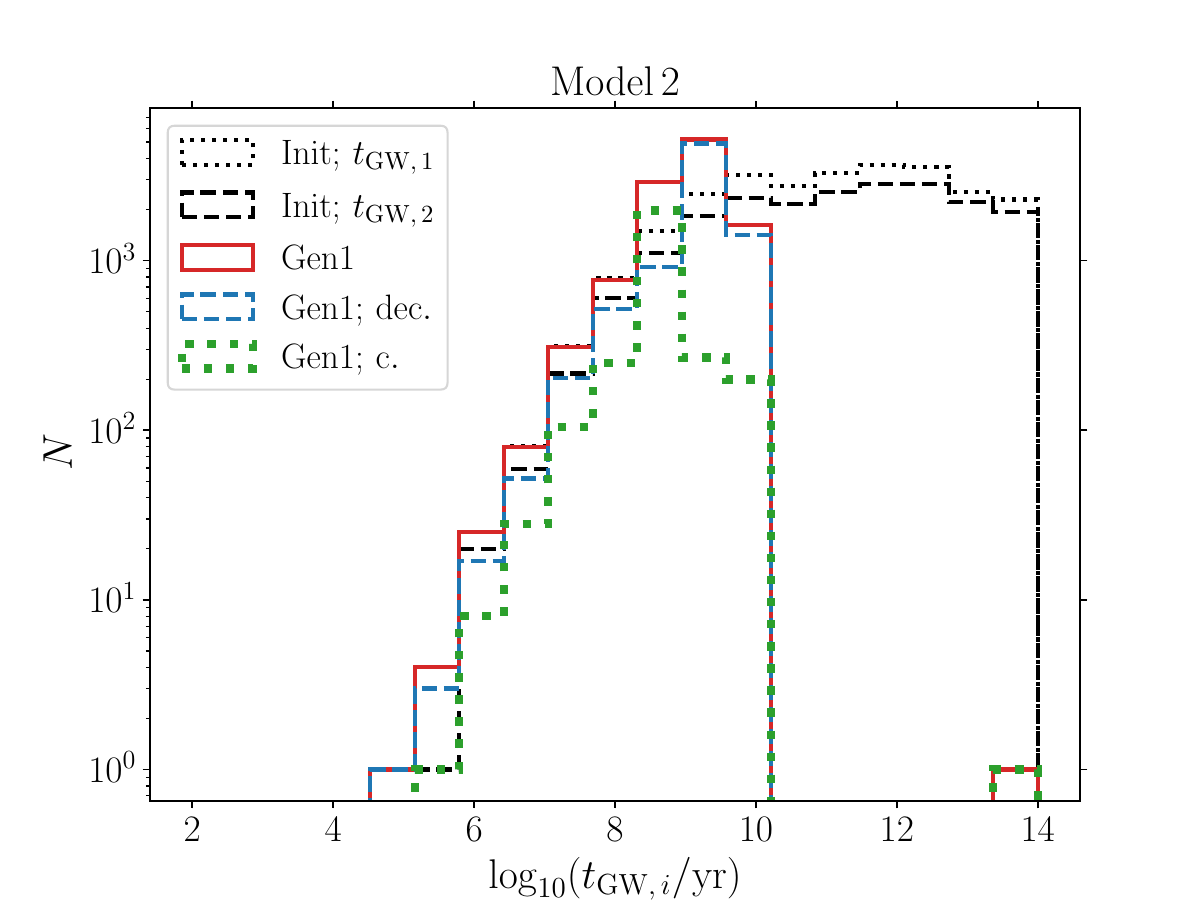}
\caption{ Similar to \MergerTimesMOne, here for Model 2. }
\label{fig:t_GW_m2}
\end{figure}

\begin{figure}
\center
\includegraphics[scale = 0.45, trim = 5mm 10mm 0mm 0mm]{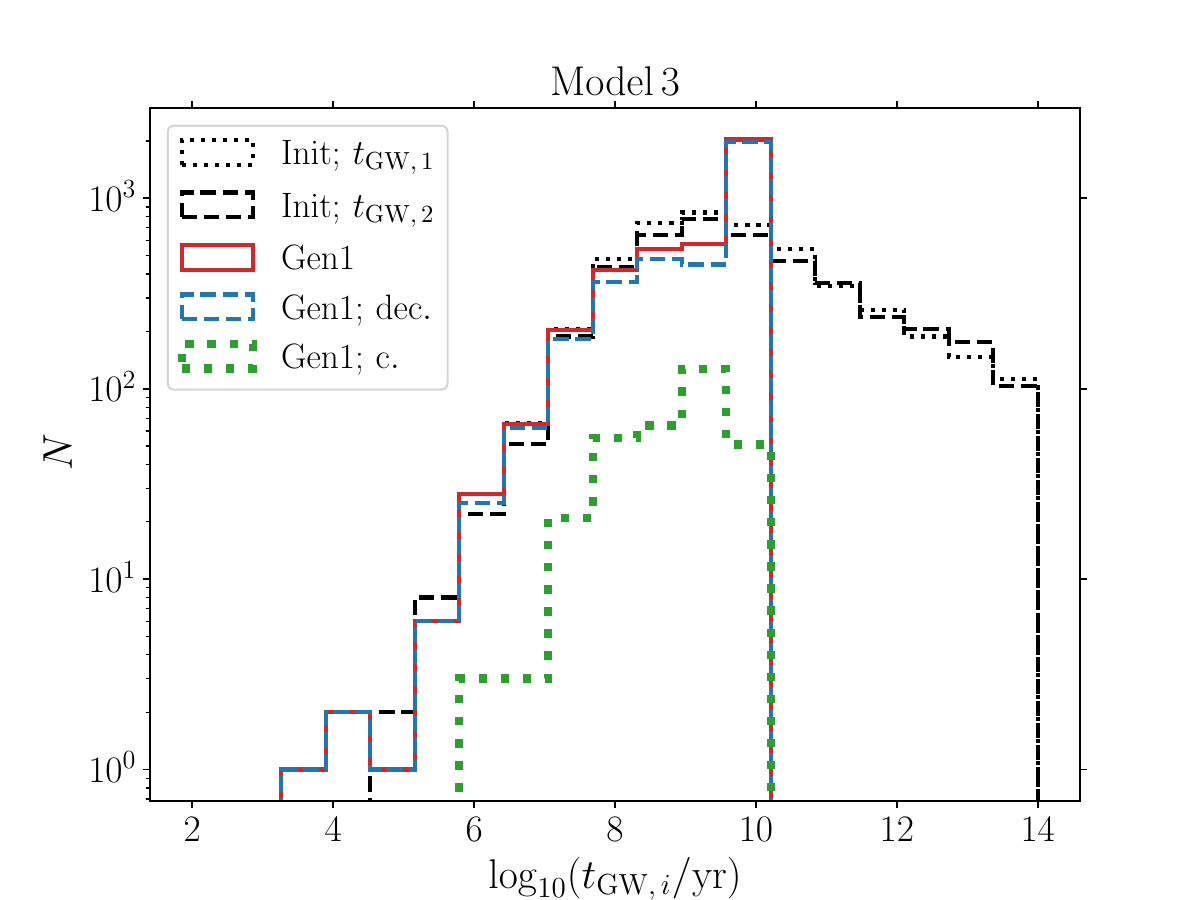}
\caption{ Similar to \MergerTimesMOne, here for Model 3. }
\label{fig:t_GW_m3}
\end{figure}

\begin{figure}
\center
\includegraphics[scale = 0.45, trim = 5mm 10mm 0mm 0mm]{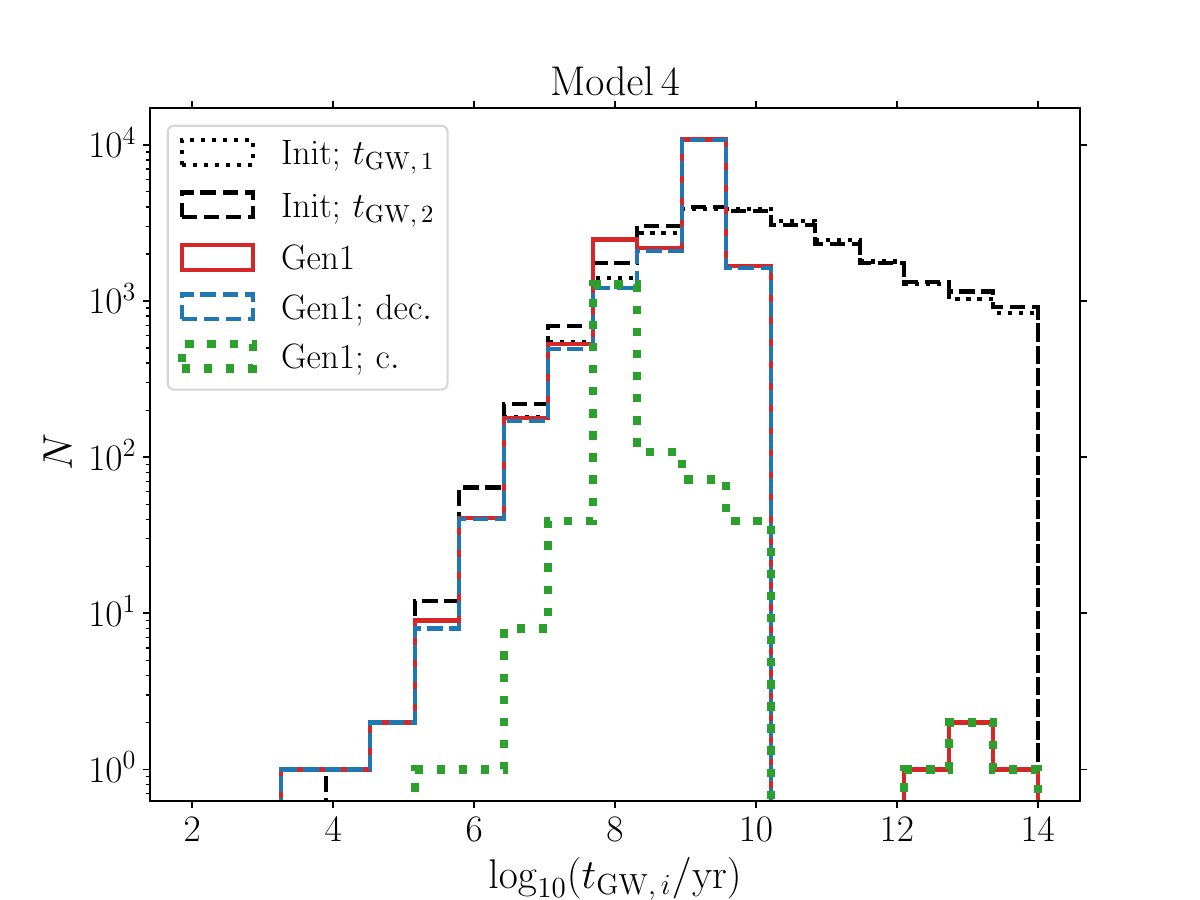}
\caption{ Similar to \MergerTimesMOne, here for Model 4. }
\label{fig:t_GW_m4}
\end{figure}

\begin{figure}
\center
\includegraphics[scale = 0.45, trim = 5mm 10mm 0mm 0mm]{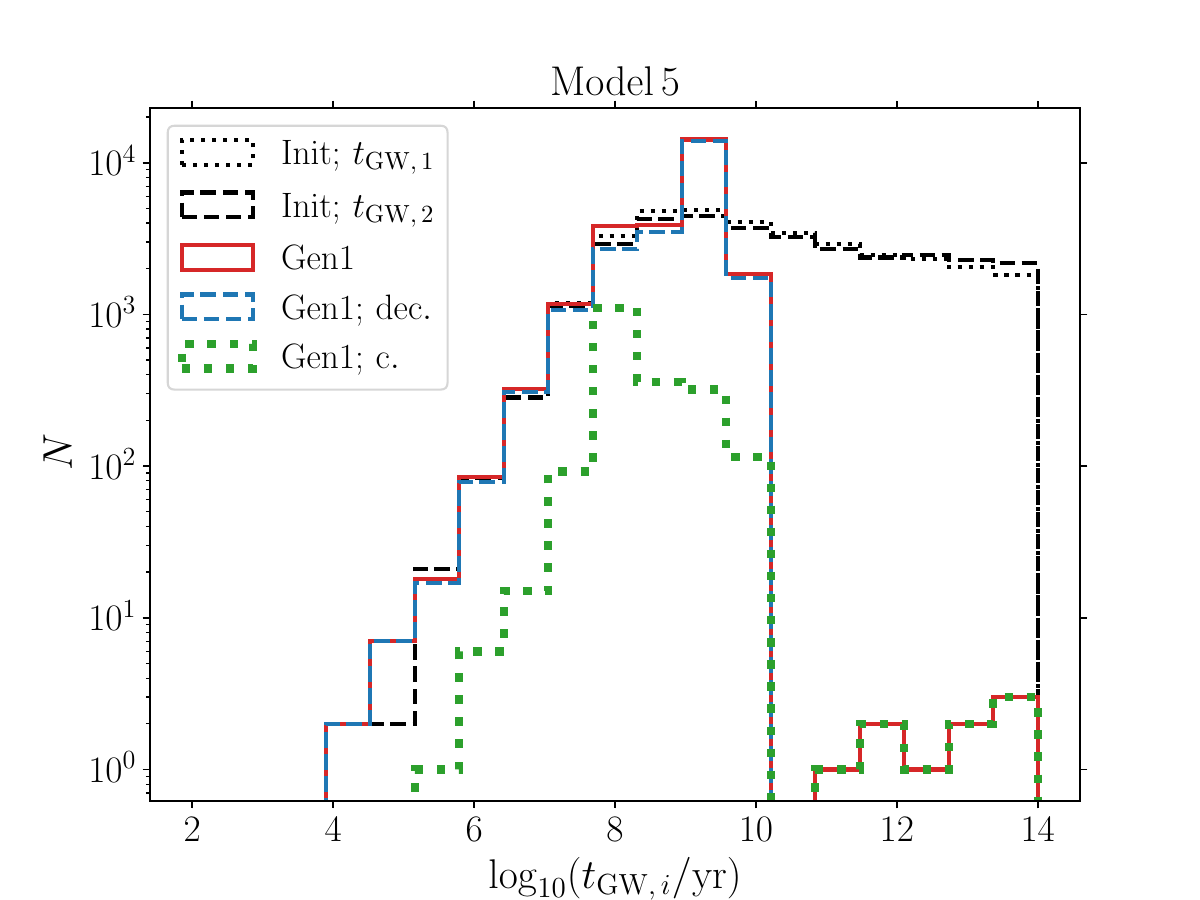}
\caption{ Similar to \MergerTimesMOne, here for Model 5. }
\label{fig:t_GW_m5}
\end{figure}

\begin{figure}
\center
\includegraphics[scale = 0.45, trim = 5mm 10mm 0mm 0mm]{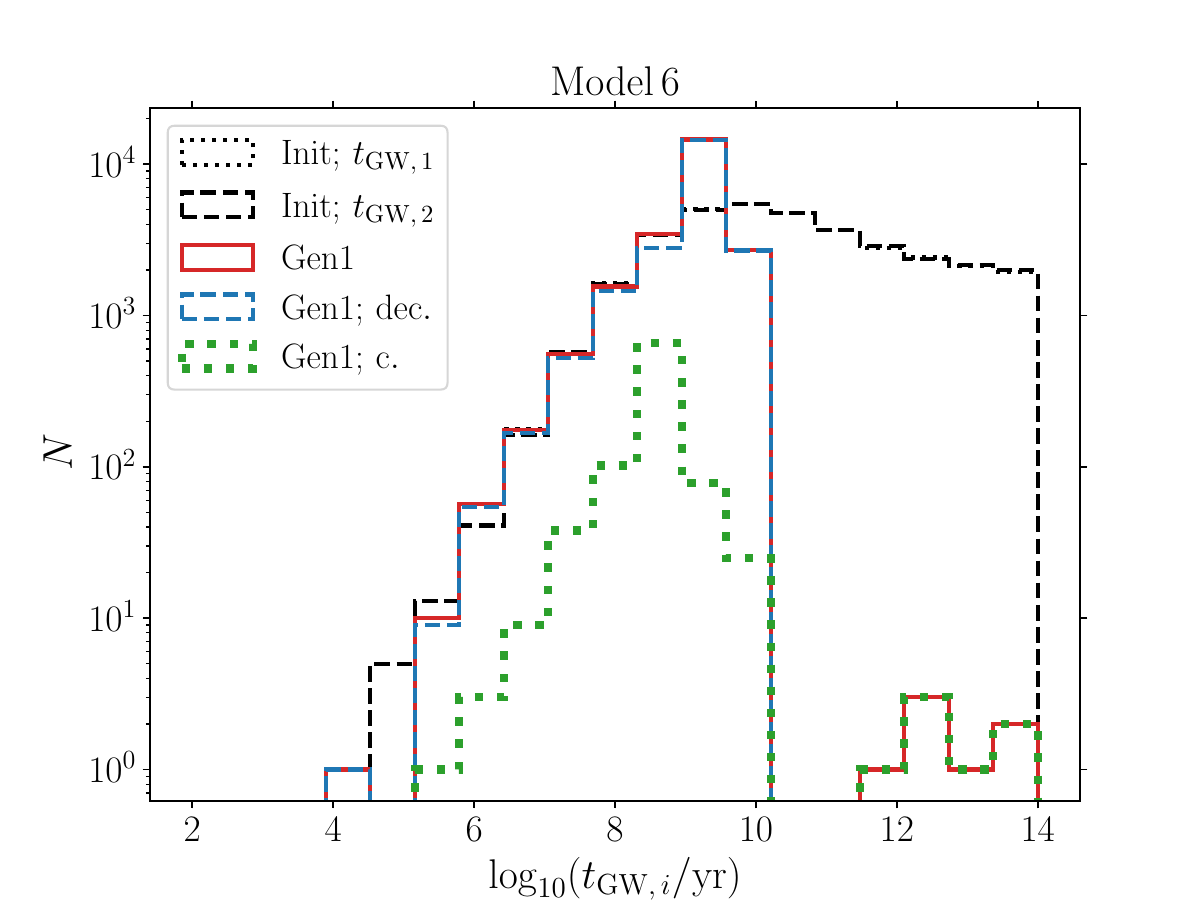}
\caption{ Similar to \MergerTimesMOne, here for Model 6. }
\label{fig:t_GW_m6}
\end{figure}

In \Fs~\ref{fig:e_LIGO_m2} through \ref{fig:t_GW_m6}, we show distributions of the eccentricity in the LIGO band, $\chi_\eff$, masses, orbital parameters, and GW merger times for models 2 through 6. These figures are complementary to \SectionFirst. 






\section{Second-generation merger properties}
\label{app:gen2}
\begin{figure}
\center
\includegraphics[scale=0.55]{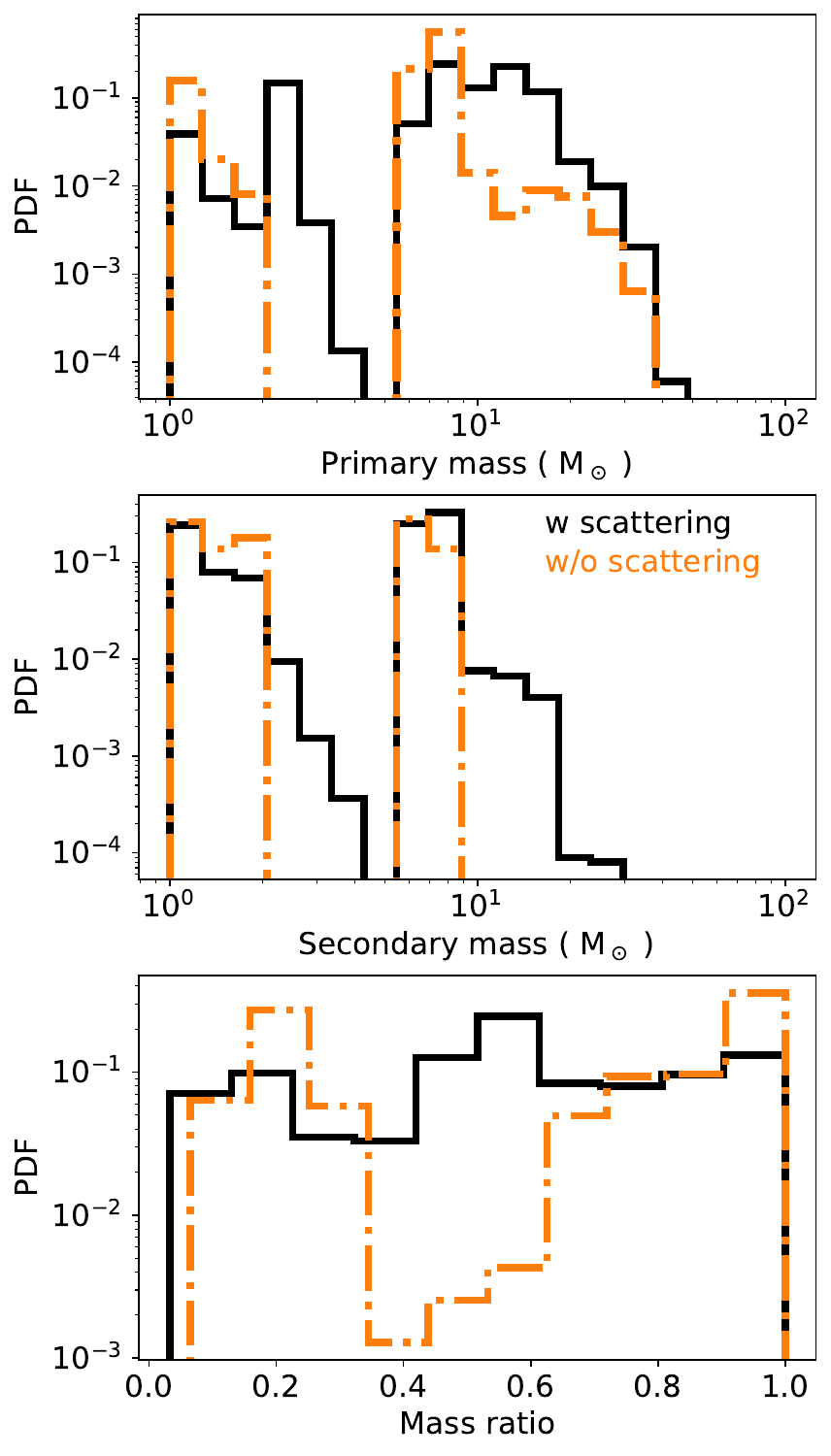}
\caption{ Similar to \SecondGenModelOneOneMass, here for Model 1 and $0<\chi_i<0.1$. }
\label{fig:2g_mod12_mass}
\end{figure}

\begin{figure}
\center
\includegraphics[scale=0.5]{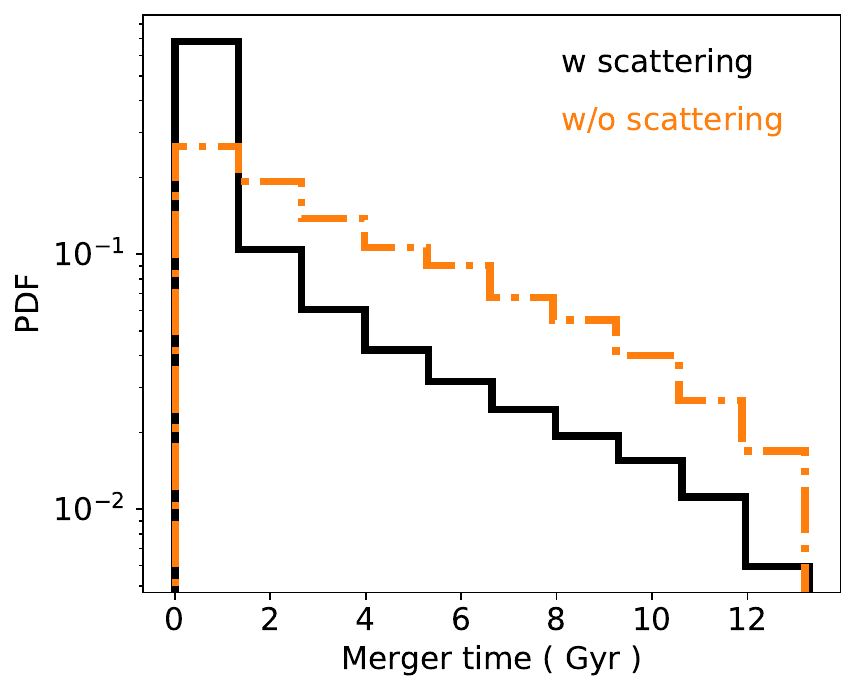}
\caption{ Similar to \SecondGenModelOneOneTime, here for Model 1 and $0<\chi_i<0.1$. }
\label{fig:2g_mod12_time}
\end{figure}

\begin{figure}
\center
\includegraphics[scale=0.5]{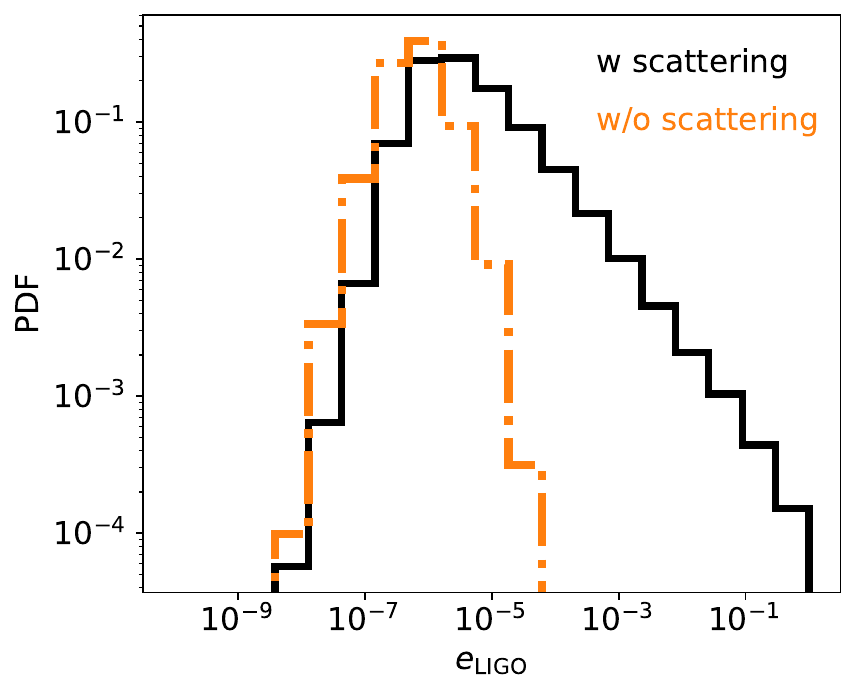}
\caption{ Similar to \SecondGenModelOneOneEcc, here for Model 1 and $0<\chi_i<0.1$. }
\label{fig:2g_mod12_ecc}
\end{figure}

\begin{figure}
\center
\includegraphics[scale=0.5]{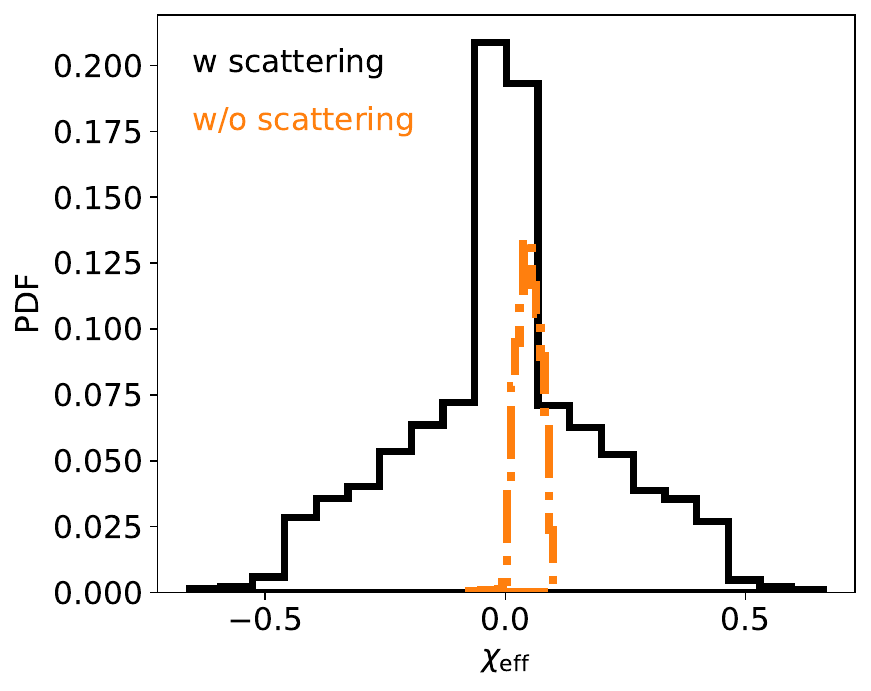}
\caption{ Similar to \SecondGenModelOneOneSpin, here for Model 1 and $0<\chi_i<0.1$. }
\label{fig:2g_mod12_spin}
\end{figure}

\begin{figure}
\center
\includegraphics[scale=0.55]{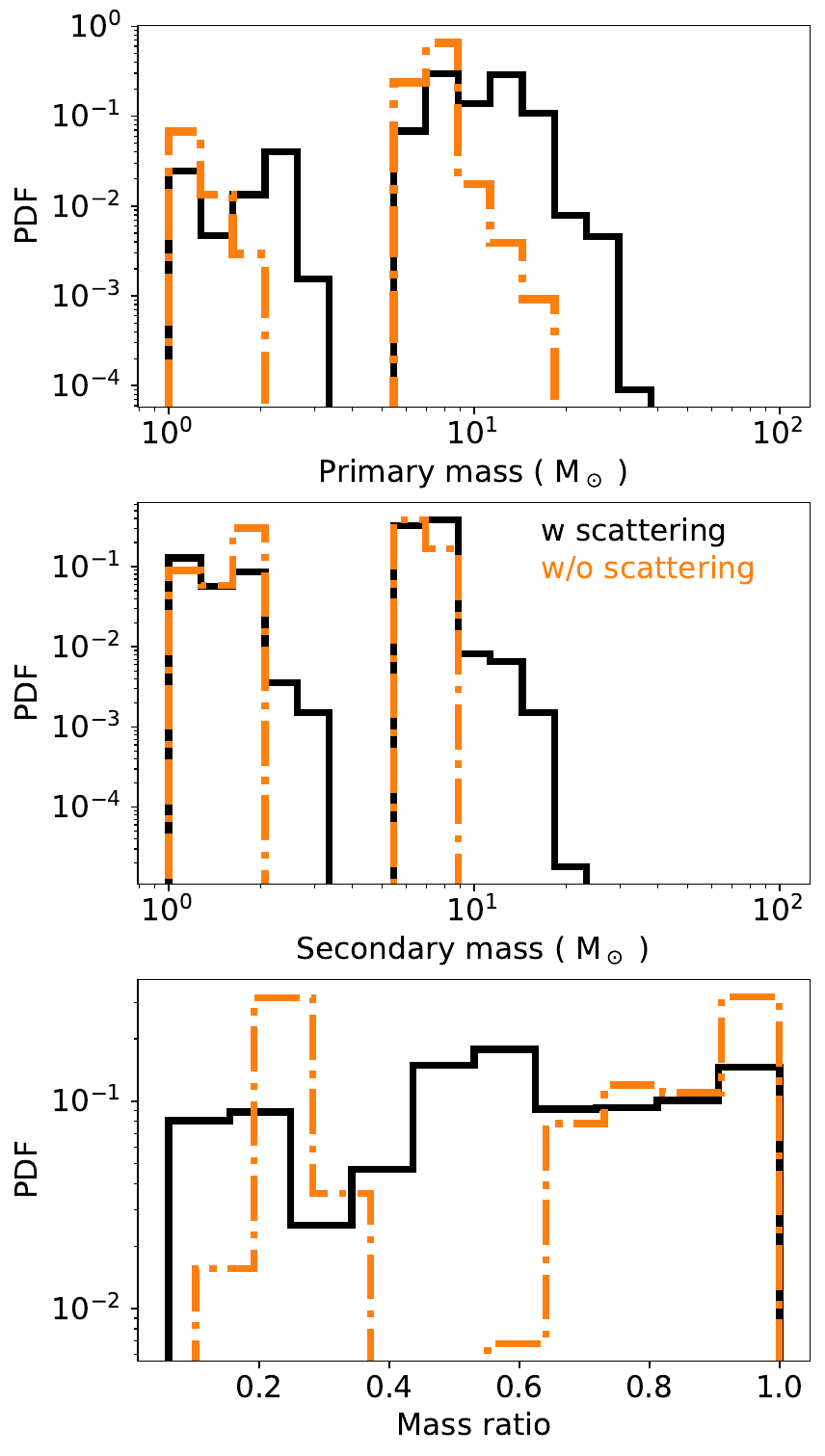}
\caption{ Similar to \SecondGenModelOneOneMass, here for Model 2 and $0<\chi_i<1$. }
\label{fig:2g_mod21_mass}
\end{figure}

\begin{figure}
\center
\includegraphics[scale=0.5]{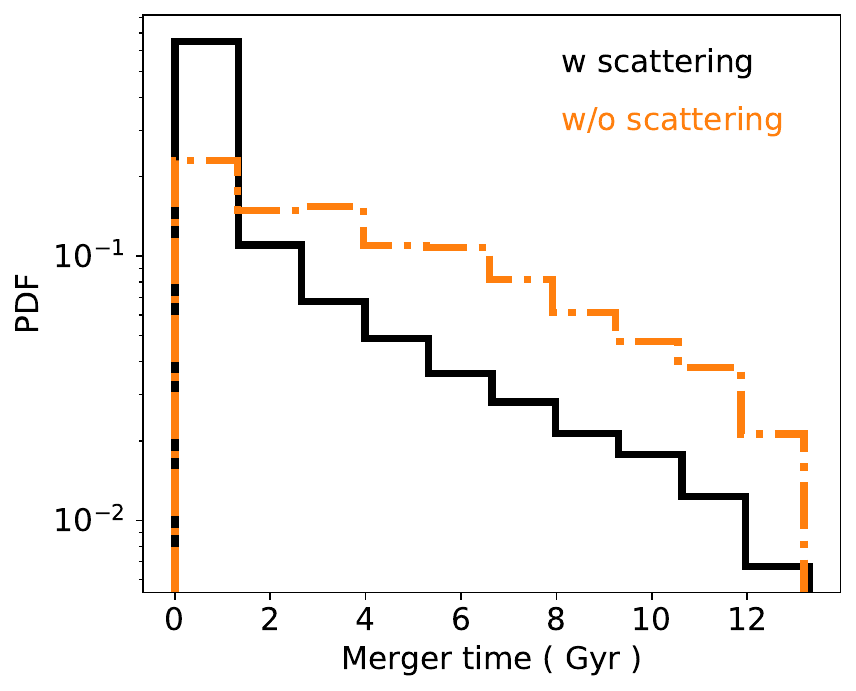}
\caption{ Similar to \SecondGenModelOneOneTime, here for Model 2 and $0<\chi_i<1$. }
\label{fig:2g_mod21_time}
\end{figure}

\begin{figure}
\center
\includegraphics[scale=0.5]{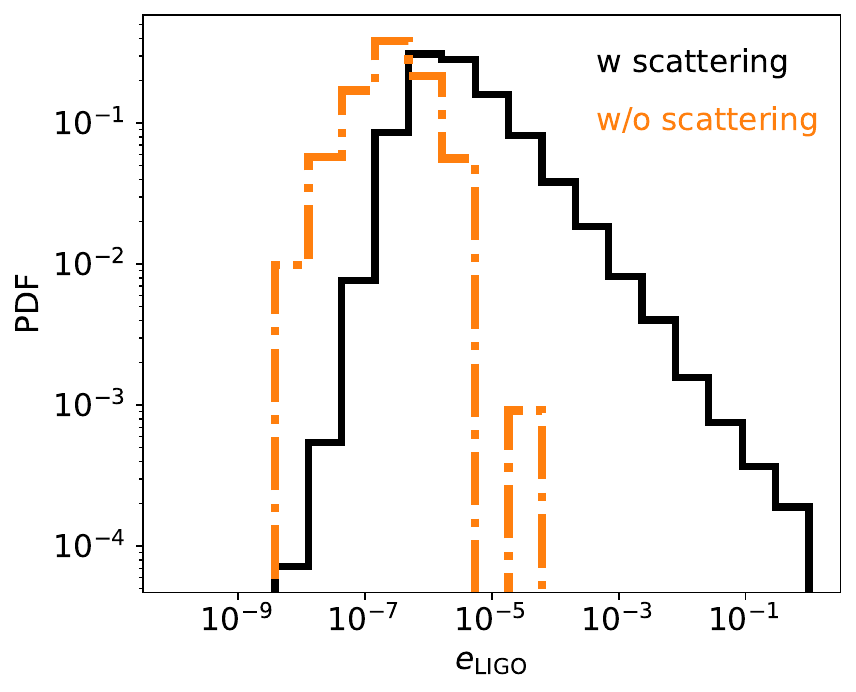}
\caption{ Similar to \SecondGenModelOneOneEcc, here for Model 2 and $0<\chi_i<1$. }
\label{fig:2g_mod21_ecc}
\end{figure}

\begin{figure}
\center
\includegraphics[scale=0.5]{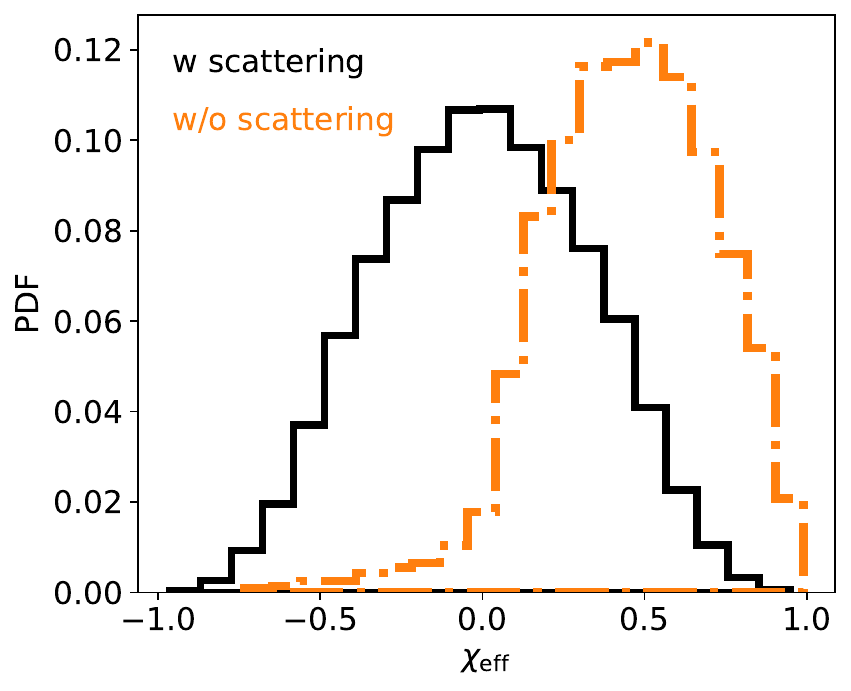}
\caption{ Similar to \SecondGenModelOneOneSpin, here for Model 2 and $0<\chi_i<1$. }
\label{fig:2g_mod21_spin}
\end{figure}

\begin{figure}
\center
\includegraphics[scale=0.55]{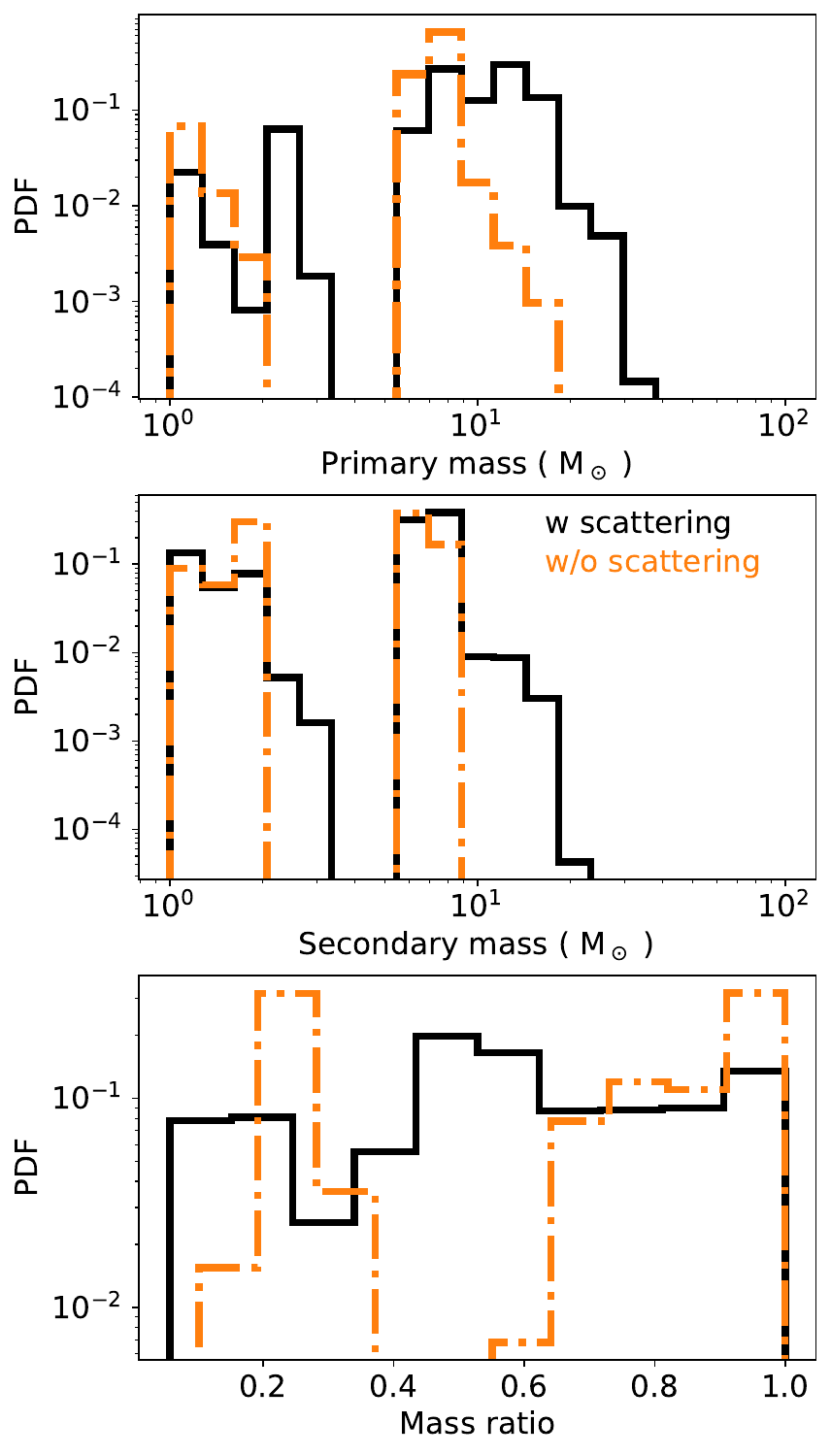}
\caption{ Similar to \SecondGenModelOneOneMass, here for Model 2 and $0<\chi_i<0.1$. }
\label{fig:2g_mod22_mass}
\end{figure}

\begin{figure}
\center
\includegraphics[scale=0.5]{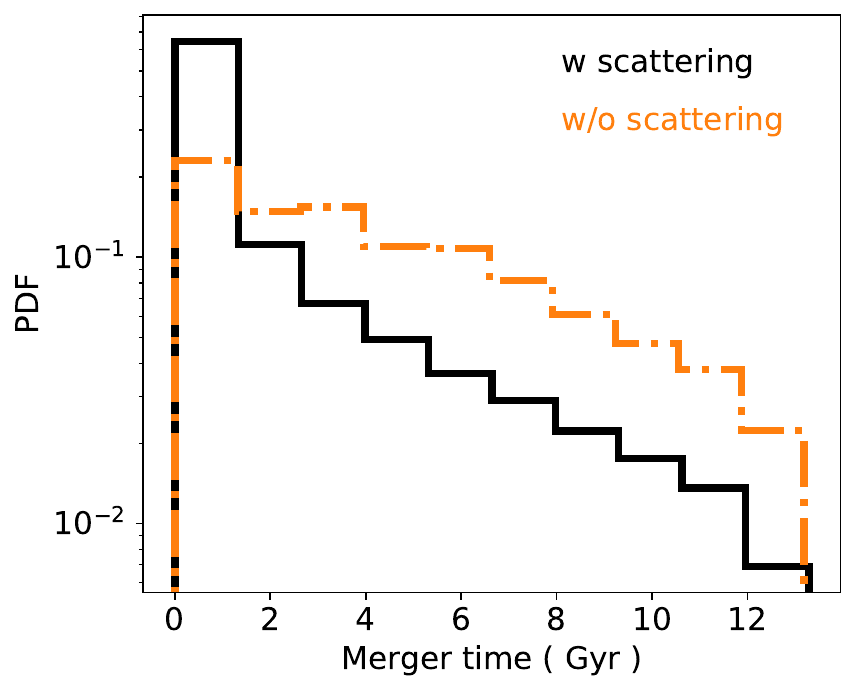}
\caption{ Similar to \SecondGenModelOneOneTime, here for Model 2 and $0<\chi_i<0.1$. }
\label{fig:2g_mod22_time}
\end{figure}

\begin{figure}
\center
\includegraphics[scale=0.5]{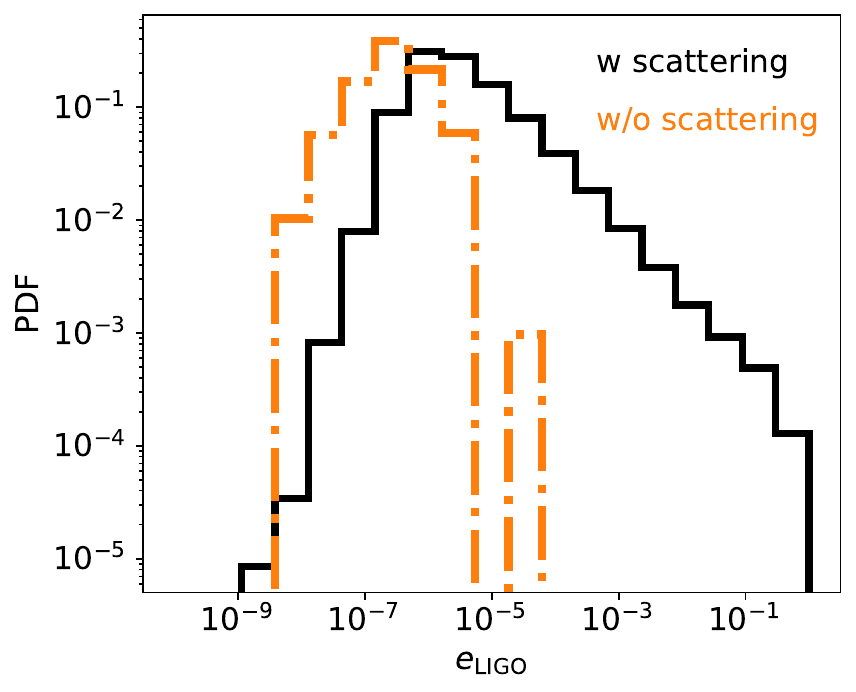}
\caption{ Similar to \SecondGenModelOneOneEcc, here for Model 2 and $0<\chi_i<0.1$. }
\label{fig:2g_mod22_ecc}
\end{figure}

\begin{figure}
\center
\includegraphics[scale=0.5]{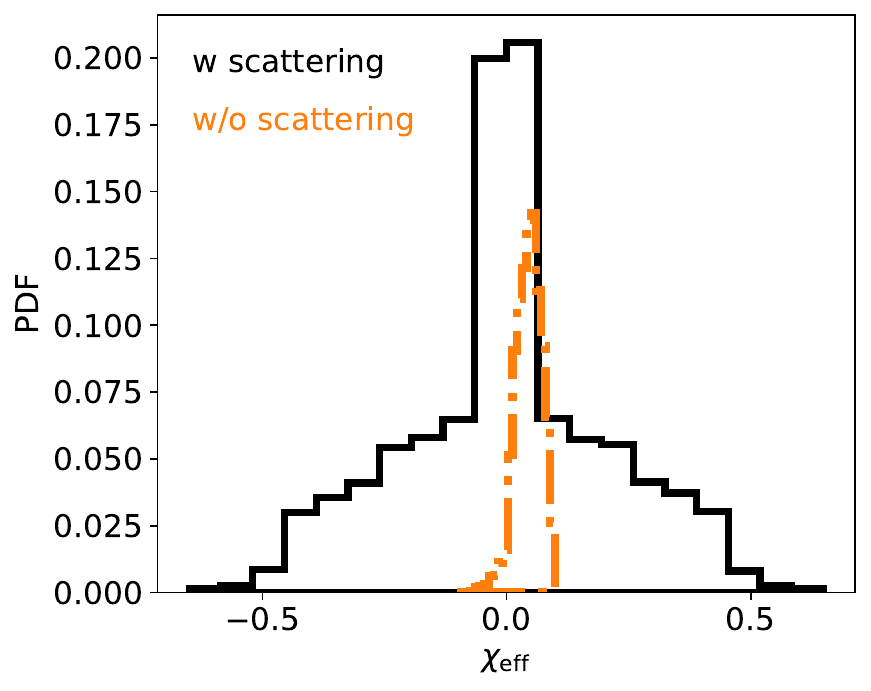}
\caption{ Similar to \SecondGenModelOneOneSpin, here for Model 2 and $0<\chi_i<0.1$. }
\label{fig:2g_mod22_spin}
\end{figure}

\begin{figure}
\center
\includegraphics[scale=0.55]{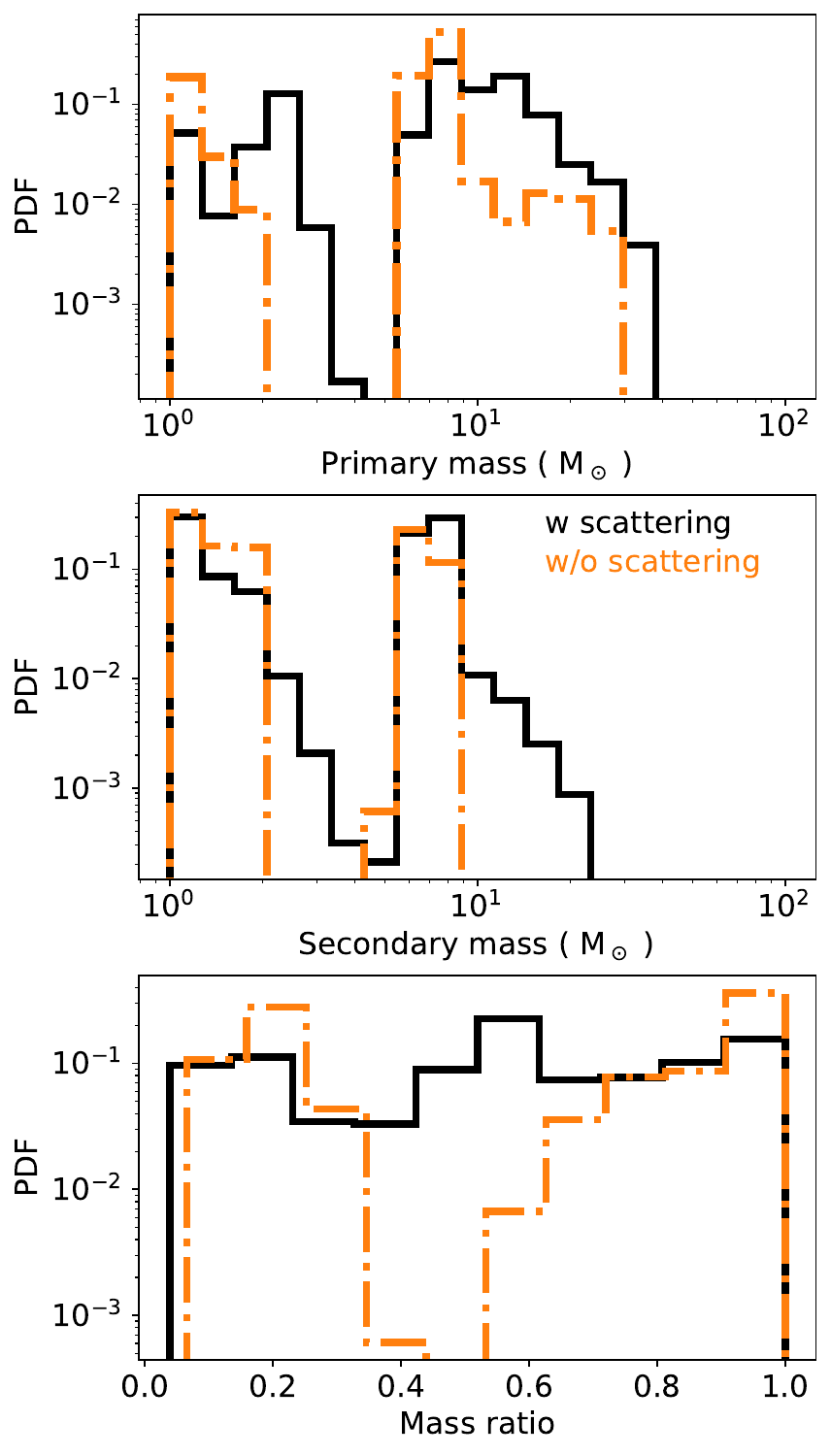}
\caption{ Similar to \SecondGenModelOneOneMass, here for Model 3 and $0<\chi_i<1$. }
\label{fig:2g_mod31_mass}
\end{figure}

\begin{figure}
\center
\includegraphics[scale=0.5]{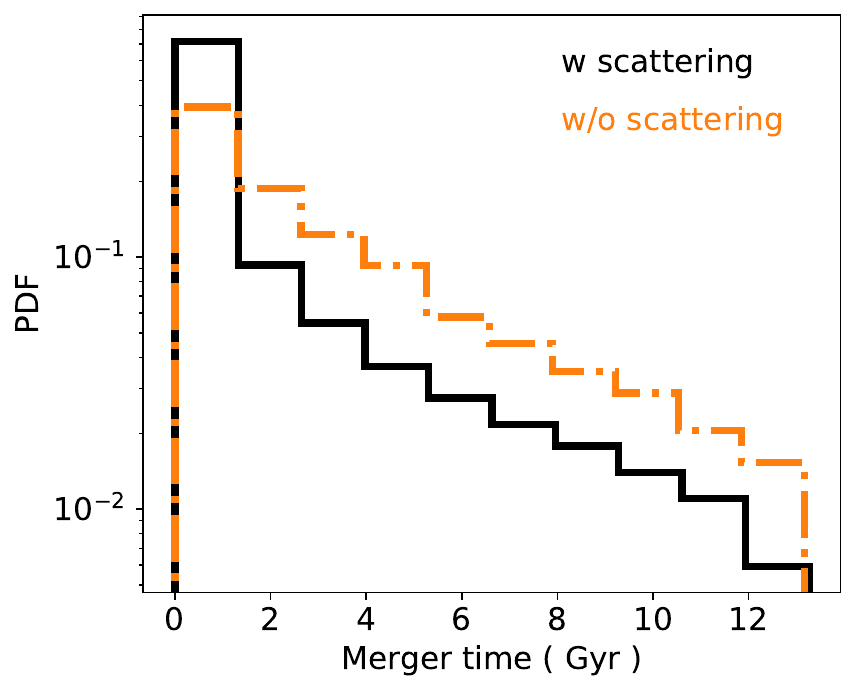}
\caption{ Similar to \SecondGenModelOneOneTime, here for Model 3 and $0<\chi_i<1$. }
\label{fig:2g_mod31_time}
\end{figure}

\begin{figure}
\center
\includegraphics[scale=0.5]{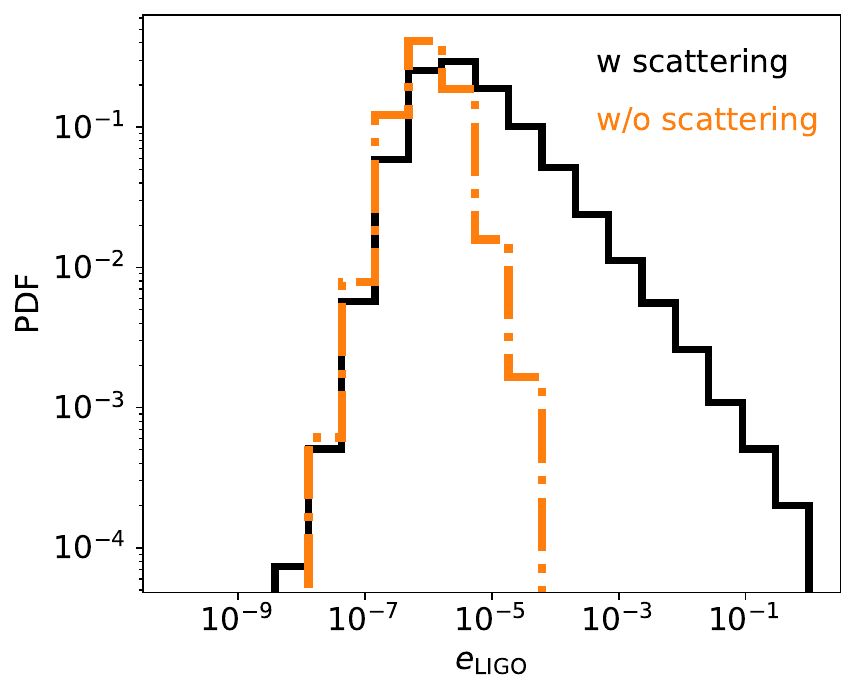}
\caption{ Similar to \SecondGenModelOneOneEcc, here for Model 3 and $0<\chi_i<1$. }
\label{fig:2g_mod31_ecc}
\end{figure}

\begin{figure}
\center
\includegraphics[scale=0.5]{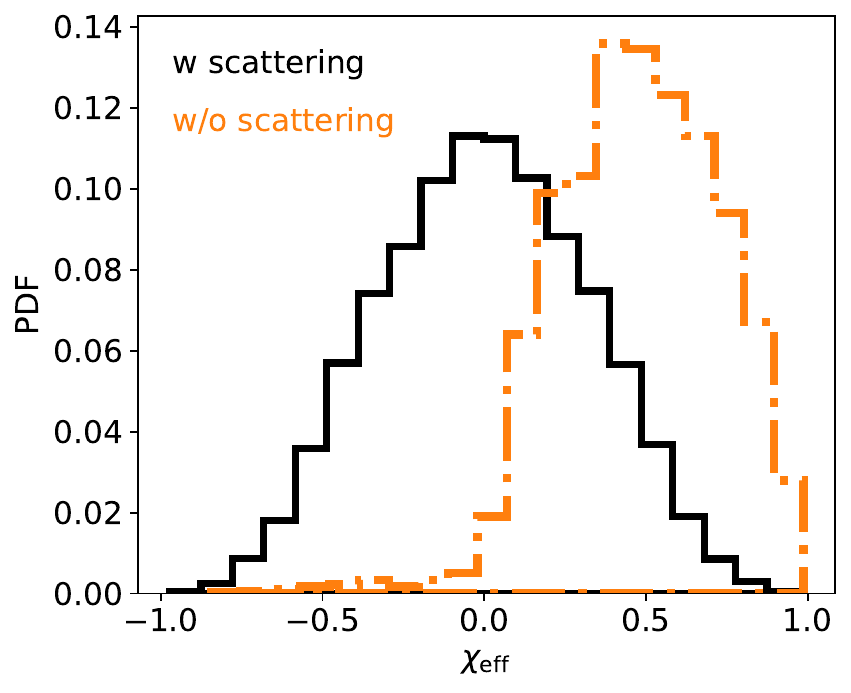}
\caption{ Similar to \SecondGenModelOneOneSpin, here for Model 3 and $0<\chi_i<1$. }
\label{fig:2g_mod31_spin}
\end{figure}

\begin{figure}
\center
\includegraphics[scale=0.55]{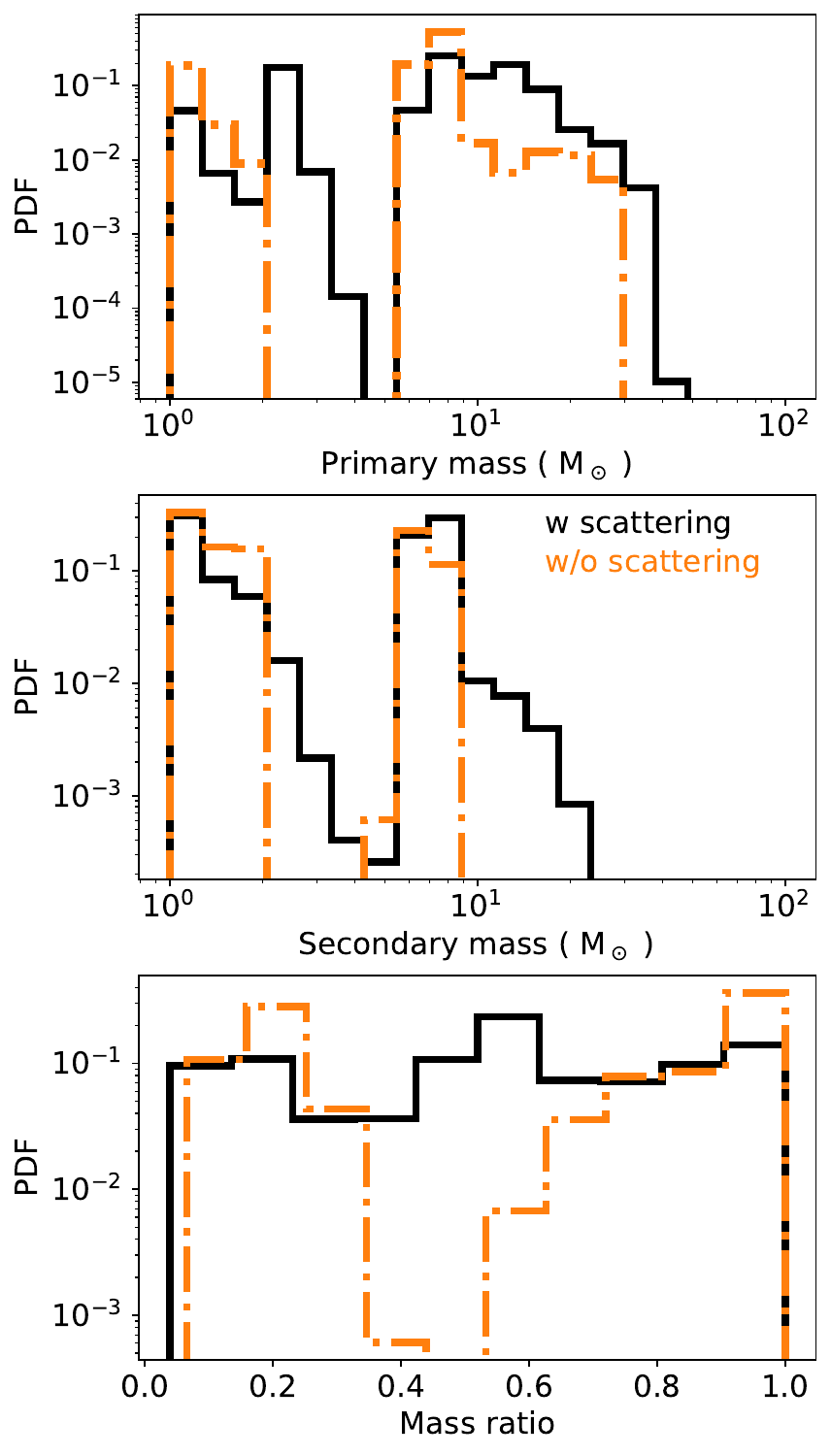}
\caption{ Similar to \SecondGenModelOneOneMass, here for Model 3 and $0<\chi_i<0.1$. }
\label{fig:2g_mod32_mass}
\end{figure}

\begin{figure}
\center
\includegraphics[scale=0.5]{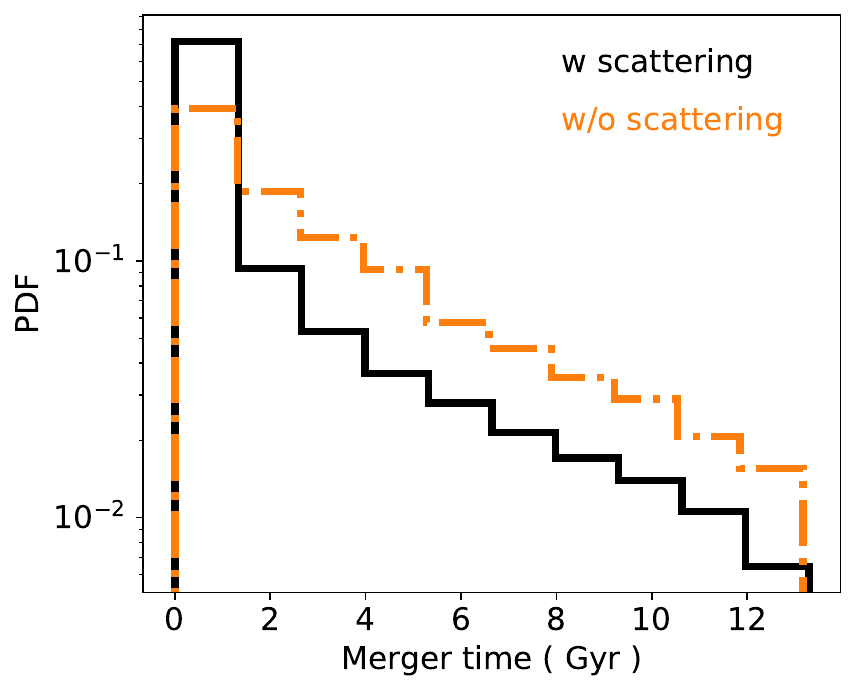}
\caption{ Similar to \SecondGenModelOneOneTime, here for Model 3 and $0<\chi_i<0.1$. }
\label{fig:2g_mod32_time}
\end{figure}

\begin{figure}
\center
\includegraphics[scale=0.5]{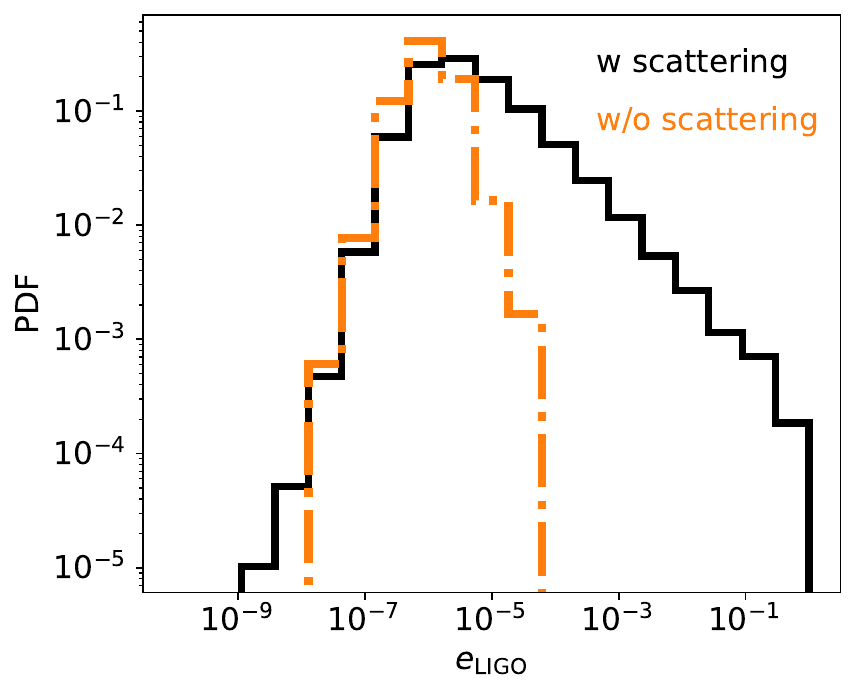}
\caption{ Similar to \SecondGenModelOneOneEcc, here for Model 3 and $0<\chi_i<0.1$. }
\label{fig:2g_mod32_ecc}
\end{figure}

\begin{figure}
\center
\includegraphics[scale=0.5]{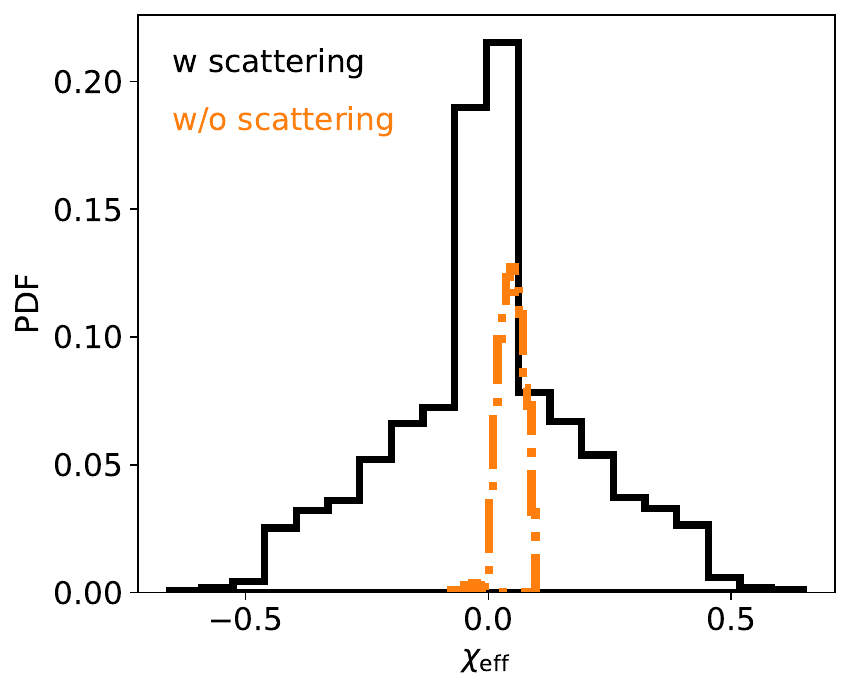}
\caption{ Similar to \SecondGenModelOneOneSpin, here for Model 3 and $0<\chi_i<0.1$. }
\label{fig:2g_mod32_spin}
\end{figure}

\begin{figure}
\center
\includegraphics[scale=0.55]{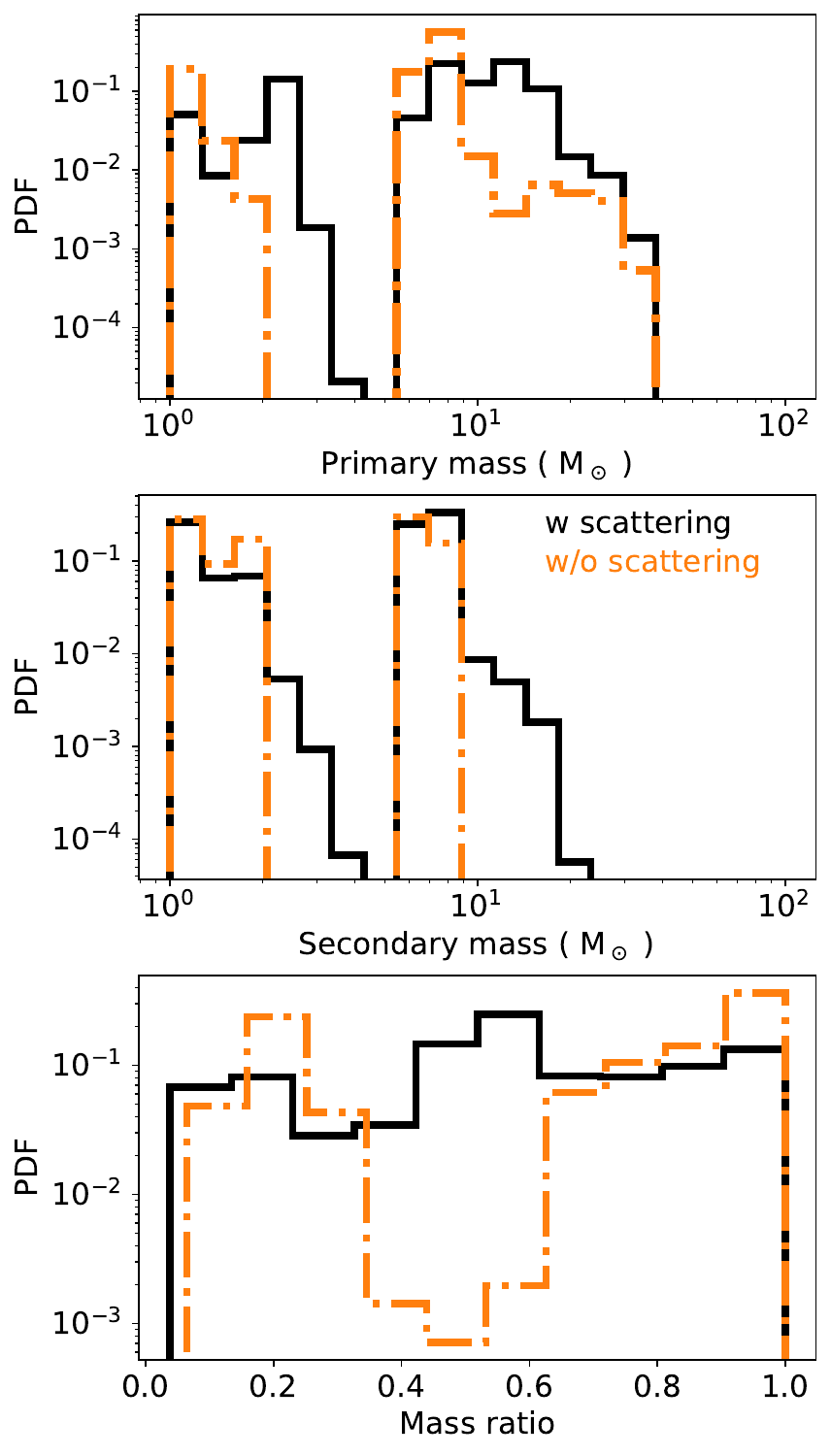}
\caption{ Similar to \SecondGenModelOneOneMass, here for Model 4 and $0<\chi_i<1$. }
\label{fig:2g_mod41_mass}
\end{figure}

\begin{figure}
\center
\includegraphics[scale=0.5]{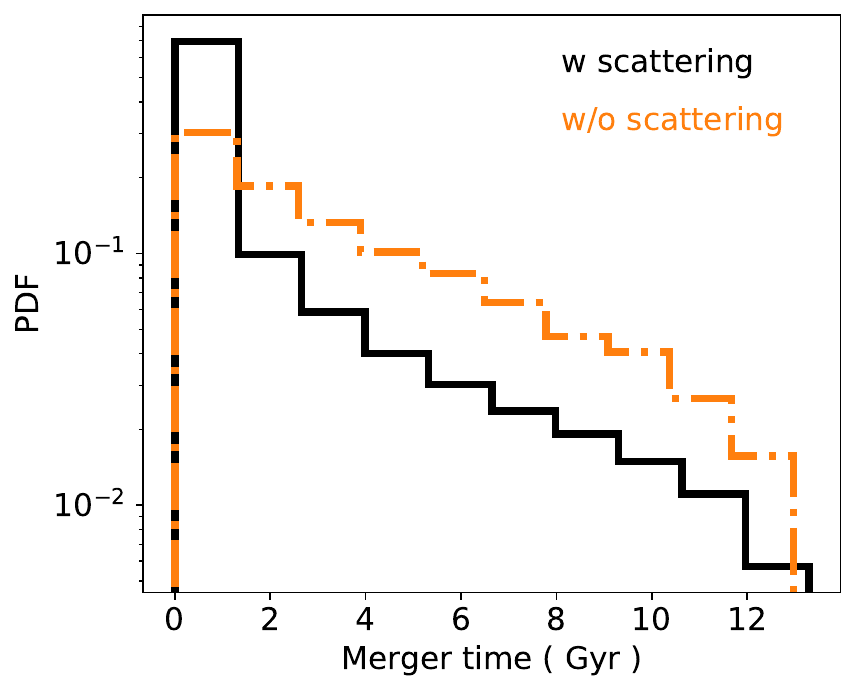}
\caption{ Similar to \SecondGenModelOneOneTime, here for Model 4 and $0<\chi_i<1$. }
\label{fig:2g_mod41_time}
\end{figure}

\begin{figure}
\center
\includegraphics[scale=0.5]{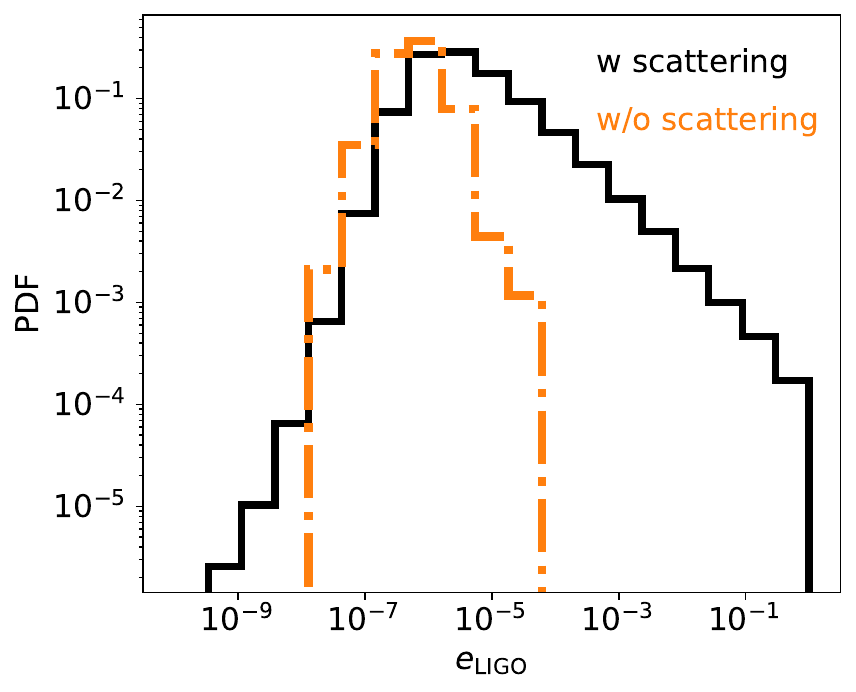}
\caption{ Similar to \SecondGenModelOneOneEcc, here for Model 4 and $0<\chi_i<1$. }
\label{fig:2g_mod41_ecc}
\end{figure}

\begin{figure}
\center
\includegraphics[scale=0.5]{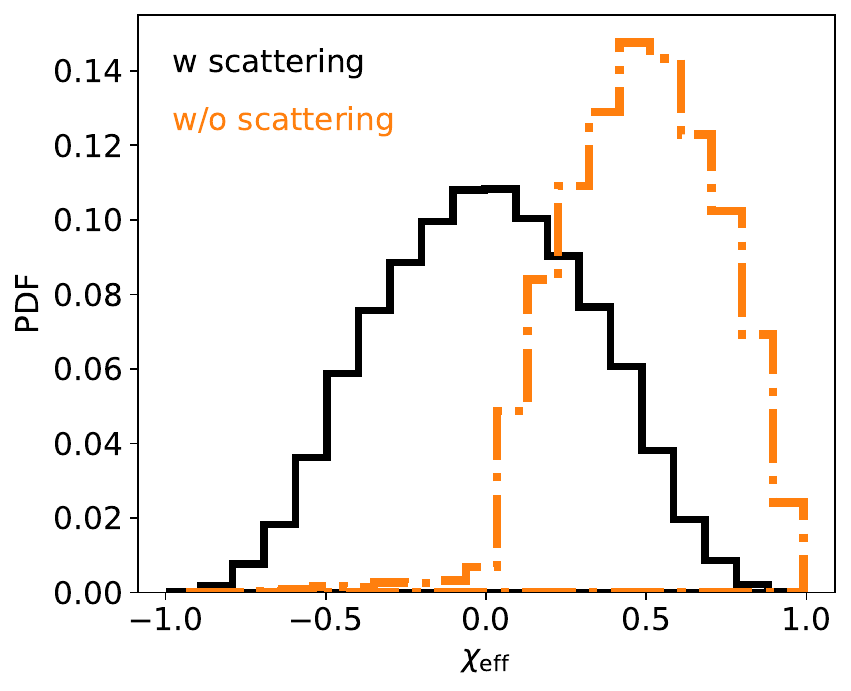}
\caption{ Similar to \SecondGenModelOneOneSpin, here for Model 4 and $0<\chi_i<1$. }
\label{fig:2g_mod41_spin}
\end{figure}

\begin{figure}
\center
\includegraphics[scale=0.55]{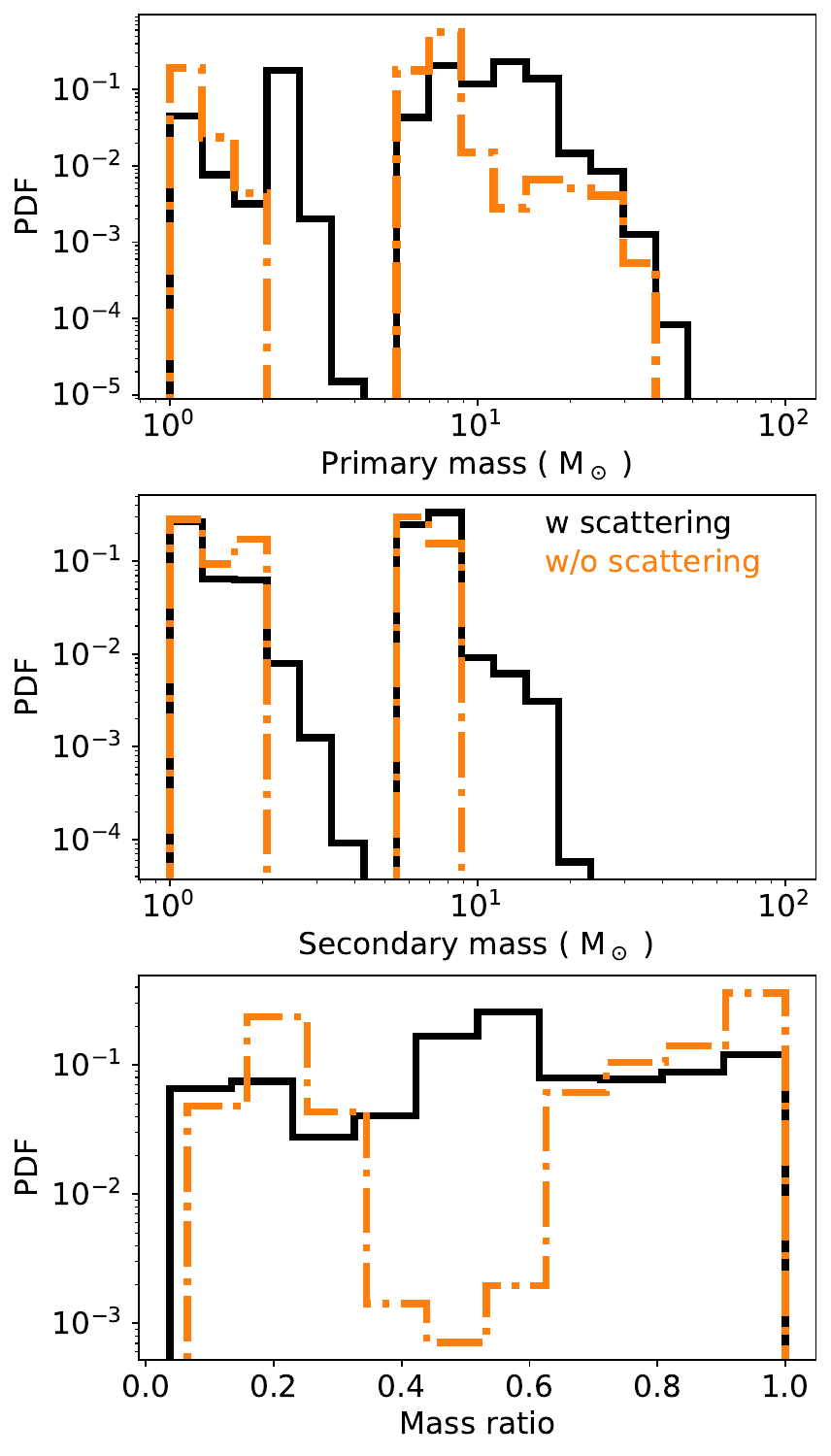}
\caption{ Similar to \SecondGenModelOneOneMass, here for Model 4 and $0<\chi_i<0.1$. }
\label{fig:2g_mod42_mass}
\end{figure}

\begin{figure}
\center
\includegraphics[scale=0.5]{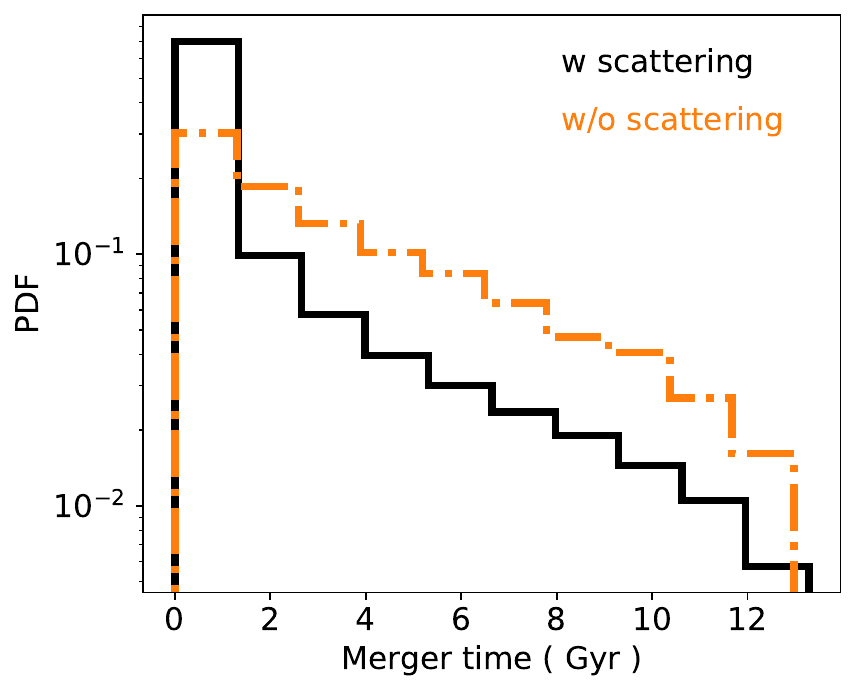}
\caption{ Similar to \SecondGenModelOneOneTime, here for Model 4 and $0<\chi_i<0.1$. }
\label{fig:2g_mod42_time}
\end{figure}

\begin{figure}
\center
\includegraphics[scale=0.5]{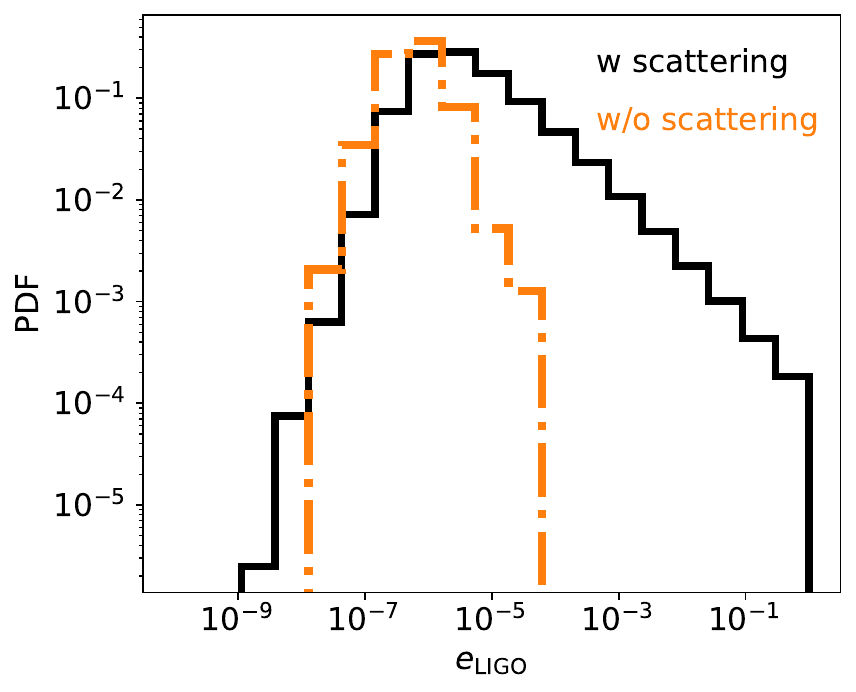}
\caption{ Similar to \SecondGenModelOneOneEcc, here for Model 4 and $0<\chi_i<0.1$. }
\label{fig:2g_mod42_ecc}
\end{figure}

\begin{figure}
\center
\includegraphics[scale=0.5]{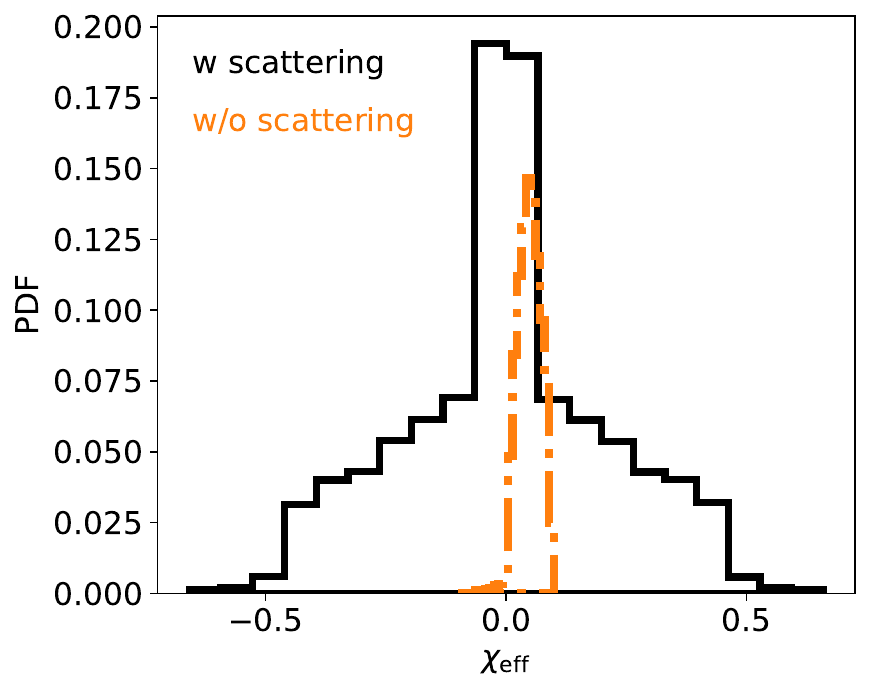}
\caption{ Similar to \SecondGenModelOneOneSpin, here for Model 4 and $0<\chi_i<0.1$. }
\label{fig:2g_mod42_spin}
\end{figure}

\begin{figure}
\center
\includegraphics[scale=0.55]{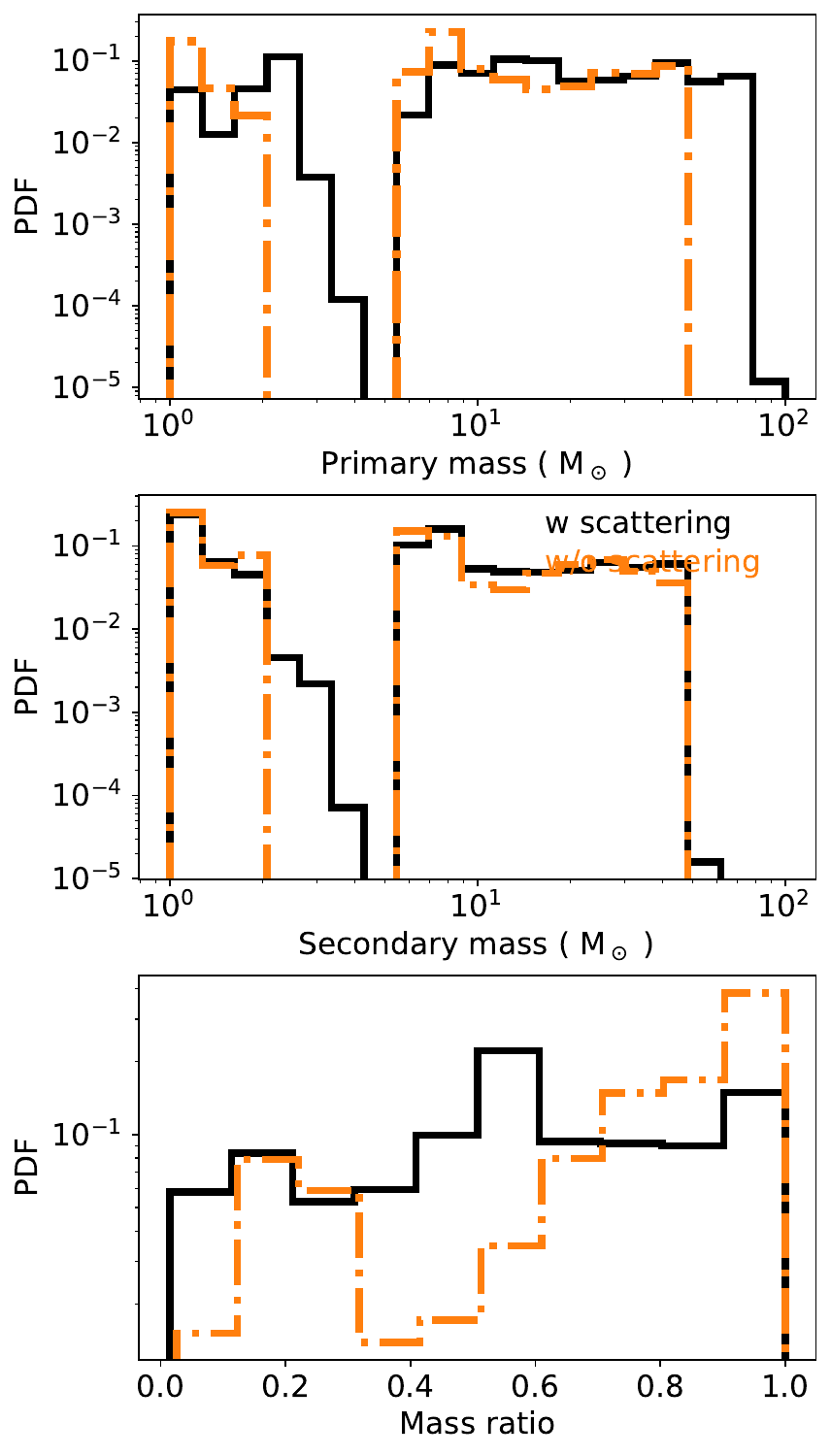}
\caption{ Similar to \SecondGenModelOneOneMass, here for Model 5 and $0<\chi_i<1$. }
\label{fig:2g_mod51_mass}
\end{figure}

\begin{figure}
\center
\includegraphics[scale=0.5]{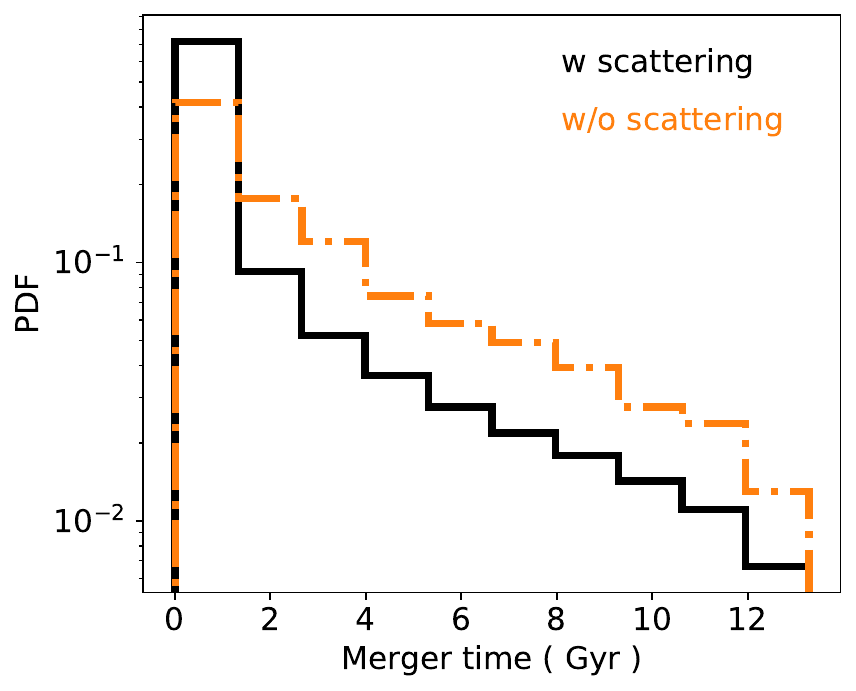}
\caption{ Similar to \SecondGenModelOneOneTime, here for Model 5 and $0<\chi_i<1$. }
\label{fig:2g_mod51_time}
\end{figure}

\begin{figure}
\center
\includegraphics[scale=0.5]{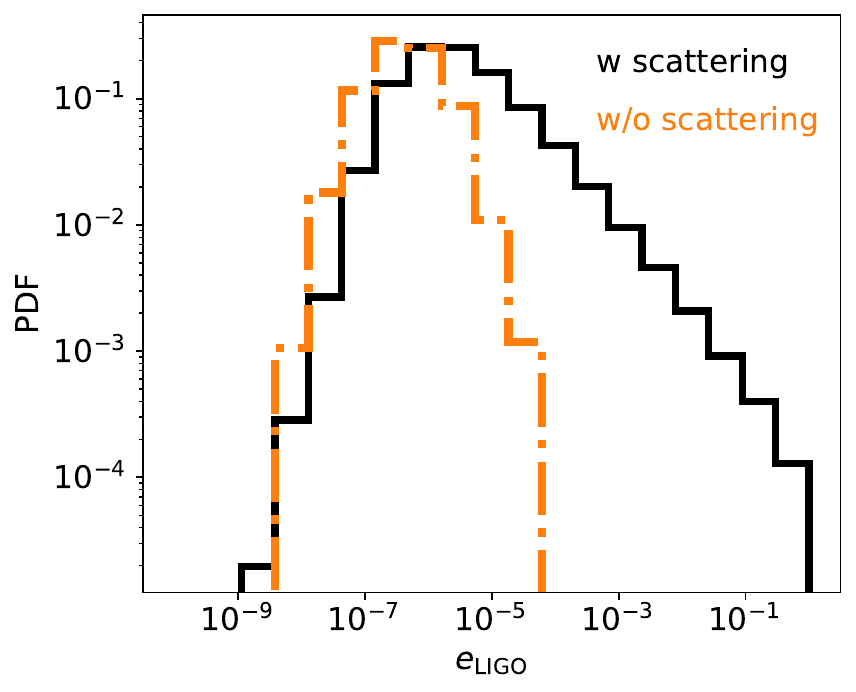}
\caption{ Similar to \SecondGenModelOneOneEcc, here for Model 5 and $0<\chi_i<1$. }
\label{fig:2g_mod51_ecc}
\end{figure}

\begin{figure}
\center
\includegraphics[scale=0.5]{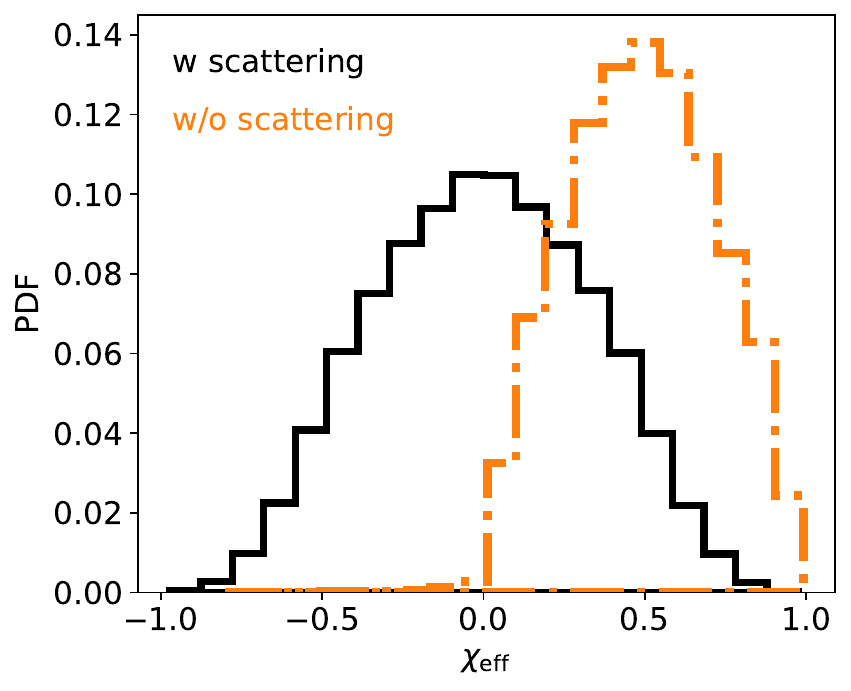}
\caption{ Similar to \SecondGenModelOneOneSpin, here for Model 5 and $0<\chi_i<1$. }
\label{fig:2g_mod51_spin}
\end{figure}

\begin{figure}
\center
\includegraphics[scale=0.55]{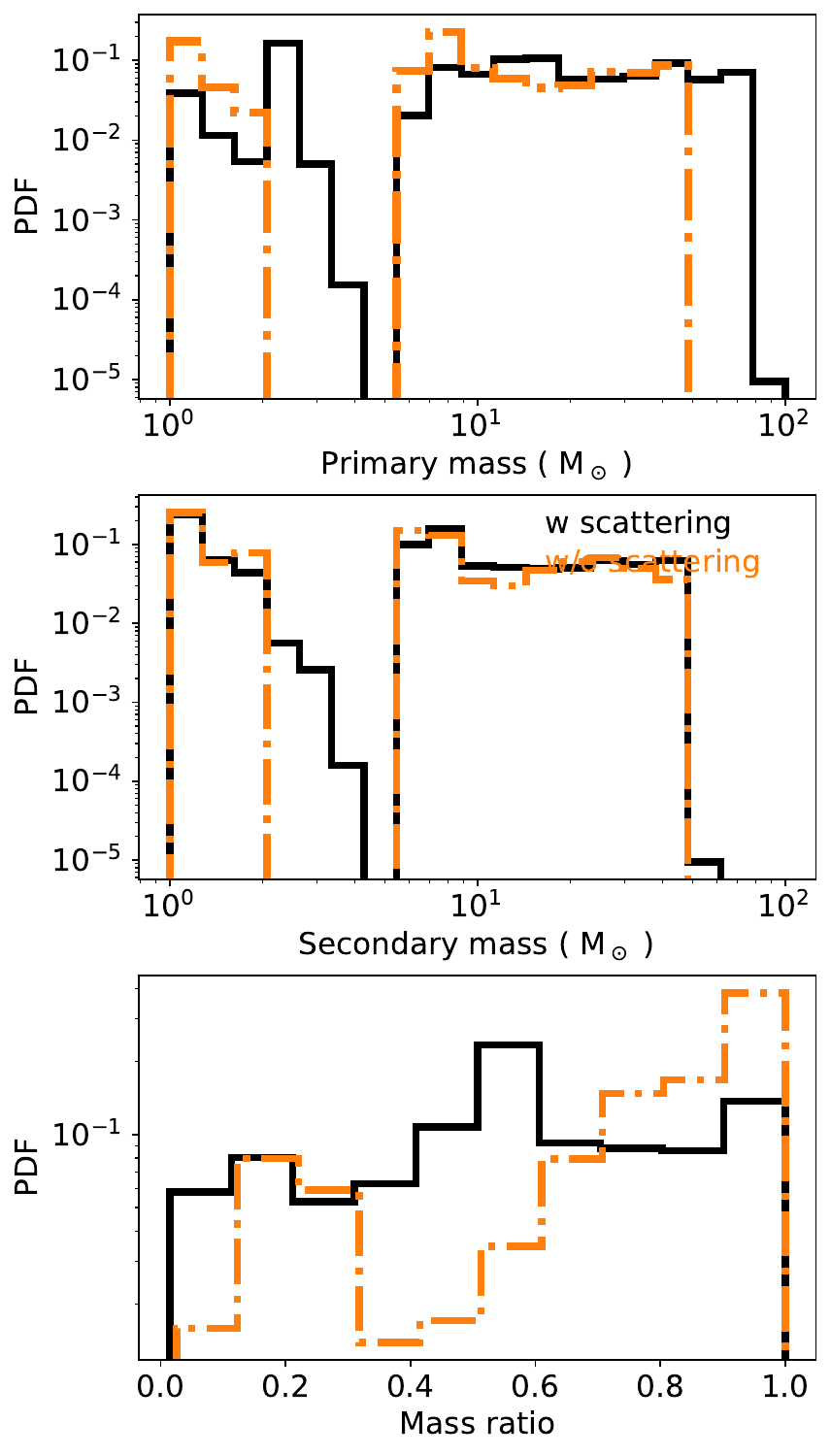}
\caption{ Similar to \SecondGenModelOneOneMass, here for Model 5 and $0<\chi_i<0.1$. }
\label{fig:2g_mod52_mass}
\end{figure}

\begin{figure}
\center
\includegraphics[scale=0.5]{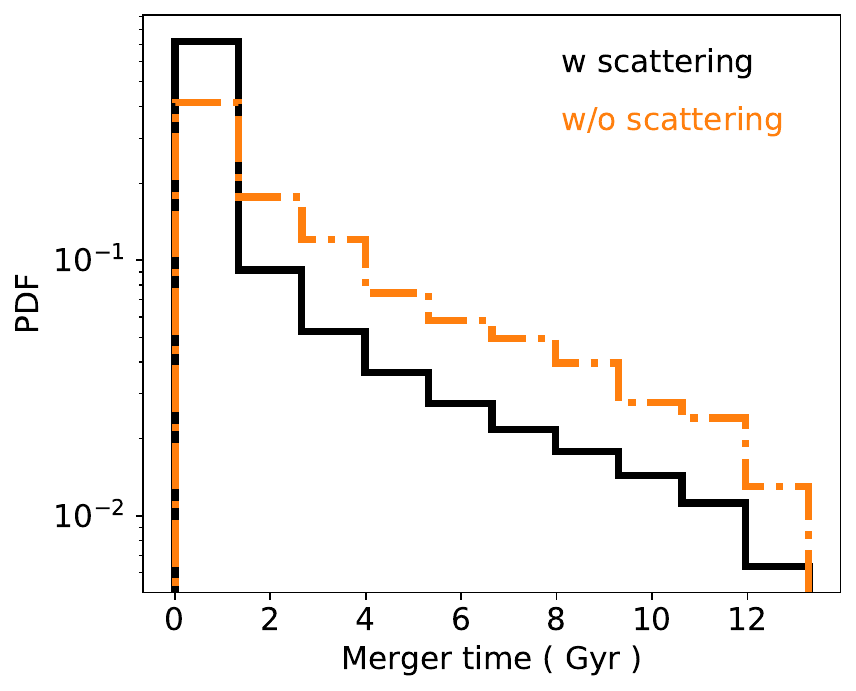}
\caption{ Similar to \SecondGenModelOneOneTime, here for Model 5 and $0<\chi_i<0.1$. }
\label{fig:2g_mod52_time}
\end{figure}

\begin{figure}
\center
\includegraphics[scale=0.5]{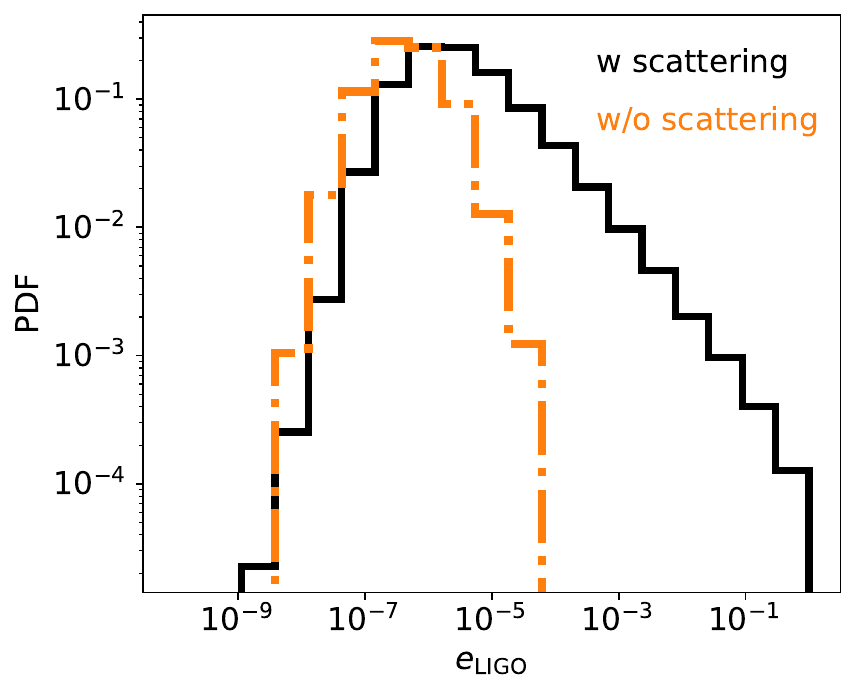}
\caption{ Similar to \SecondGenModelOneOneEcc, here for Model 5 and $0<\chi_i<0.1$. }
\label{fig:2g_mod52_ecc}
\end{figure}

\begin{figure}
\center
\includegraphics[scale=0.5]{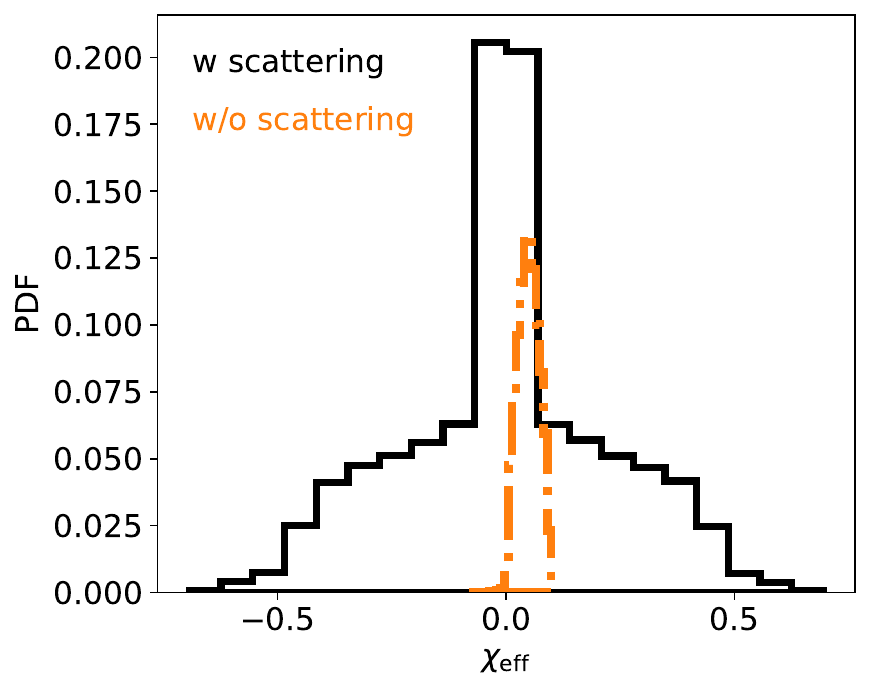}
\caption{ Similar to \SecondGenModelOneOneSpin, here for Model 5 and $0<\chi_i<0.1$. }
\label{fig:2g_mod52_spin}
\end{figure}

\begin{figure}
\center
\includegraphics[scale=0.55]{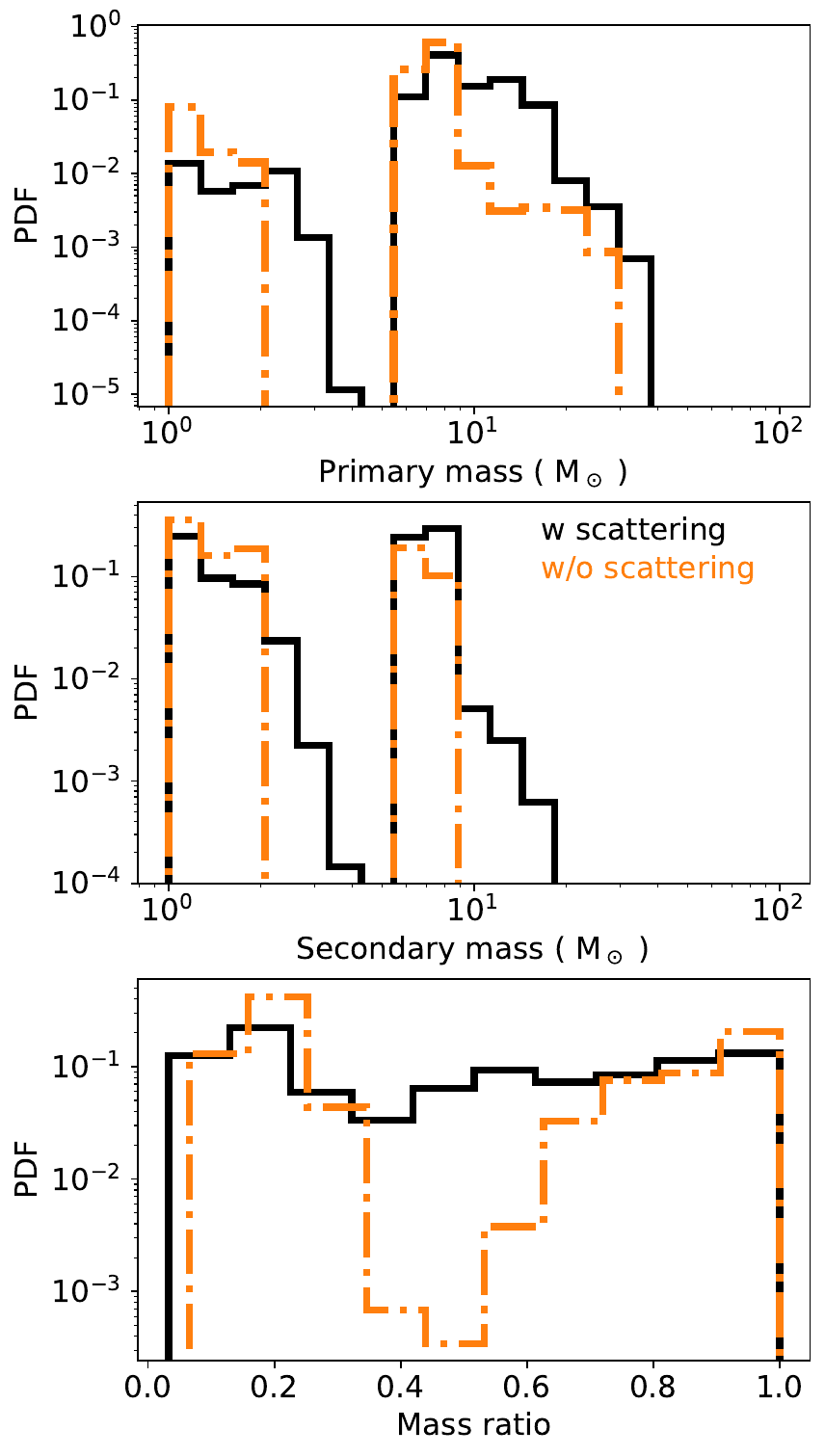}
\caption{ Similar to \SecondGenModelOneOneMass, here for Model 6 and $0<\chi_i<1$. }
\label{fig:2g_mod61_mass}
\end{figure}

\begin{figure}
\center
\includegraphics[scale=0.5]{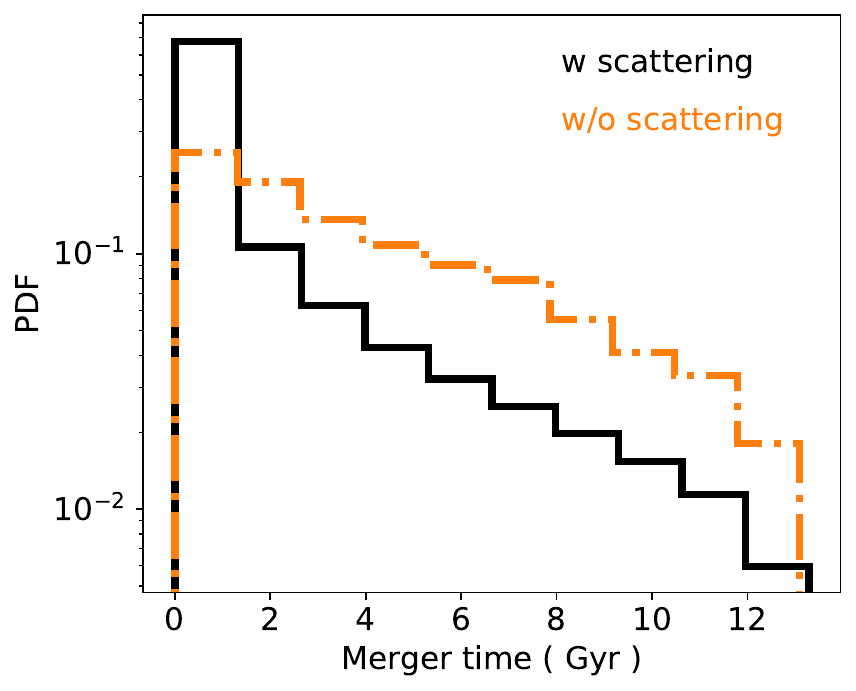}
\caption{ Similar to \SecondGenModelOneOneTime, here for Model 6 and $0<\chi_i<1$. }
\label{fig:2g_mod61_time}
\end{figure}

\begin{figure}
\center
\includegraphics[scale=0.5]{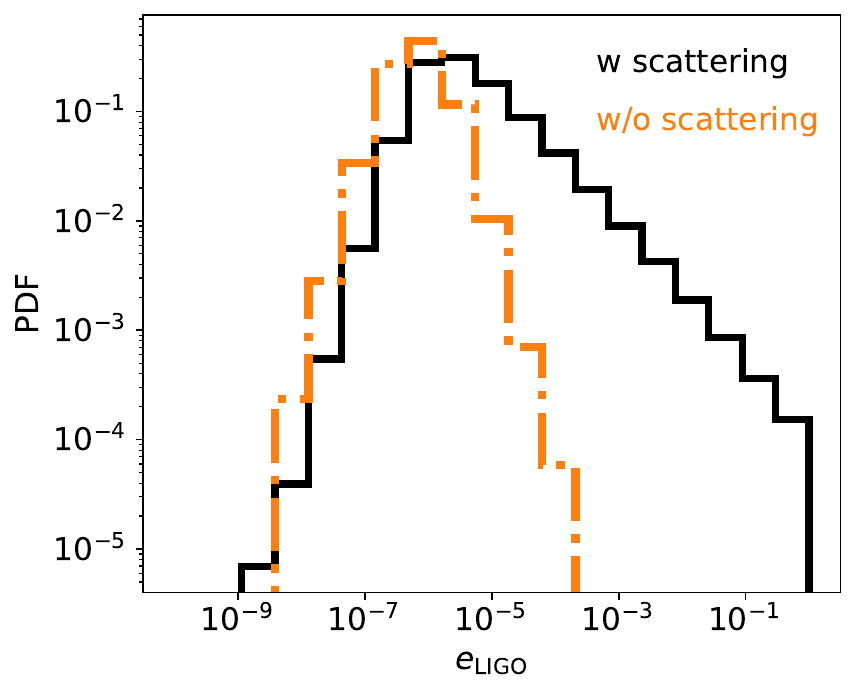}
\caption{ Similar to \SecondGenModelOneOneEcc, here for Model 6 and $0<\chi_i<1$. }
\label{fig:2g_mod61_ecc}
\end{figure}

\begin{figure}
\center
\includegraphics[scale=0.5]{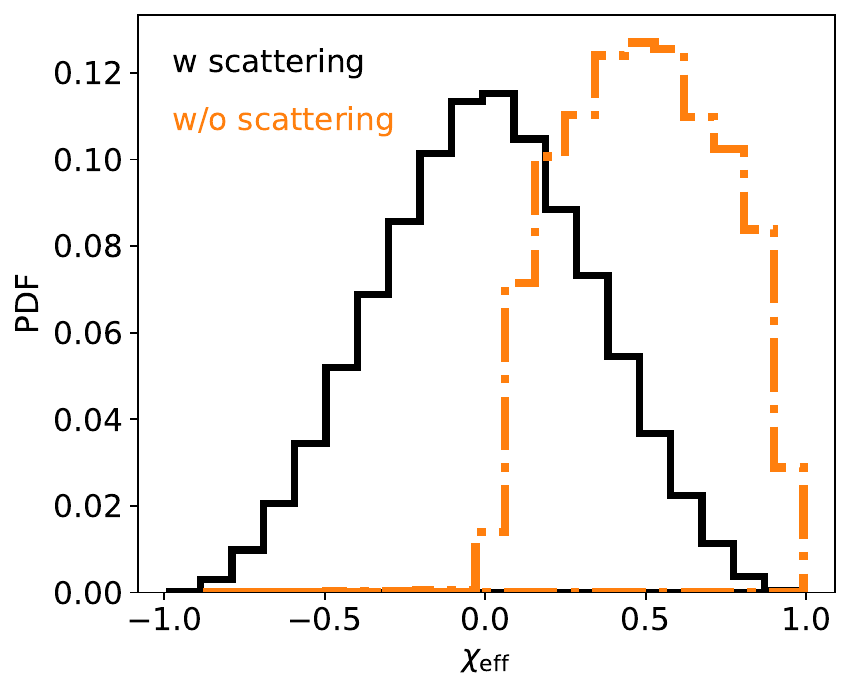}
\caption{ Similar to \SecondGenModelOneOneSpin, here for Model 6 and $0<\chi_i<1$. }
\label{fig:2g_mod61_spin}
\end{figure}

\begin{figure}
\center
\includegraphics[scale=0.55]{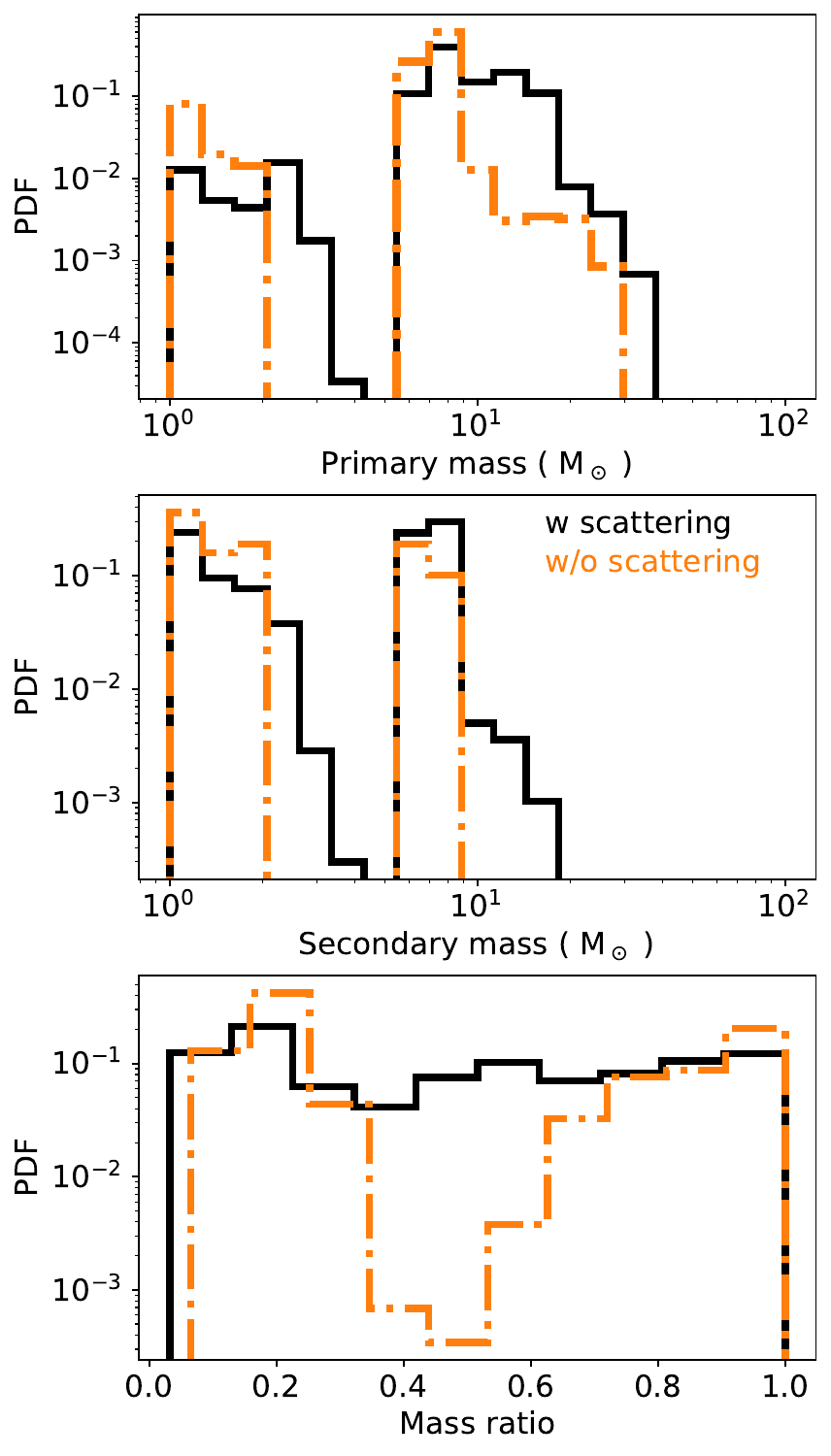}
\caption{ Similar to \SecondGenModelOneOneMass, here for Model 6 and $0<\chi_i<0.1$. }
\label{fig:2g_mod62_mass}
\end{figure}

\begin{figure}
\center
\includegraphics[scale=0.5]{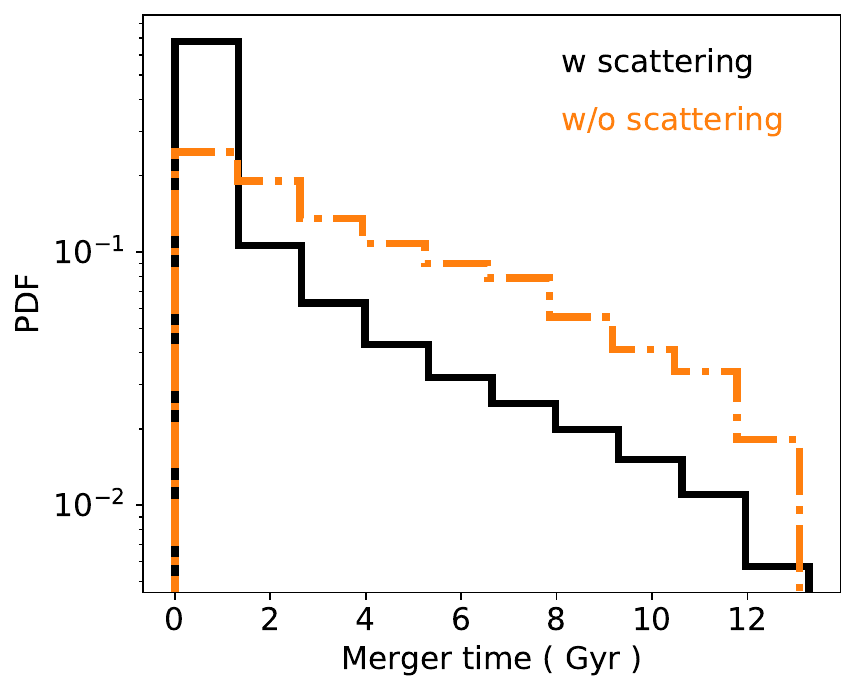}
\caption{ Similar to \SecondGenModelOneOneTime, here for Model 6 and $0<\chi_i<0.1$. }
\label{fig:2g_mod62_time}
\end{figure}

\begin{figure}
\center
\includegraphics[scale=0.5]{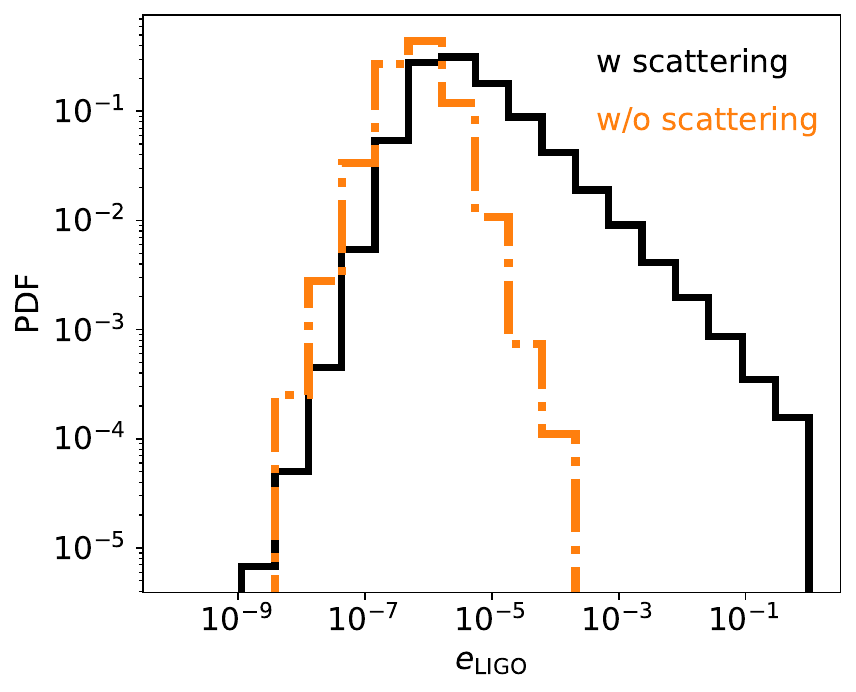}
\caption{ Similar to \SecondGenModelOneOneEcc, here for Model 6 and $0<\chi_i<0.1$. }
\label{fig:2g_mod62_ecc}
\end{figure}

\begin{figure}
\center
\includegraphics[scale=0.5]{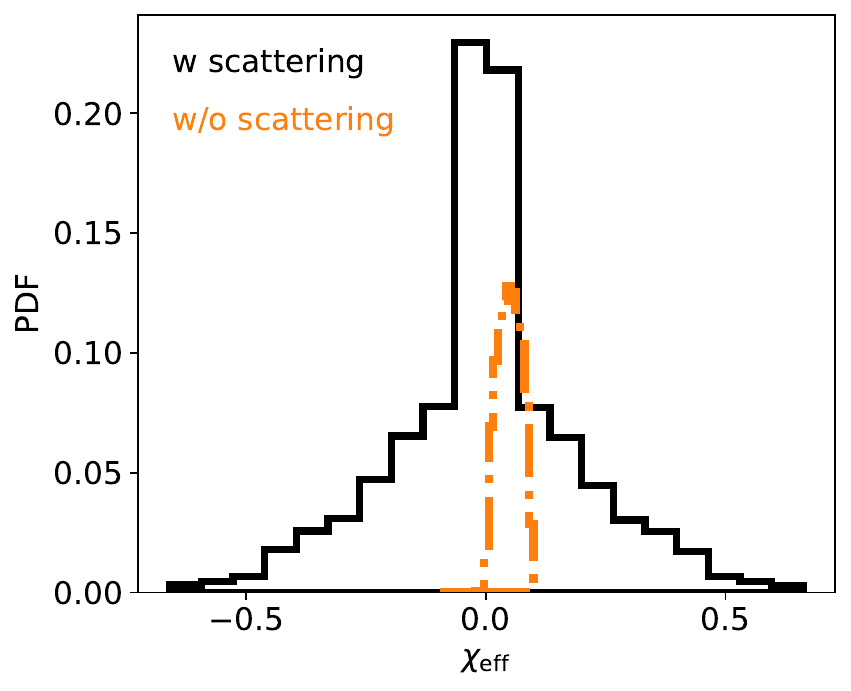}
\caption{ Similar to \SecondGenModelOneOneSpin, here for Model 6 and $0<\chi_i<0.1$. }
\label{fig:2g_mod62_spin}
\end{figure}

In \Fs~\ref{fig:2g_mod12_mass} through \ref{fig:2g_mod62_spin}, we show distributions of the masses, GW merger times, eccentricity in the LIGO band, and $\chi_\eff$ for models 2 through 6. These figures are complementary to \SectionSecond. 

\label{lastpage}